%% file: main.tex
\PassOptionsToPackage{table,dvipsnames}{xcolor}

\documentclass{thesisclass}

\import{Prestuff/}{Acronyms.tex}

\import{Prestuff/}{Personal.tex}

\begin{document}

\include{Prestuff/titlepage}


\include{Prestuff/abstract}

\include{Prestuff/Table_of_contents}
\include{Prestuff/List_of_figures}
\include{Prestuff/List_of_tables}
\include{Prestuff/List_of_acronyms}

\include{1Introduction}


\include{2Basics}

\include{3Method_of_investigation}

\include{4Experimental_Results}


\include{5Discussion}

\include{6Conclusion}


\include{Poststuff/References}


\include{Poststuff/Appendix}


\end{document}

%% file: Prestuff/Acronyms.tex
\makeglossaries

\newacronym{aert}{A-ERT}{automated-ERT}
\newacronym{ert}{ERT}{electrical resistivity tomography}
\newacronym{er}{ER}{electrical resistivity}
\newacronym{fdm}{FDM}{finite difference method}
\newacronym{rmse}{RMSE}{root-mean-squared error}
\newacronym{lma}{LMA}{Levenberg-Marquardt algorithm}
\newacronym{pcf}{PCF}{permafrost carbon feedback}
\newacronym{vwc}{VWC}{volumetric water content}
\newacronym{ec}{EC}{electrical conductivity}
\newacronym{mae}{MAE}{mean absolute error}
\newacronym{mre}{MRE}{mean relative error}
\newacronym{mare}{MaRE}{maximum relative error}
\newacronym{dem}{DEM}{digital elevation model}
\newacronym{rf}{RF}{random forest}

%% file: Prestuff/Personal.tex
\newcommand{\myname}{Constantin Schorling}
\newcommand{\birthcity}{Soltau (Germany)}
\newcommand{\mytitle}{Estimating volumetric water content from electrical resistivity using a random forest model}

\newcommand{\myinstitute}{Institute of Environmental Physics in Heidelberg in cooperation with the University Center in Svalbard (UNIS)}

\newcommand{\mytype}{Bachelor's Thesis in Physics} 
\newcommand{\reviewerone}{Prof. Dr. Norbert Frank}
\newcommand{\reviewertwo}{Prof. Dr. Werner Aeschbach}
\newcommand{\advisor}{Knut Ivar Lindland Tveit, MSc.}

\newcommand{\submissiontime}{2024}

\hypersetup{
 pdfauthor={\myname},
 pdftitle={\mytitle},
 pdfsubject={\mytype},
 pdfkeywords={\mytype}
}

%% file: Prestuff/titlepage.tex
\begin{titlepage}
\thispagestyle{empty}
\begin{center}
\Large\textbf{Department of Physics and Astronomy\\
Heidelberg University}\\
\vspace{12cm}
\normalsize
\mytype \\
submitted by\\
\vspace{0.5cm}
\Large\textbf{\myname}\\
\normalsize
\vspace{0.5cm}
born in \birthcity \\
\vspace{0.5cm}
\Large\textbf{\submissiontime}
\normalsize

\newpage\null\thispagestyle{empty}\newpage
\newpage\null\thispagestyle{empty}

\Large\textbf{\mytitle}\\
\vspace{12cm}
\normalsize
This Bachelor's Thesis has been carried out by \myname\ at the \\
\myinstitute\\
\begin{center}
    \begin{tabular}[ht]{l c l}
        Supervision and review: & \hfill  & \reviewerone \\
        Supervision and second review: & \hfill  & \reviewertwo \\
        Supervision: & \hfill  & \advisor \\
    \end{tabular}
\end{center}
\vfill
\end{center}

\pagenumbering{gobble}
\end{titlepage}

\newpage\null\thispagestyle{empty}\newpage

%% file: Prestuff/Table_of_contents.tex
\begingroup
\pagestyle{plain}

\tableofcontents

\endgroup

%% file: Prestuff/List_of_figures.tex
\begingroup
\pagestyle{plain}
\renewcommand{\listfigurename}{Figures}
\listoffigures

\endgroup

%% file: Prestuff/List_of_tables.tex
\begingroup

\pagestyle{plain}

\renewcommand{\listtablename}{Tables}
\listoftables

\endgroup

%% file: Prestuff/List_of_acronyms.tex




%% file: 1Introduction.tex
\chapter{Introduction}
\label{ch:introduction}
\pagenumbering{arabic}

Rising global temperatures pose severe challenges, and the necessity for adaptation is becoming increasingly apparent \cite{29_Rantanen_2022_arctic_warming}. This is particularly evident for Arctic regions, where the mean temperature increase is 3.8 times as high as observed globally between 1971-2021 \cite{29_Rantanen_2022_arctic_warming,0_jonassen2023improved}. This yields an enhanced thawing of the prevailing permafrost. The active-layer -- the upper most part of permafrost that seasonally freezes and thaws -- thickens while the permafrost body underneath continuously melts \cite{24_anisimov_1995_pemrafrostinclimatechange,25_Christinansen_2003_activelayer}. The permafrost shrinking presents challenges both globally and locally. The increased risk of debris flows and landslides threatens communities and infrastructure in its immediate vicinity \cite{0_PermaMeteoCommunity}. 
\newline
Understanding and predicting impacts of these changes require a comprehensive assessment and monitoring of sub-surface structures. Today, there is a number of different geophysical methods employed to investigate lithosphere structures. Electrical resistivity tomography (ERT) can be used to indirectly obtain hydrological information of large areas. A two-dimensional resistance cross-section is generated, which can then be associated with hydrogeological properties calibrated with nearby boreholes and sensors \cite{3_knödel_2005_basics, 36_meng2024rf_ert_boss}. Especially for distinguishing active-layer and ground ice in permafrost this minimally invasive method can be useful as frozen and unfrozen areas vary in resistance in several orders of magnitude \cite{1_herring_2023_bestpractice}. Furthermore, there is a strong correlation between \acrfull{vwc}, which is the ratio of water volume divided by the total sample volume, and \acrshort{er} offering the possibility to indirectly estimate \acrshort{vwc} in a more cost-effective and extensive way \cite{36_meng2024rf_ert_boss, 35_Wicki_2022_landslides}. 
\newline
This thesis is part of the PermaMeteoCommunity Project \cite{0_PermaMeteoCommunity} and focuses on the investigation of sub-surface structures using \acrshort{ert} in Longyearbyen, Svalbard, an Arctic settlement situated at 78°N 15°E \cite{0_NPI2014terreng_map}. The objective of the PermaMeteoCommunity Project is to develop a coupled permafrost and meteorological climate change response system, yielding an enhanced understanding and ability to predict and respond to permafrost-related changes in an Arctic environment. 
\newline
The scope of the thesis includes a systematic contribution of \acrshort{ert} data observing active layer development as well as a novel approach to estimate and predict \acrshort{vwc} from \acrshort{er} measurements (sensor and tomography) for enhanced landslide hazard assessment. 

%% file: 2Basics.tex
\chapter{Basics}
\label{ch:basics}

\section{Permafrost characteristics}
Sub-surface material and soil is considered permafrost if it remains at or below 0~°C for at least two consecutive years \cite{24_anisimov_1995_pemrafrostinclimatechange,SESS_2019}. The state can therefore solely be described thermally. As the temperatures, especially in polar latitudes, continue to rise \cite{0_jonassen2023improved,13_Ranganathan_2004_LMA}, the permafrost proceeds to thaw \cite{18_hu_2021_temp_rise_permafrost}. The stability of permafrost is determined almost exclusively by the ice content of the soil whereas the active layer -- the layer above the permafrost which seasonally thaws and freezes -- is also prone to \acrfull{vwc} level elevations and sudden increases followed by subsidence land detachments \cite{0_bogaard2016landslide}. 
It is therefore a vital endeavor to quantify water and ice content seasonally, spatially and directly after heavy rainfalls \cite{35_Wicki_2022_landslides, 0_bogaard2016landslide, 11_Tomaskovicova_2023_timelaps_ert} to detect areas of instability. 


\subsection{Permafrost structure}
Vertically, permafrost soil can be characterized by three zones \cite{19_Schirrmeister_2012_permafrost_charakterisierung}: An upper seasonally freezing and thawing active layer, a transition zone underneath, in which the mean temperature is slightly negative but fluctuating, and an isothermic (perennially below zero) permafrost body. Below the lower limit of the permafrost base is a subpermafrost region with low positive temperatures. 
\newline
The depth of the \textbf{active layer} is typically influenced by ground temperatures, but also several other environmental factors are relevant including topography, prolonging snow cover, vegetation, soil moisture and material \cite{19_Schirrmeister_2012_permafrost_charakterisierung}. The active layer thickness in Longyearbyen reaches up to 200-300~cm  \cite{25_Christinansen_2003_activelayer,32_hanssen_2019_climate_svalbard}.
Christiansen et. al \cite{0_christiansen2020permafrost} showed also for two sites in Svalbard (Adventdalen area and Ny-Ålesund) a continuous increase (0.6-1.6~cm/a, 5~cm/a) in the active layer thickness over the last 30 years. Another article \cite{25_Christinansen_2003_activelayer} suggested a high influence of local parameters on the active layer thickness. 
\newline
If a negative heat balance persists perennially, aggradation of the \textbf{transition zone} and also the \textbf{ground ice} increases \cite{18_hu_2021_temp_rise_permafrost}. Furthermore, a mean annual temperature decrease for deeper layers is observed \cite{0_christiansen2020permafrost}. Conversely, the permafrost volume decreases and its temperature increases with a positive heat balance. Thus, the shrinking or growing rate heavily depends on the annual mean heat balance \cite{17_Dobinski_2011_Permafrost}. Nevertheless, also other physical properties as heat capacity and permeability or a potential snow cover are significant \cite{17_Dobinski_2011_Permafrost}. 

\subsection{Physical properties}
Besides the stability, also other physical parameters of the sub-surface structures are heavily influenced by the active layer \acrshort{vwc}. In the following section, the key parameter for observing water and ice as well as geological distributions with \acrfull{ert} will be discussed. 

\subsubsection*{Electrical conductivity and resistivity}
The \acrfull{ec} $\sigma$ or resistivity (ER) $\rho$ (reciprocal) of the soil is generally determined by three factors \cite{3_knödel_2005_basics}.  
The electronic or metallic matrix conductivity (1) is influenced by the presence of free electrons within rocks (mostly ore minerals or carbons such as graphite). The in situ accumulation of water and ions from dissolved salts within the rock pores results in an electrolytical contribution (2) to the conductivity \cite{2_matthes_2012_ertPhD}.  Conversely, the presence of air acts as an isolator, thereby reducing the conductivity \cite{3_knödel_2005_basics}. The conductivity of clay-free rocks can be well approximated by the Archie's formula \cite{28_archie_1942_beginning_ert}: 
\begin{equation}
    \rho_0 = F \cdot \rho_w \cdot S^{-n}
    \label{eq:archie}
\end{equation}
where 
\begin{conditions}
    \rho_{0/w} & electrical resistivity of the sediment/pore water \\
     F  & formation factor \\
     S/n  & degree/exponent of saturation
    factor.
\end{conditions}
For clay rich conditions, a double layer forms between the pore water and the rock yielding current flow on the grain surfaces \cite{3_knödel_2005_basics, 2_matthes_2012_ertPhD}. An enhanced electrical conductivity due to interface contributions (3) is measured. 
\newline

\subsubsection*{Rain, volumetric water content (VWC) and landslide risk} 
Heavy rainfall and subsequent infiltration can trigger dangerous slope failure due to a decrease of shear strength below a certain threshold \cite{0_bogaard2016landslide, 35_Wicki_2022_landslides, 0_patton2019landslide_permafrost}. Furthermore hydrological preconditions such as elevated moisture levels may reduce this threshold \cite{0_bogaard2016landslide}. One way to measure moisture is based on the \acrfull{vwc}. Piciullo et al. \cite{0_segoni2018review_landslides} summarized precipitation intensity, \acrshort{vwc} elevation and sudden increases for various landslides. These thresholds are often multidimensional and require a sophisticated meteorological and soil observation.

\section{Electrical resistivity tomography (ERT)}

\subsection{Physical background}
\acrshort{ert} is one of the most commonly used geophysical techniques \cite{1_herring_2023_bestpractice} and exploits the great heterogeneity of permafrost and active layer using basic electrical phenomena. In particular, the potential differences $\Delta \Phi$ and thus the apparent resistivity $\rho$ is measured between metallic electrodes penetrating the ground minimally invasive with direct current \cite{3_knödel_2005_basics}. As the electrical field is static and vortex-free ($rot \vec{E}= 0 $) it is also a gradient field ($\vec{E} = - \nabla \Phi$) with a scalar potential $\Phi$. The electrical field can be expressed with the electrical density $\vec{j}$ or the electrical resistance $\sigma$ as $\vec{E} = \frac{1}{\sigma}\vec{j} = \rho \vec{j}$. With this, the continuity equation:  
\begin{equation}
    div\vec{j} + \frac{\partial \rho_{el}}{\partial t} = 0
\end{equation}
yields 
\begin{equation}
    -\nabla \cdot [\sigma (x, y, z)\nabla \Phi (x, y, z)] = I\delta (x) \delta (y) \delta (z)
    \label{eq:potentialdistribution}
\end{equation}
for a point source electrode at $(0, 0, 0)$ \cite{0_loke1996quasi_newton_first}. Assuming a homogeneous, endless ground confined by a straight surface (yielding a perfect half space), the radial potential between a feeding electrode A and a measuring electrode M is given by
\begin{equation}
    \Phi (r) = \rho \frac{I}{2 \pi r}
    \label{eq:pot_hom_soil}
\end{equation}
with the distance $\overline{AM} = r$ \cite{3_knödel_2005_basics}. For common setups a current ($\pm I$) is applied between two feeding electrodes A and B while the potential difference (voltage) is measured between two other electrodes M and N (see Fig.~ \ref{fig:wenner_basics} A). All individual potentials superimpose and result in the Neumann equation \cite{3_knödel_2005_basics}
\begin{equation}
    U = \Delta \Phi = \rho I \left[\frac{1}{2\pi}\left(\left(\frac{1}{r_1}-\frac{1}{r_3}\right)-\left(\frac{1}{r_2}-\frac{1}{r_4}\right)\right)\right] = \frac{\rho I}{K}.
\end{equation}
Setting the distance dependent term $K$ and current $I$ and measuring the voltage $U$ yields the resistivity $\rho$
\begin{equation}
    \rho = \frac{U K}{I}.
\end{equation}
For simplicity, a point shaped electrode was assumed \cite{2_matthes_2012_ertPhD}. The potential for a rod current source requires a small correction, however this is negligible at distances exceeding  50~cm (<~5\% deviation) \cite{26_Oeser_Skript_lesen_2020}.
\newline
Soil inhomogeneities and the transition from active layer to permafrost result in different electrical conductivities and therefore change equation \ref{eq:pot_hom_soil}. The current is refracted at material boundaries according to the general law of refraction $\frac{\rho_1}{\rho_2} = \frac{tan(\theta_1)}{tan(\theta_2)}$ \cite{3_knödel_2005_basics}. 

\subsection*{Wenner array}
\begin{figure}[h!]
    \centering
    \includegraphics[width=\linewidth]{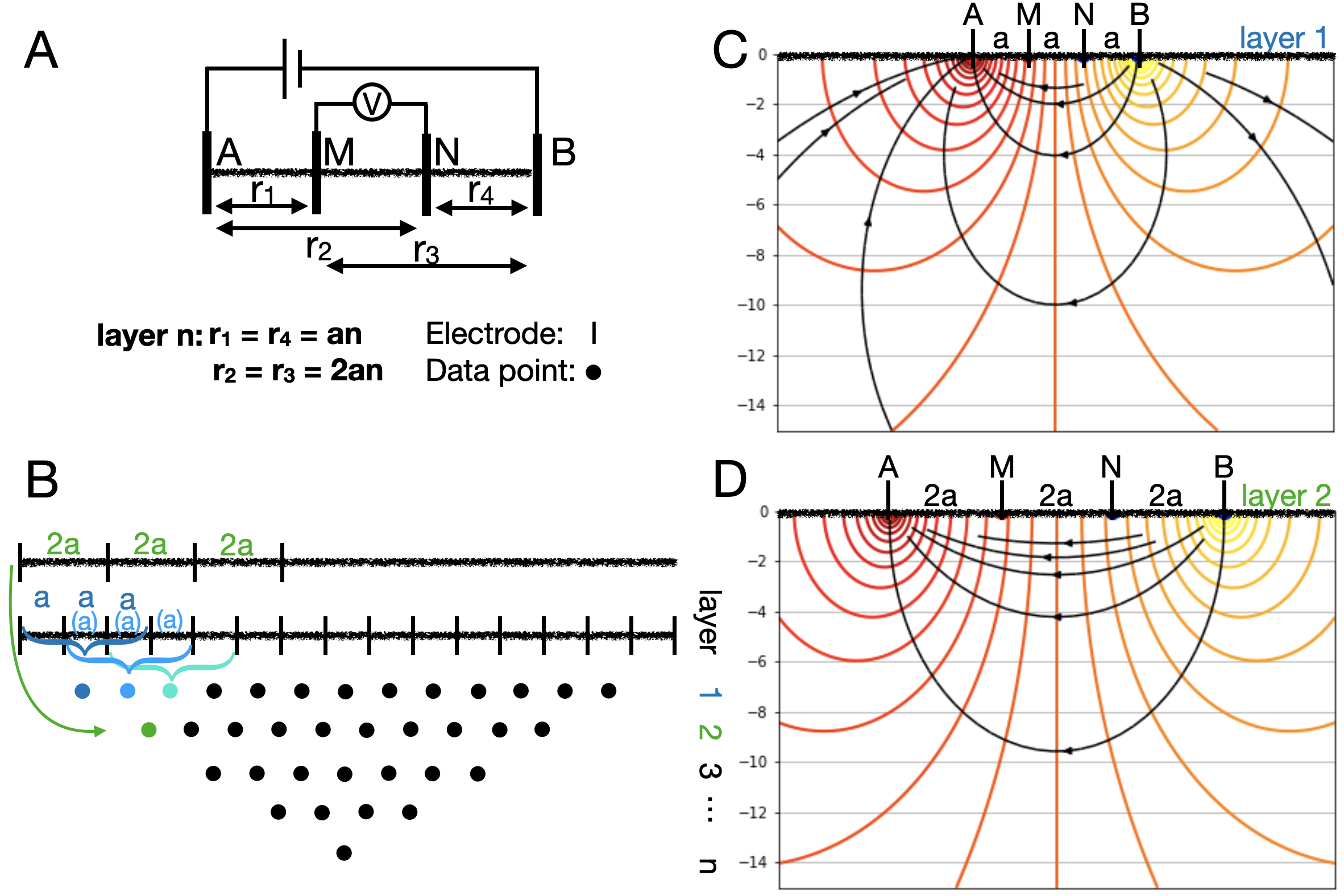}
    \caption[\acrshort{ert} basics and background]{A characterization of Wenner array under assumption of homogeneous ground; A: Schematic of the feeding and measuring electrodes and intermediate distances, B: The overall data acquisition and penetration depth, C: The current flow lines (black) and equipotential lines (heat colors) for layer one with electrode spacings of a, D: Layer two, electrode spacing: 2a; A is adapted from \cite{2_matthes_2012_ertPhD} and B from \cite{3_knödel_2005_basics}.}
    \label{fig:wenner_basics}
\end{figure}

The electrodes can be controlled individually depending on the desired resolution. In this thesis, a Wenner array was chosen as it provides a high Signal-to-Noise-Ratio and layer resolution \cite{3_knödel_2005_basics}. The feeding electrodes (A, B) surround the measurement electrodes (M, N) with an equidistant electrode spacing $r = n \cdot a$ ($n \in \mathbb{N}$) for all electrodes (see Fig.~ \ref{fig:wenner_basics} A). Given that $r_1 = r_4 = a n$ and $r_2 = r_3 = 2 a n$, this yields $\rho = 2\pi a \frac{U}{I}$ \cite{2_matthes_2012_ertPhD}. 

After the first measurement using the first four electrodes, the feeding and read out is shifted altogether by one electrode. Shifting the electrode tasks is continued until the last electrode (layer 1 is then finished). Thereafter, the distance between each utilized electrode is increased by $a$, yielding an electrode spacing of $2a$ for the second layer (see Fig.~ \ref{fig:wenner_basics} B). This is continued until the $\overline{AB}$ spacing is larger than the array and the lowest reachable layer $n$ is observed. In Fig.~ \ref{fig:wenner_basics} C and D for the first two layers the current flow lines and equipotential lines are displayed in black and heat colors, respectively. The final result is obtained by combining all separate measurements constructively \cite{3_knödel_2005_basics}. The Wenner array is superior in lateral resolution compared to other array types \cite{3_knödel_2005_basics}. However, it lacks depths resolution. The initial electrode spacing can be varied to gain near-surface resolution with shorter spacing or deeper investigations with wider spacing \cite{1_herring_2023_bestpractice}. The exploration depth is constrained by the array length and can be approximated for the Wenner array as $0.17 \overline{AB}$ \cite{3_knödel_2005_basics}. Other commonly used arrays like Dipole-dipole, Wenner-Schlumberger or Gradient feature different sensitivity profiles and penetration depths as they have other electrode configurations \cite{1_herring_2023_bestpractice}.

\subsection*{Data acquisition prerequisites}
To generate statistically relevant data, a series of measurements is usually recorded successive as a stack \cite{1_herring_2023_bestpractice} avoiding electrode location errors because of repositioning. 
\newline
There are multiple prerequisites for obtaining useful data. After choosing a specific \acrshort{ert} array length and type, all electrode positions need to be recorded with GPS to facilitate further data evaluation \cite{3_knödel_2005_basics,1_herring_2023_bestpractice}. Other external influences as topography, meteorological circumstances and surface cover must be addressed and considered for evaluation. When measuring the resistivity, it is crucial to guarantee a sufficient galvanic coupling between all electrodes and the soil \cite{5_doetsch_2015_ert}. Depending on the above mentioned factors it can be challenging to secure this. Prior knowledge of soil conditions is generally not required, but may be valuable for selecting appropriate currents to produce voltage magnitudes within a detectable range \cite{1_herring_2023_bestpractice}. 

\subsection{Data processing}
The acquired data is first arranged in a pseudosection \cite{0_loke_1996_ert_inversion}, a rough schematic of the sub-surface resistivity profile. To obtain a more accurate description of the resistivity and potential soil composition, a series of data analysis must be conducted.

\subsubsection*{Data filtering}
The propagation of errors introduced by systematic measurement faults is mitigated using three filtering processes \cite{1_herring_2023_bestpractice}. 
In the first step, all data points indicating negative voltages or currents are removed. Furthermore, if multiple surveys were performed (stacking), data revealing a variance between the measurements above an adjustable threshold are also eliminated (e.g. >~5\% \cite{35_Wicki_2022_landslides}). The same applies to data that deviate very strongly from all others. Thresholds are often chosen to be around 9 times the standard deviation \cite{1_herring_2023_bestpractice}.
Secondly, at each depth a horizontally sliding array of each five values is used to calculate for every position the mean value and detect and delete anomalies. The threshold for acceptable deviations from the mean is approx. 5-10\% \cite{1_herring_2023_bestpractice}. 
Lastly, an electrode filter eliminates all data points associated with electrodes, from which in previous steps, much data was removed (above an adjustable percentage).

\subsubsection*{Inversion}
After filtering the measured data a first resistivity model (step one) is generated. A heuristic approach is employed setting an initially homogeneous model $m_0$ modified by prior knowledge \cite{2_matthes_2012_ertPhD}. Thereafter, with forward modeling (step two), the partial differential equation \ref{eq:potentialdistribution} is solved numerically using the \acrfull{fdm} \cite{0_virieux2011fdm} leading to an ambiguous pseudosection and adapted model parameter $m$. 
\newline
Furthermore, an objective function $\phi (m)$ (step three) is calculated incorporating both, the data misfit $\phi_d$ and the model misfit $\phi_m$ \cite{1_herring_2023_bestpractice}. A deviation description from model to the data ($\phi_d$) is combined with a smoothness constraint ($\phi_m$) ensuring realistic values. A regularization parameter $\beta$ is introduced to allow for a weighting of the constraint \cite{1_herring_2023_bestpractice}. 
\begin{align}
    \phi (m) & = \phi_{d}(m) + \beta \phi_{m}(m_0, m) \\
    & = \left\lVert d_{meas} - d_{model}(m) \right\rVert_p + \beta \left\lVert  d_{model}(m_0) - d_{model}(m) \right\rVert_q
\end{align}
The data is delineated by $d$, while $p$ and $q$ describe the order of the norm. L1, L2 or a combination are the common for both \cite{2_matthes_2012_ertPhD, 30_glazer_2020_svalbard, 27_Farquharson_1998_nonlinear_inversion}. Loke et al. \cite{9_loke_2003_l1norm_sharpboundaries} argued that employing an L1 norm for data misfit modeling yields sharper boundaries as it penalizes outliers less. Furthermore, Herring et al. \cite{1_herring_2023_bestpractice} conclude its benefits especially for permafrost investigations as the lateral boundaries of frozen and unfrozen ground is considered sharp. This was also shown by Glazer et al. \cite{30_glazer_2020_svalbard} who investigated spatial permafrost distributions in Spitsbergen. 
\newline
To minimize $\phi (m)$ the model parameters $m$ are varied (step four). Commonly, a form of the gradient based Gauss-Newton algorithm or \acrfull{lma} \cite{2_matthes_2012_ertPhD,13_Ranganathan_2004_LMA,10_loke_2002_newton_vs_quasi_newton,12_gavin_2024_LM} is used to update the parameters. The \acrshort{lma} shifts the initial parameters depending on the Jacobian matrix $J_r$ and residuum $r$.
\begin{align}
    m_{k+1} & = m_k - (J_r^TJ_r + \lambda I)^{-1}J_r^T r (m_k)\\
    r & = d_{meas} - d_{model}(m)
\end{align}
Moreover, a damping coefficient $\lambda$ is introduced yielding the Gauss-Newton algorithm for small values of $\lambda$ and an iterative gradient descent for high values \cite{12_gavin_2024_LM}. Thus, $\lambda$ should initially be chosen large and decreasing with lower model-data-deviations \cite{13_Ranganathan_2004_LMA,12_gavin_2024_LM}. 
\newline
Calculating Jacobian matrices for each iteration of $m_k$ is time-consuming. Loke and Barker \cite{0_loke1996quasi_newton_first} introduced a so called quasi-Newton method to reduce the computation significantly. First, the Jacobian matrix $J_0$ for the initial homogeneous model $m_0$ is solved analytically. All following Jacobians are approximated iteratively, thereby obviating the necessity to solve all differentials exactly \cite{0_loke1996quasi_newton_first}. However, especially for large resistivity contrasts Loke and Dahlin \cite{10_loke_2002_newton_vs_quasi_newton} showed a significant accuracy drop for quasi-Newton compared to Gauss-Newton. For such cases -- including permafrost -- a potential compromise between accuracy and computing effort might be achieved, by combining two recalculations of the exact Jacobian in the first iterations and then approximating subsequent ones \cite{2_matthes_2012_ertPhD, 0_loke1996quasi_newton_first}. 
\newline
Steps three and four are repeated iteratively until a terminating convergence threshold is reached. This is not necessarily a global minimum of the objective function as multiple solutions to the nonlinear least squares problem exist and depend on the initial parameters \cite{1_herring_2023_bestpractice}. Further, the initial parameters can also be varied finding better convergences \cite{12_gavin_2024_LM}. 
\newline
The converged model provides insights about sub-surface resistivities and can help to understand the underlying structures. Nevertheless, it is necessary to obtain further field investigations, including but not limited to sensors, surface observations, borehole analysis, ground-penetrating radar and frost table depths \cite{1_herring_2023_bestpractice}. The combination of multiple survey forms then yields a sophisticated understanding of the sub-surface \cite{11_Tomaskovicova_2023_timelaps_ert,35_Wicki_2022_landslides, 30_glazer_2020_svalbard, 20_Gilbert_2019_lonyearbyen_testsite}.

\section{Random forests}

For the correlation of \acrfull{er} and \acrfull{vwc} a random forest regressor is used in this thesis. The regressor consists of $n$ decision trees \cite{0_murphy2012machine_learning}. For each tree an input space with \acrshort{vwc} as weights and the observed features (\acrshort{er}, temperature, precipitation) is split into different regions by successive statements about the features. The splits are chosen to minimize the mean square error of \acrshort{vwc} prediction and measurement. \cite{0_breiman2001random_forests}. At each node, the best feature and the best value of the feature are chosen as the split. For a random forest, the $n$ regression trees are constructed from subsets of the input space, randomly selected with replacement, and with $i$ randomly selected features for each node. The prediction for \acrshort{vwc} is then the mean of the predictions from all trees (bagging) \cite{0_breiman1996bagging}. 
\newline
The concept of trees is generally easy to interpret, since only logical rules (for example: \acrshort{er} value <~300~$ \Omega \cdot m$; yes: \acrshort{vwc} value = 0.24 $cm^3 cm^{-3}$; no: \acrshort{vwc} value = 0.22 $cm^3 cm^{-3}$) are derived as decisions, and an automatic selection of these decisions is performed, which is robust to outliers \cite{0_murphy2012machine_learning}. Nevertheless, a tree is often unstable and of high variance due to a greedy minimization of the splits. This is taken into account by averaging many trees for the random forest regression, which leads to better accuracy \cite{0_murphy2012machine_learning}.

\newpage

%% file: 3Method_of_investigation.tex
\chapter{Materials and methods}
\label{ch:materials_methods}
\section{Study area description}


The study area is located in Longyearbyen (78°13'N, 15°28'E \cite{20_Gilbert_2019_lonyearbyen_testsite}) on the Spitsbergen island (Svalbard, Norway) \cite{0_NPI2014terreng_map}. 
\begin{figure}[h!]
    \centering
    \includegraphics[width=0.75\linewidth]{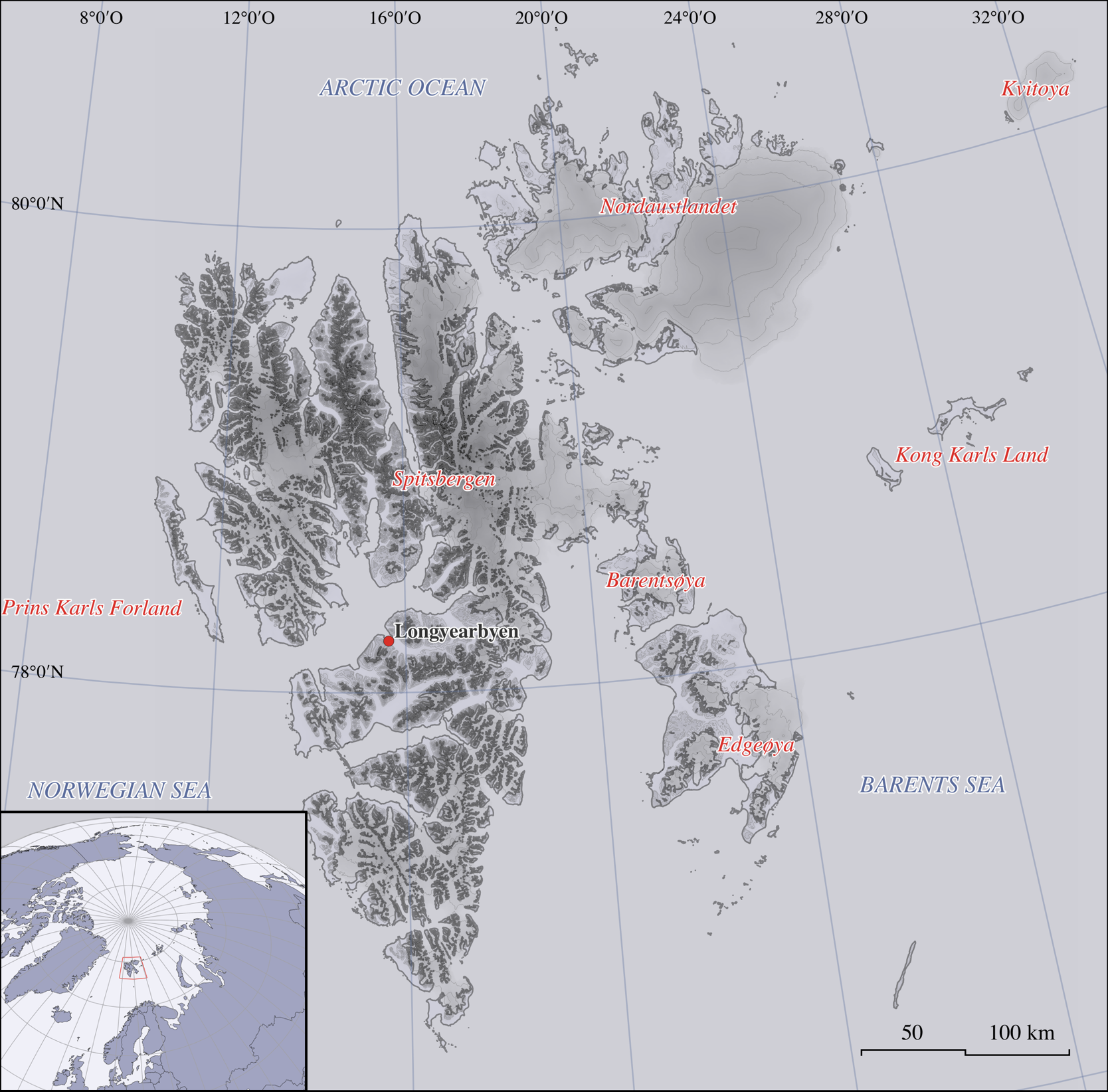}
    \caption[Map of the Svalbard archipelago]{Map of the Svalbard archipelago.}
    \label{fig:maps:svalbard}
\end{figure}
The study area in central Svalbard is located in the continuous permafrost region \cite{19_Schirrmeister_2012_permafrost_charakterisierung, 23_gilbert_2018_cryostratigraphyPhD}. 
Typical permafrost thicknesses in Svalbard range from few meters to more than 100~m in valleys and up to 450~m at mountains around Longyearbyen \cite{SESS2020, humlum2005holocene_permafrost_thickness}. Active layer thicknesses reach approx. 200-300~cm in Longyearbyen \cite{25_Christinansen_2003_activelayer,32_hanssen_2019_climate_svalbard}. Spitsbergen is influenced by a polar-tundra climate \cite{22_Rouyet_2019_remotesensing_svalbard} according to the Köppen-Geiger classification. The annual mean air temperature in Longyearbyen (Airport) for the observation period 1971-2000 was approx. -6~°C thus well below 0~°C \cite{32_hanssen_2019_climate_svalbard}. Nevertheless, a continuous increase particularly during the most recent decades (1.0~°C-1.2~°C per decade between 1980-2010) is visible and an even more drastic rise is observed comparing the winter seasons \cite{32_hanssen_2019_climate_svalbard,0_forland_2011_svalbard_temp}. The 30-year temperature mean in Longyearbyen was -3.9~°C in 2018 \cite{20_Gilbert_2019_lonyearbyen_testsite}. Further drillhole measurements in Svalbard suggested a continuous mean ground temperature increase of 0.06-0.15~°C/a and 0.07-0.08~°C/a between 2008 and 2016 for depths of 10~m and 20~m, respectively \cite{32_hanssen_2019_climate_svalbard}. 


As Longyearbyen is surrounded by plateau mountains up to 500~m a.s.l. a natural rain shadow is created \cite{32_hanssen_2019_climate_svalbard} and the mean annual precipitation is about 200~mm for the time period 1971-2000 in the Longyearbyen area. Approximately half of the yearly precipitation occurs as snow in winter and spring \cite{0_forland_2011_svalbard_temp}. Furthermore, since 1912 the mean annual precipitation increase was about 2\%-4\% per decade \cite{32_hanssen_2019_climate_svalbard, 0_forland_2011_svalbard_temp} with higher increases in autumn. For a sophisticated investigation other parameter as topography, prolonging snow cover, vegetation, soil moisture and material need to be consolidated \cite{19_Schirrmeister_2012_permafrost_charakterisierung}.

\newpage
\begin{sidewaystable}[thbp] 
    \centering
    \caption[Measurements at each study sites]{Specifications for the study sites. T-depths includes the depths [0, 10, 20, 40, 60, 80, 100, 120, 140, 160, 180, 200, 250, 300, 350, 400]~cm while T stands for temperature measurements which is measured at each sensor location.}
    \begin{spacing}{1.5}
    \begin{tabular}{ccccccc}
        \toprule
        \multirow{10}{*}{\textbf{\rotatebox{90}{\acrshort{ert}}}} & Location ID & Start Coordinates & End Coordinates & Length [m] & Elevation range [m a.s.l.] \\ \midrule
        & 1.1 & 78° 12' 59,39" N 15° 39' 34,04" E & 78° 12' 58,91" N 15° 39' 46,39" E & 80 & 98.53 - 97.46 \\ 
        & 1.2 & 78° 12' 59,20" N 15° 39' 45,91" E & 78° 13' 02,22" N 15° 39' 47,94" E & 100 & 96.20 - 64.77 \\ 
        & 2   & 78° 12' 45,82" N 15° 38' 01,64" E & 78° 12' 47,66" N 15° 37' 53,40" E & 80 & 77.28 - 58.42 \\ 
        & 3   & 78° 12' 53,86" N 15° 36' 17,39" E & 78° 12' 52,79" N 15° 36' 28,17" E & 80 & 82.66 - 60.41 \\ 
        & 4   & 78° 12' 34,91" N 15° 37' 22,53" E & 78° 12' 36,65" N 15° 37' 14,38" E & 80 & 93.47 - 65.94 \\ 
        & 5   & 78° 12' 40,51" N 15° 35' 43,62" E & 78° 12' 38,40" N 15° 35' 49,82" E & 80 & 94.56 - 72.40 \\ 
        & 6   & 78° 12' 06,15" N 15° 35' 40,98" E & 78° 12' 08,03" N 15° 35' 32,71" E & 80 & 124.11 - 108.67 \\ \midrule
    \end{tabular}

    \begin{tabular}{cccccc}
        \midrule
        \multirow{14}{*}{\textbf{\rotatebox{90}{Sensors}}} & Location ID & Coordinates & Ground height [m a.s.l.] & Depths [cm] & Measurements \\ \midrule
        &S 1.11 & 78° 12' 59,14" N 15° 39' 35,65" E & 99,46 & T-depths & T \\ 
        &S 1.21 & 78° 13' 02,44" N 15° 39' 48,68" E & 62,45 & 50, 100, 150 & \acrshort{vwc}, \acrshort{er}, T \\ 
        &S 1.22 & 78° 13' 01,91" N 15° 39' 48,20" E & 67,26 & 10, 90, 140, 190 & \acrshort{vwc}, \acrshort{er}, T \\ 
        &S 2.1  & 78° 12' 46,37" N 15° 37' 58,12" E & 101,10 & 10, 50, 100, 150, 190 & \acrshort{vwc}, \acrshort{er}, T \\ 
        &S 2.2  & 78° 12' 46,81" N 15° 37' 56,07" E & 64,73 & 50, 100, 150, 160 & \acrshort{vwc}, \acrshort{er}, T \\ 
        &S 3.11  & 78° 12' 53,29" N 15° 36' 22,66" E & 69,89 & T-depths, 450, 500 & T \\         
        &S 3.12  & 78° 12' 53,29" N 15° 36' 22,66" E & 69,89 & 10, 50, 100, 150, 200 & \acrshort{vwc}, \acrshort{er}, T \\ 
        &S 3.2  & 78° 12' 52,97" N 15° 36' 25,08" E & 64,58 & 10, 50, 100, 150 & \acrshort{vwc}, \acrshort{er}, T \\ 
        &S 4 & 78° 12' 36,13" N 15° 37' 16,46" E & 73,73 & T-depths excluding 350 and 400 & T \\ 
        &S 5    & 78° 12' 38,86" N 15° 35' 49,02" E & 74,78 & T-depths, 450, 500, 550, 600, 650, 700 & T \\ 
        &S 6    & 78° 12' 06,94" N 15° 35' 34,88" E & 114,73 & T-depths, 450 & T \\ \bottomrule

    \end{tabular}

    \end{spacing}

    \label{tab:ert_locations}
\end{sidewaystable}

\begin{figure}[h!]
    \centering
    \includegraphics[width=\linewidth]{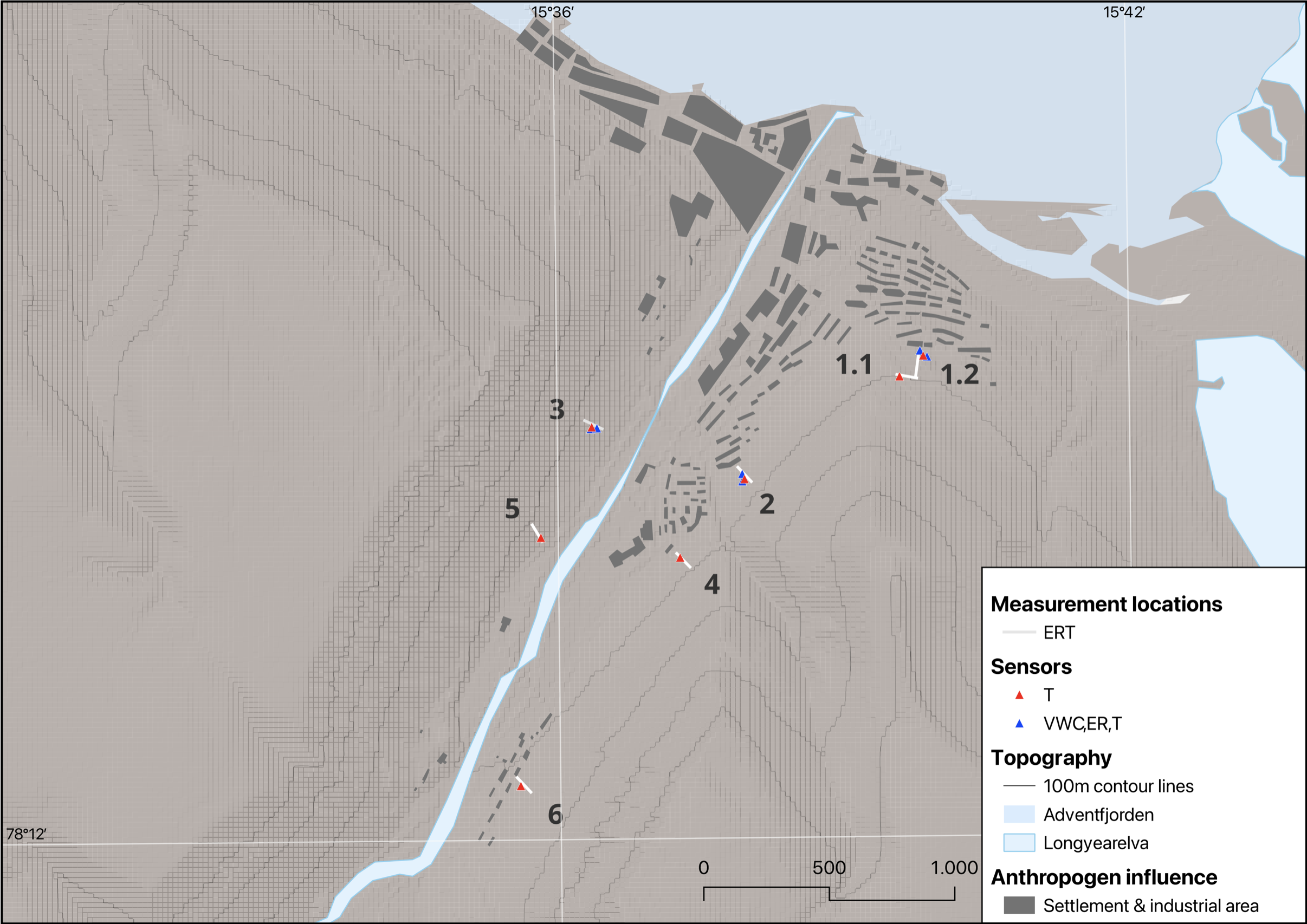}
    \caption[Structural map of Longyearbyen with all study sites]{A structural map of Longyearbyen showing settlements and height contours. The study areas are marked, T means temperature while VWC, ER and T depict combined known abbreviations.}
    \label{fig:maps:lyr}
\end{figure}

Six test areas (see Fig.~ \ref{fig:maps:lyr} and Tab. \ref{tab:ert_locations} for further details) were selected as representatives for present geomorphological and slope conditions in Svalbard as well as in other Arctic locations \cite{20_Gilbert_2019_lonyearbyen_testsite}. Further, all locations are situated on slopes and directed along the gradient  (see also Fig.~ \ref{fig:maps:lyr}) except \acrshort{ert} 1.1, which is situated transverse it.  

The investigated slopes face the settlements and pose a considerable danger on nearby infrastructure and buildings in forms of avalanche, debris flow and landslide risk. 
\newline
At study area 1 two \acrshort{ert} lines were installed to cover a broader slope area. The northern slope of the Sukkertoppen mountain above the highest building line was selected for this study site as in October 2016 a landslide was nearly caused after a warm and wet summer, followed by a heavy rainfall event \cite{christiansen2016report}. This is also visible on aerial pictures taken from that area. 
Study area 1 covers approximately 0.8~ha and ends about 20~m above the first houses. Line 1.1 was chosen perpendicular and line 1.2 parallel to the slope direction. Next to line 1.2, temperature sensors were already measuring every 6~h at depths between 10~cm and 400~cm. Further down, besides line 1.1 \acrshort{vwc}, \acrshort{ec} and temperature sensors in depths of 10 to 190~cm were just installed in early October 2024. 
\newline
Site 2 was chosen on the southwards slope of Sukkertoppen. Nearby this measurement station there was an avalanche incident in December 2015 displacing several houses and taking two lives \cite{0_avalanche_2015}. Thereupon, multiple fences on the slopes of Sukkertoppen and an avalanche embankment underneath were build to prevent avalanches, landslides and mud slushes in the most endangered regions. For further ground observation especially connected to landslides, location 2 was chosen on the slope approximately 30~m above the very top houses and about 200~m north-east to Vannledningsdalen. 
There, temperature, \acrshort{vwc} and \acrshort{ec} sensors were already installed in December 2023 in depths of 10~cm to 190~cm. The subsurface for the top 30~m of the \acrshort{ert} line consists of colluvial material, heterogeneous, loose and unconsolidated rocks. Below this, the geomorphology is dominated by solifluction, a slow mass wasting process due to freeze-thaw action \cite{0_matsuoka2001solifluction, 0_Christiansen_2014_geomorph_map}.
\newline
Study sites 3 and 5 are located west of the village on the slopes of Sverdruphamaren. These sites are also characterized by colluvial material (the top 20~m at site 5) and solifluction (the top 60~m at site 3). The contact resistance for site 5 (and 6) was significantly higher (mean around 2~k$\Omega$, sometimes up to 5~k$\Omega$) compared all other sites (0-2~k$\Omega$) because of the colluvium. Site 5 is additionally monitored by temperature sensors down to 700~cm, at site 3 temperature, \acrshort{vwc} and \acrshort{ec} sensors are installed at two locations.  
\newline
Study site 4 is located about 30~m above the Haugen district. Continuous temperature data for depths of up to 300~cm is available here. 
\newline
Study site 6 is located on the south-eastern slope above Nybyen, a district of Longyearbyen, The \acrshort{ert} line was constructed about 15~m above the upper most houses next to the temperature sensors (down to 450~cm). This slope is completely dominated by colluvium. The contact resistance was correspondingly high at $2-4 \ k \Omega$ as site 5. 
\newline
In general, the measuring positions were selected based on the occurrence of the greatest number of avalanches and landslides with damage to infrastructure or people in the recent years. 

\section{Observation period}
The overall investigation period began in mid August 2024 and ended in late October 2024. Three surveys at different locations and times were conducted to assess spatial, long-term temporal and short-term temporal variations. 
\newline
The short-term variation observations (survey one) were designed to test the influence of small mean temperature fluctuations, but primarily precipitation variations on the accuracy and robustness of the models. Therefore, two time intervals were chosen, one without significant rainfall and one shortly after a heavy rainfall event after which resulting infiltration was to be observed. Survey one was carried out in early October to ensure air temperatures above zero degrees.  
\newline
The spatial observation survey (survey two) encompasses seven single measurements performed from 02.09.2024 until 06.09.2024. The temporal proximity of these measurements was designed to minimize seasonal temperature changes as well as precipitation deviations. To observe the highest depth dependent variations, i.e. the highest active layer thickness, the measurement period was set to early September. 
\newline
Long-term seasonal changes especially in the uppermost active layer were observed by conducting weekly measurements (survey three). In total 12 weeks were investigated resulting in a time laps ranging from summer (August) until the complete freezing of the ground and a termination of the measurement period in the beginning of winter (October). In Tab. \ref{tab:overview_measurements} all surveys and dates of measurements are listed. 
\begin{table}[h!]
    \centering
    \caption[Observation periods]{Three different surveys were performed to guarantee insights over short-term temporal variations (survey 1), spatially (survey 2) and long-term temporal variations (survey 3).}
    \begin{tabular}{cccc}
        \toprule
        Survey & Location & Time & Frequency \\ \hline
        \multirow{2}{*}{1}
        & 2 & 29.08.-01.09.2024 & daily: 6:30, 12:30, 18:30 \\
        & 2 & 06.09.2024 & 9:00, 12:00, 15:00 \\ \midrule
        \multirow{7}{*}{2}
        & 1.1 & 03.09.2024 & once \\
        & 1.2 & 03.09.2024 & once \\
        & 2 & 04.09.2024 & once \\
        & 3 & 04.09.2024 & once \\
        & 4 & 05.09.2024 & once \\
        & 5 & 05.09.2024 & once \\
        & 6 & 02.09.2024 & once \\ \midrule
        \multirow{1}{*}{3}
        & 2 & 14.08.-30.10.2024 & weekly \\ \bottomrule
    \end{tabular}

    \label{tab:overview_measurements}
\end{table}

\section{ERT and sensor equipment and setup}
The \acrshort{ert} were conducted using an ABEM Terrameter SL 2 (Guideline Geo AB, Sweden) driven by an embedded ARM 9 processor and powered by an external car battery. For each \acrshort{ert} 81 rod electrodes (40~cm length, 3.5~cm diameter) were placed at 1~m equidistance and were used with a Wenner array to ensure high lateral resolution, but also to obtain some depth resolution into the permafrost. They were hammered about 10~cm into the ground to achieve sufficient galvanic coupling. Additional salt water was introduced for electrodes with high contact resistances, which was found efficient. For study site 1.2 a 100~m line was set up with an equal spacing. The measurement of 445 points took around 2~h. 
\newline
For all sites, both resistivity and induced polarization were measured while only resistivity was currently used for the analysis. The number of stacking was set from 2 to 4 which ensures small errors -- the standard deviation in a stack divided by the mean value for a data point was set to 1\% -- but also keeps the measurement time short. The delay and acquisition time were also set to achieve a good compromise between above criteria. The transmit parameters were chosen to secure the instrument while improving the signal-to-noise ratio. All important settings for receiving and transmitting are summarized in Tab. \ref{tab:ert_parameter}. 
\begin{table}[h!]
    \centering
    \caption[ERT parameters]{Important receive and transmit parameter used to obtain the \acrshort{ert}.}
    \begin{tabular}{cccc}
        \toprule
        \multicolumn{2}{c}{Receive} & \multicolumn{2}{c}{Transmit}\\ \midrule
        Measurement mode & Res, IP & Min Current & 1~mA\\
        Min Stacking & 2 & Max Current & 1000~mA\\
        Max Stacking & 4 & Max Power & 250~W\\
        Error limit & 1.0 \% & Max output voltage & 600~V\\
        Delay time & 0.6~s & Load variation margin & 10\%\\
        Acquisition time & 0.4~s &&\\
        Power line frequency & 50~Hz &&\\
        Sample rate & 1000/1200~Hz &&\\ \bottomrule
    \end{tabular}

    \label{tab:ert_parameter}
\end{table}

To install the sensors, a hole was excavated and sensors were inserted at different depths. The sensor needles were placed horizontally to allow a more discreet depth for all sensors. \acrshort{vwc} is measured by applying a 70~MHz oscillation and a read out of the subsequent charge time of the 1,010~mL sensitive volume around the sensors. The charge time is proportional to the dielectric properties and \acrshort{vwc} of the substrate. A thermistor in the center needle is used for temperature measurement. The \acrfull{ec} is obtained by running an alternating current between the outermost needles and gauging the resistance between them. The \acrshort{er} as reciprocal \acrshort{ec} is then obtained. All sensors measure every 6~h and are connected to a data logger. An already existing sensor setup was used for the research described in this thesis. 
\begin{figure}[h!]
    \centering
    \includegraphics[width=\linewidth]{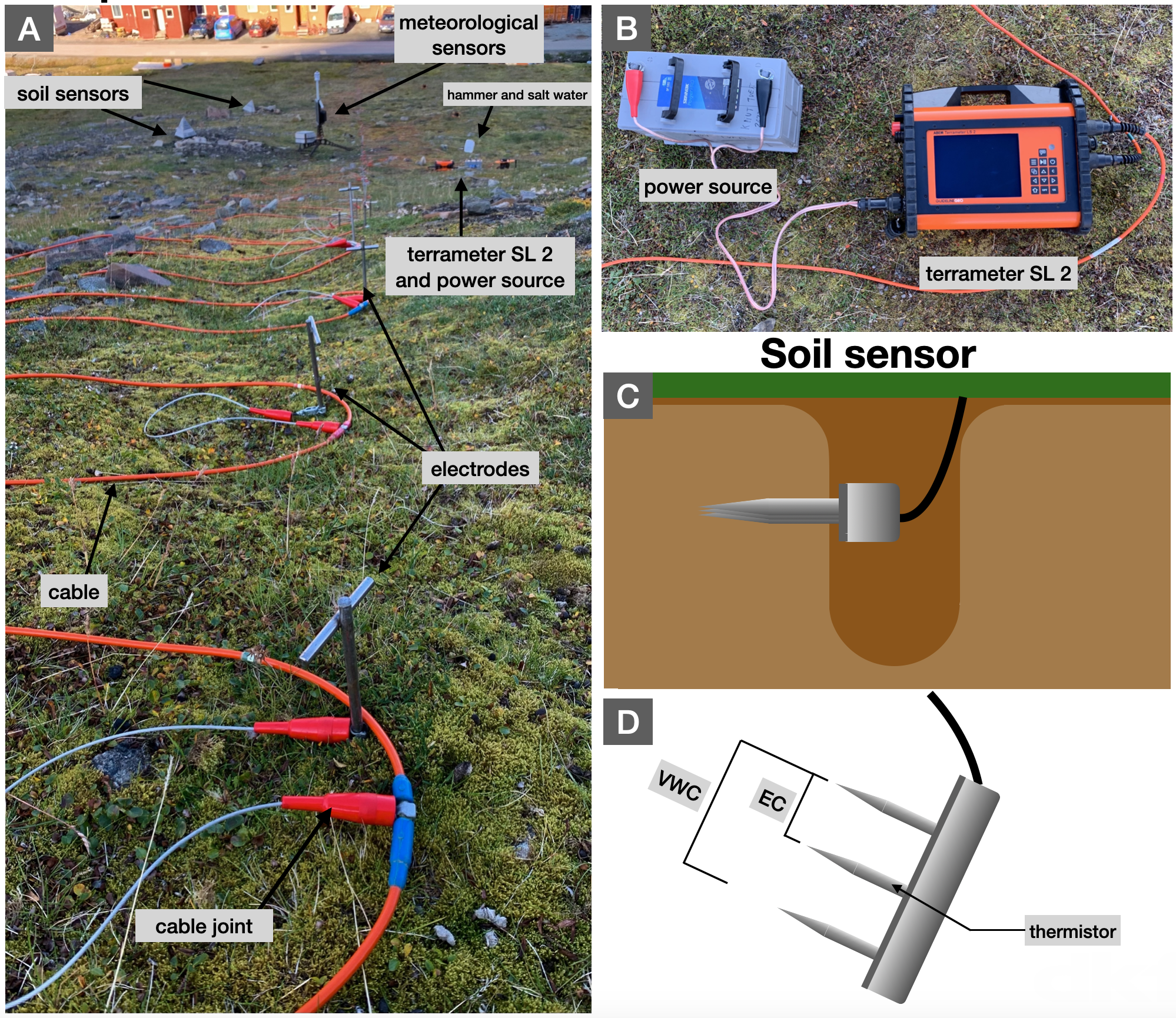}
    \caption[Measurement setup for ERT and sensors]{The overall measurement set up including sensor measurements and \acrshort{ert}; A: The electrodes were connected by cable joints to the cable and the Terrameter SL2, B: The Terrameter SL 2 and a car battery as external power source, C: Sensors are brought invasively into a borehole at different depths, D: The Teros 12 (METER Group, Inc. USA) sensor, it is equipped with one thermistor, the \acrshort{vwc} is measured between the outer most needles and the \acrshort{ec} is measured between the upper and middle needle. }
    \label{fig:set_up_ert_sensor}
\end{figure}

\section{GPS and height correction}
The GPS data acquisition for the sensor points and the individual electrodes of all \acrshort{ert} was recorded with a Leica Viva GS16 GNSS rover (Leica Geosystems AG, Switzerland). Subsequent corrections were performed by combining the individual data further improved the accuracy. The achieved average accuracy was within centimeter range. However, the height data of the electrodes showed large fluctuations of neighboring electrodes, which was not realistic. For this reason, the coordinates were combined with a \acrfull{dem} for height calculations. A \acrshort{dem} developed from the Norwegian Polar Institute \cite{0_NPI2014terreng_map} with a 2.5~m resolution was used. The resulting heights were interpolated and flattened by taking the average of three adjacent electrodes and applying it to the center electrode. 
\newline
In an effort to better illustrate the layering and infiltration, the analysis was applied to the non-height corrected \acrshort{ert}. The data points were interpolated linearly in a trapezoidal pattern. Nonetheless, as the slope is not perfectly linear, errors are introduced.

\section{Filtering and inversion scheme}

The recorded data points were processed further using the open-source software ResIPy (version: 3.3.5) \cite{0_blanchy_resipy_2020}
and the previously determined topography was applied. Three filters were employed to remove outliers. 

Initially, erroneous data points, defined as those with transfer resistance values less than zero ohms, were identified and removed. Additionally, data points exhibiting unusually high deviations (>~10\%) from the transfer resistance values of neighboring points within a given layer were also filtered out. The same methodology was employed for apparent resistivity values. 
A maximum tolerance of 5\% was established for the staking error. 
Afterwards, a triangular mesh -- accounting for complexity and flexibility in topography -- was then constructed as reference for the inversion. A regularized solution with a linear filter (model misfit term) was implemented. The relative error in magnitudes was approximated to be  2\%, while the minimum magnitude error was set to 1\%. For each iteration, the greatest misfit reduction was selected. The termination point was set to a maximum of 10 iterations or a final misfit of 1.0 as convergence threshold. In order to avoid the exclusion of any solution while remaining within the confines of reasonable physical possibility, the possible apparent resistivity range after inversion was set to a range of $(10^{-10} - 10^{10}) \ \Omega \cdot m$. In case where any normalized error exceeded an absolute value of 3 after inversion, the inversion calculations were repeated, excluding this data point.


%% file: 4Experimental_Results.tex
\chapter{Experimental Results}
\label{ch:experimental_results}

    
\section{Overall ERT and sensor data processing}
This section exemplifies the processing for an \acrshort{ert}, based on the measurement on August 28th 2024 at study site 2. Around that time the thaw depth reached this years maximum, which occurs usually around August and September. Furthermore, mainly the sensor data of one borehole (S 2.1) is processed and discussed. This implies that all subsequent analysis can be generalized, summarized and consolidated. Perspectives are presented and discussed in the Ch. \ref{ch:conclusion}. All obtained \acrshort{ert} are summarized in App. \ref{AppendixA}. 

\subsection{Basic ERT}
\begin{figure}[h!]
    \centering
    \includegraphics[width=1\linewidth]{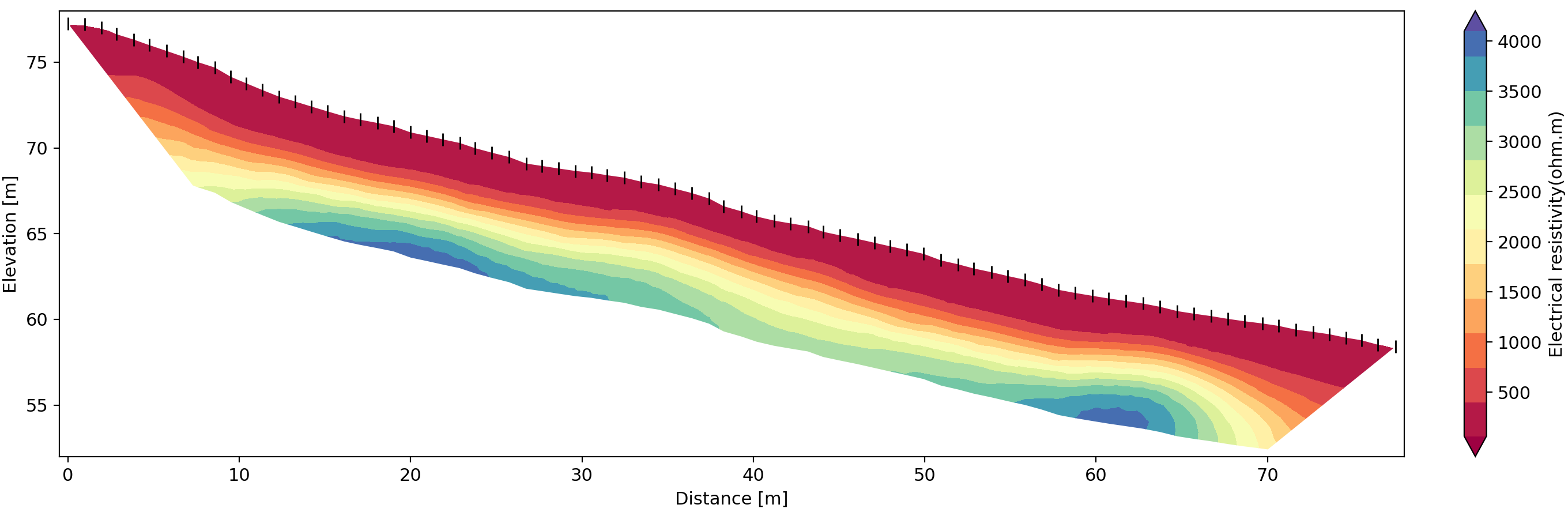}
    \caption[ERT on 28.08.2024 at study site 2]{\acrshort{ert} at study site 2 on August 28th 2024. The rod electrodes are displayed as vertical lines which reach 10~cm into the ground. The figure was created using ResIPy \cite{0_blanchy_resipy_2020}. }
    \label{fig:28_08_ert_example}
\end{figure}

The resulting filtered and inverted  \acrshort{ert} is shown in Fig.~ \ref{fig:28_08_ert_example}. The \acrshort{ert} is characterized by homogeneous stratification with increasing electrical resistivity at depth. Two spots of higher electrical resistivity lay at around 20~m and 60~m distance from the top.

The layering can also be presented numerically: In Tab. \ref{tab:resistivity_mean_meter} for every meter in depth the electrical resistivity means are shown. The means increase monotonously with the depth from (135 $\pm$ 33)~$\Omega \cdot m$ for the upper most meter to (3340 $\pm$ 490)~$\Omega \cdot m$ for a depth of 6-7.36~m. Except in the first two meters, the relative error decreases with depth, between one to three meters the relative error seems to be stable around 40\%.

\begin{figure}[h!]
    \centering
    \includegraphics[width=\linewidth]{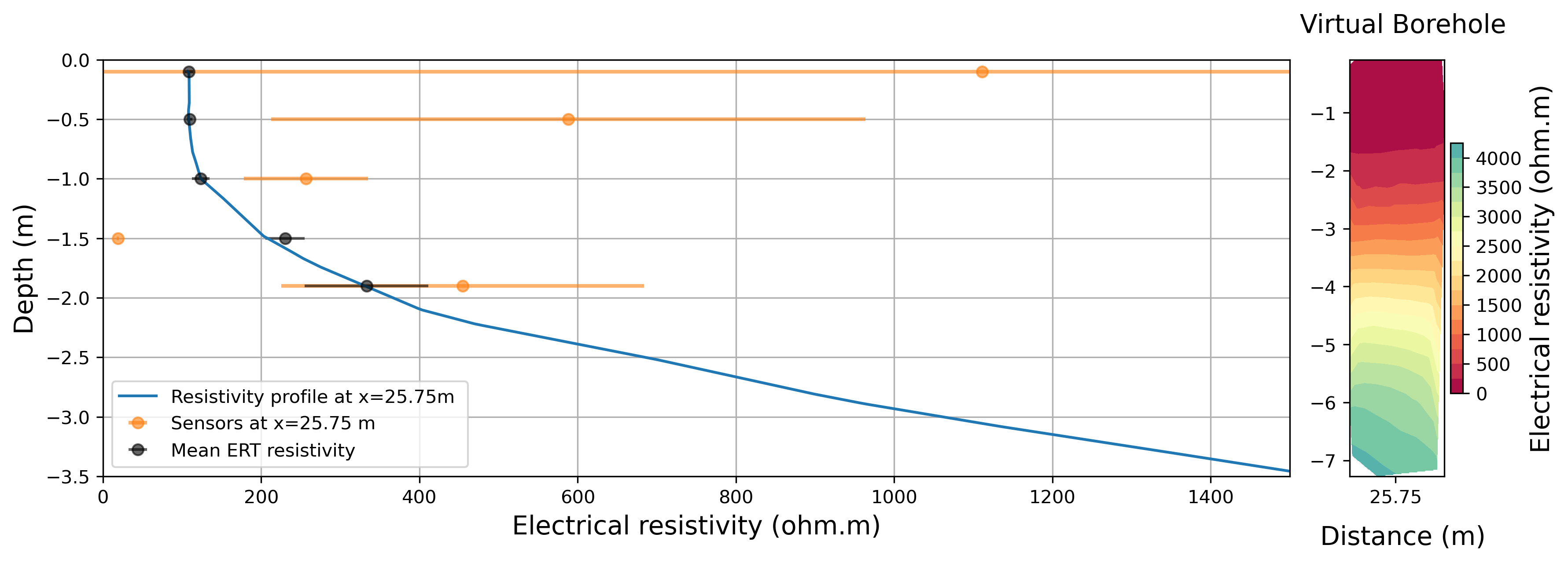}
    \caption[Virtual borehole from ERT]{Left: The electrical resistivity profile obtained from the \acrshort{ert} line close to the sensors S 2.1 with the means of the surrounding $30x30 \ cm^2$ area and sensor measurements; Right: The virtual borehole showing the electrical resistivity profile.}
    \label{fig:virtual_borehole}
\end{figure}

In App. \ref{AppendixB}, Tab. \ref{tab:dynamic_heights_mean_ert} different dynamic height ranges were calculated, all of which should be at least 5~cm great and further have a relative error of about 20\% Two regions from 7~m to 4.6~m and 1.2~m to 0.3~m could be identified as significantly homogeneous. 
\newline
The sensors, which were used for analysis, were the ones within location ID  S 2.1 (see also Tab. \ref{tab:ert_locations} for further information). A virtual borehole -- an electrical resistivity column -- was calculated for the \acrshort{ert} at the most proximate distance and at the same height as the sensors (Fig.~ \ref{fig:virtual_borehole}). Furthermore, the mean value for a $30x30~cm^2$ area around each sensor position in the \acrshort{ert} was calculated and compared to the sensors (deviation) in Tab. \ref{tab:deviation_table}. 
\begin{table}[h!]
\centering
\caption[Mean ER from ERT for sensor spots]{The electrical resistivity was calculated for a $30x30 \ cm^2$ area at the most proximate distance from the \acrshort{ert} line and the sensors and at the same height as the sensors.}
\begin{tabular}{lccc}
\toprule
Depth (cm) & Deviation($\Omega \cdot m$) & Combined error ($\Omega \cdot m$) &  \makecell{Deviation \\($\sigma)$} \\ \midrule
       10 & 1002.50 & 1290.13 & 0.78 \\
        50 & 478.78  & 375.45  & 1.28 \\
        100 & 133.00  & 79.35   & 1.68 \\
        150 & 211.33  & 24.74   & 8.54 \\
        190 & 121.45  & 242.29  & 0.50 \\
        \bottomrule
\end{tabular}

\label{tab:deviation_table}
\end{table}

The combined error includes both the standard deviation of the mean \acrshort{ert} values and  the sensor error.  The deviation between the mean \acrshort{ert} values and the sensor measurements is less than two $\sigma$ except for 1.5~m depth. The deviation decreases by depth except at 1.9~m.

\subsection{Permafrost estimation}

The concept of the virtual borehole allows for the calibration of the \acrshort{ert} using the installed temperature sensors, assuming that the subsurface structure is homogeneous. If the geological structure is similar for all depths, the thaw depth becomes very clear as there is a steep increase in electrical resistivity. The use of surrounding temperature data enhances the accuracy of the layer transition even further and serves as calibration. The sensor data is linearly interpolated to find the depth at which the temperature is zero degrees.
Due to the high temperature variability of the uppermost points (10~cm, 50~cm), these are excluded from the interpolation, as they are exposed to weather data and fluctuations rather than climate (see also Fig.~ \ref{fig:temp_cali}). The depth at which the temperature reaches 0~$^{\circ}C$ is $D_{21}(T=0^{\circ}C) = (-1.82 \pm 0.65) \ m$ for study site and sensors S 2.1. For the sensors S 2.2 $D_{22}(T=0\ ^{\circ}C) = (-1.48 \pm 0.07)\ m$. $D_{21}$ corresponds to an electrical resistivity value of $( 316 + 157 - 385 )\ \Omega \cdot m$ and $D_{22}$ to $( 357 \pm 35 )\ \Omega \cdot m$. 
\newline
    
\begin{figure}[h!]
    \centering
    \includegraphics[width=1\linewidth]{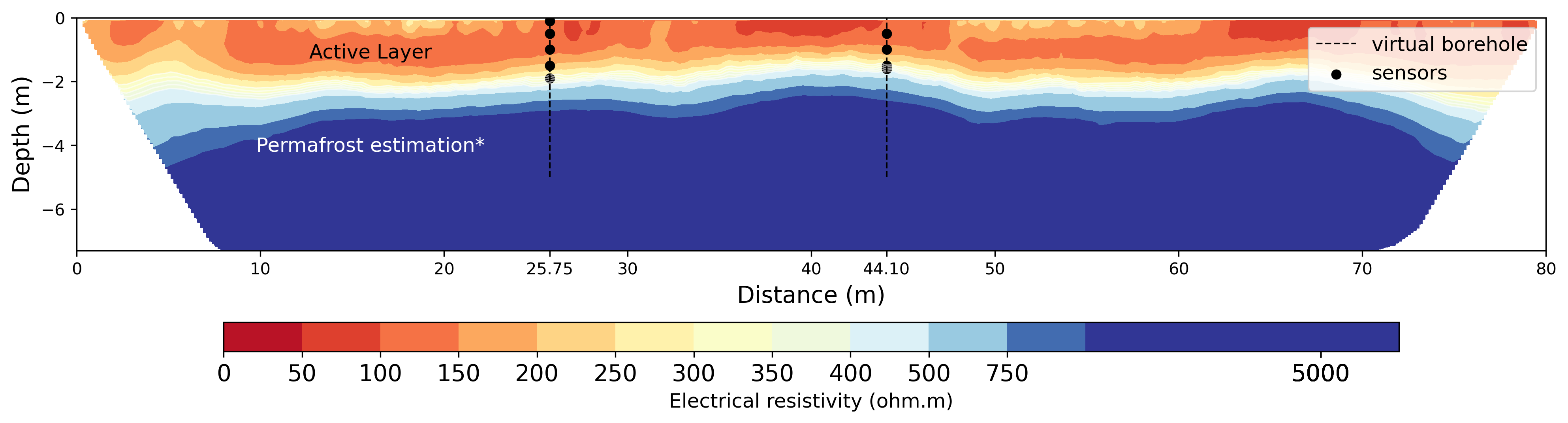}
    \caption[Active Layer and permafrost estimation]{The permafrost estimation based on temperature sensors and the \acrshort{ert}.}
    \label{fig:permafrost_est}
\end{figure}

Fig.~ \ref{fig:permafrost_est} is divided in a dark blue area indicating permafrost overlaid by a white transient layer and the active layer. The transient layer reaches from 300~$\Omega \cdot m$ to 375~$\Omega \cdot m$. In the Fig.~ it is assumed that the thaw depth ($D(T=0\ ^{\circ}C)$) also represents the border between permafrost and the active layer. This assertion is supported by the observation that, around that time, the overall temperatures for all depths were the highest recorded for this year (see also Fig.~ \ref{fig:temp_long_term}).

\subsection{Sensor data}

In the following Figs.~ [\ref{fig:temp_long_term}, \ref{fig:vwc_long_term}, \ref{fig:resis_long_term}] the temperature, \acrshort{vwc} and \acrshort{er} trends are shown. Generally the temperature at 10~cm depth is the most variable. Temperature fluctuations decrease by depth. The sensor errors become significant for deeper depths as the values are low but its errors remain high at $\pm 0.5 \ ^{\circ}C$ for values greater than zero and $\pm 0.3\ ^{\circ}C$ otherwise. For the different depths, the \acrshort{vwc} values are heterogeneous (Fig.~ \ref{fig:vwc_long_term}). There are a few prominent peaks, especially in early June 2024 for all depths simultaneously, additionally at 100~cm in mid-July and from mid-August 150~cm is constantly at a very high magnitude (0.40~$cm^3cm^{-3}$) and only decreases end-September. The peaks at 10~cm depth and at 50~cm are smaller but more frequent than for other depths. The \acrshort{vwc} at 190~cm is generally the most constant, varying only between 0.12-0.16~$cm^3cm^{-3}$ within the measurement period. For 50~cm there are only two medium peaks in early and mid August. Apart from the significant peaks at 100~cm, the \acrshort{vwc} curve is also relatively stable. Starting in October, all curves decrease.
\newline
The \acrshort{ec} and \acrshort{er} data from all sensors have disproportionate high errors. The conductivity is generally low, staying below 50 $ \frac{\mu S}{cm}$ most times for all depths except 150~cm. For these depths from the beginning of August on the \acrshort{ec} is as high as 200-600~$ \frac{\mu S}{cm}$ at an elevated constant level. For 100~cm two larger peaks reaching 75-150~$\frac{\mu S}{cm}$ are visible. The freezing and a resulting decrease of \acrshort{vwc}, \acrshort{ec} and temperature is recognized in early October. The error ranges of most depths overlap each other. This makes the results difficult to analyze and interpret.

\begin{figure}[H]
    \centering
    \includegraphics[width=0.65\linewidth]{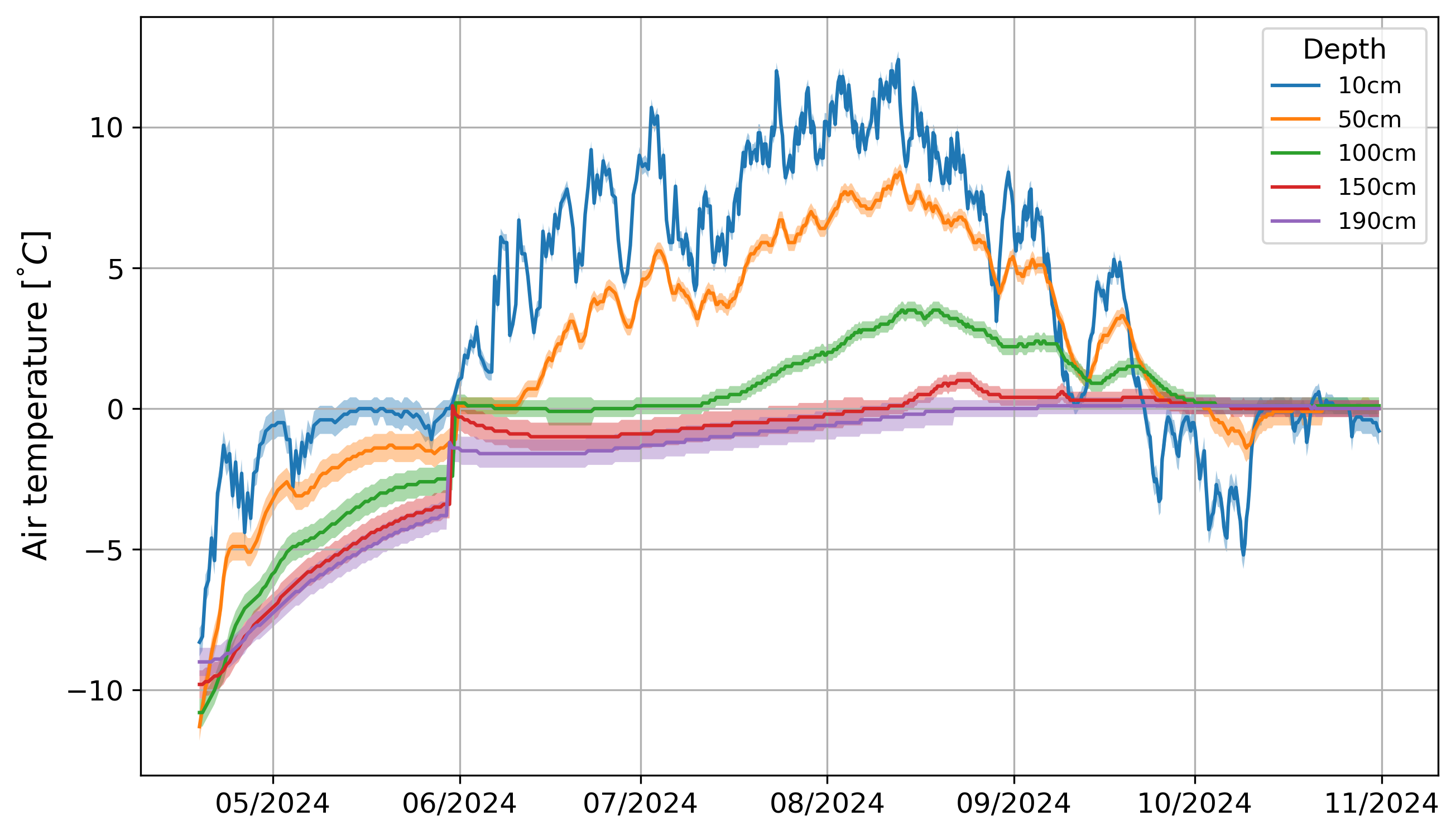}
    \caption[Long-term temperatures]{Long-term temperatures recorded by the sensors S 2.1 at depths reaching from 10~cm to 190~cm.}
    \label{fig:temp_long_term}
    
    \vspace{0.5cm} 

    \includegraphics[width=0.65\linewidth]{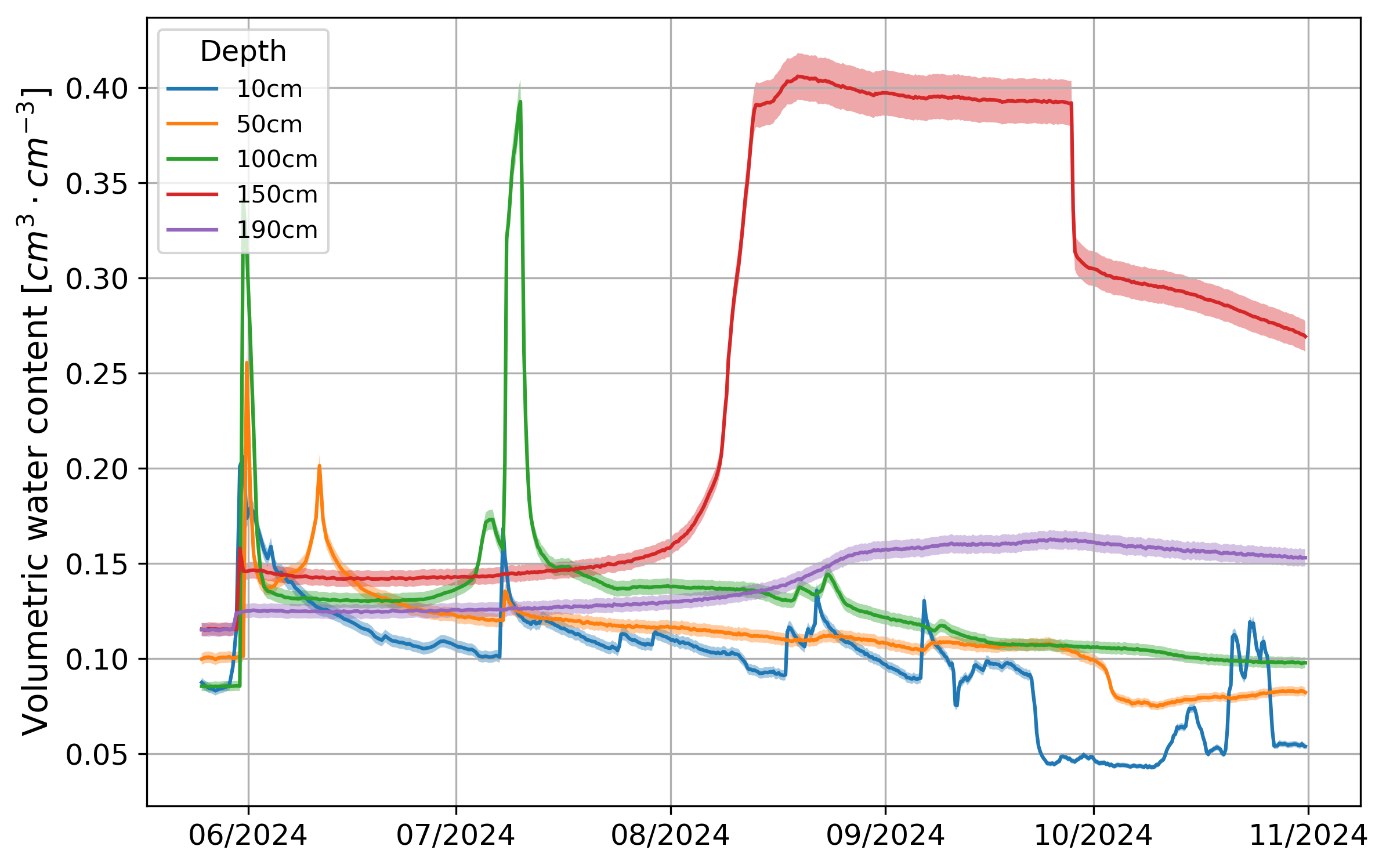}
    \caption[Long-term VWC]{Long-term \acrshort{vwc} recorded by the sensors S 2.1 at depths reaching from 10~cm to 190~cm. }
    \label{fig:vwc_long_term}
    
    \vspace{0.5cm} 

    \includegraphics[width=0.65\linewidth]{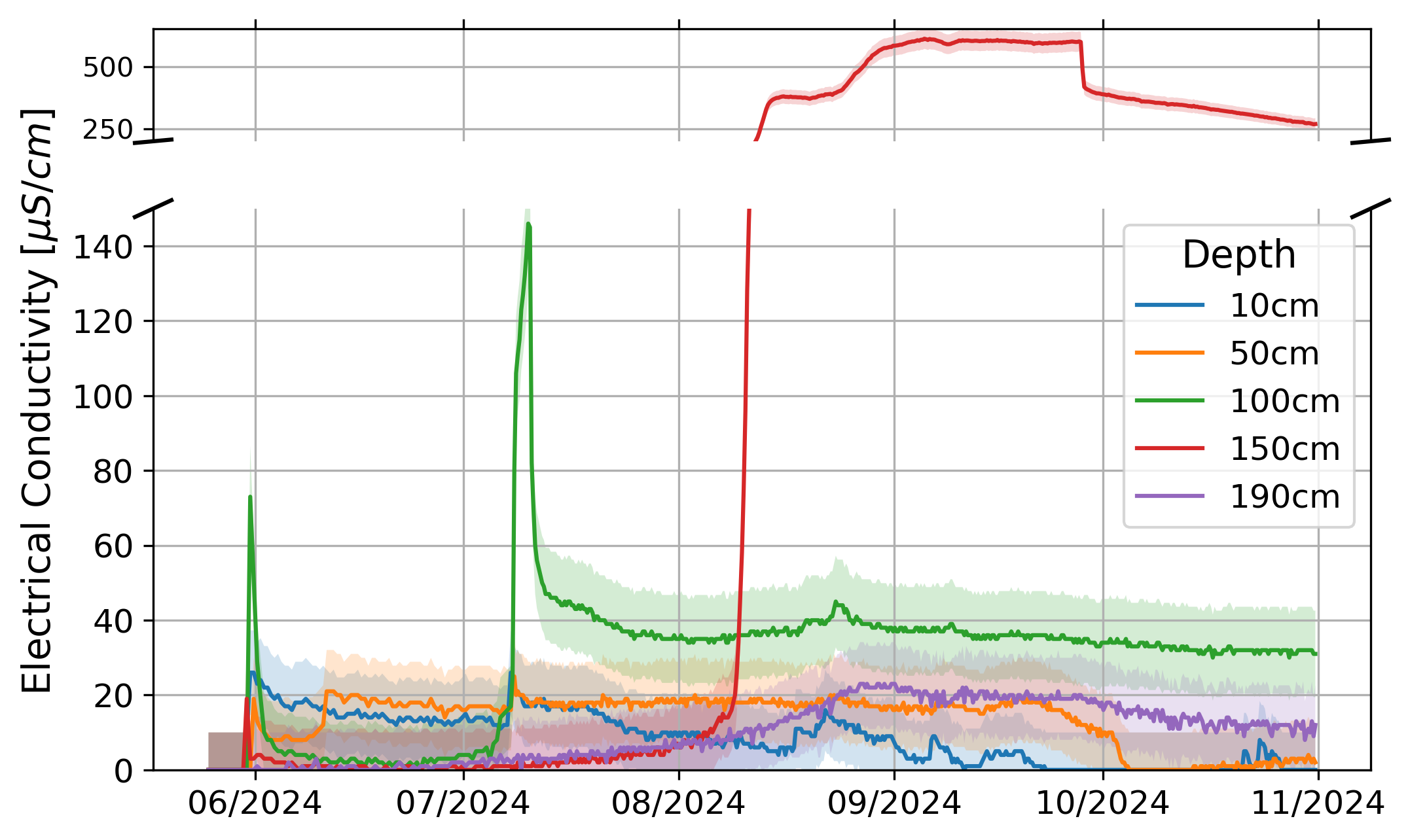}
    \caption[Long-term ER]{Long-term electrical conductivity recorded by the sensors S 2.1 at depths reaching from 10~cm to 190~cm.}
    \label{fig:resis_long_term}
\end{figure}

\section{Finding the best correlation between \acrfull{er} and \acrfull{vwc}}
\subsection{Random forest estimation}

\begin{figure}[H]
    \centering
    \includegraphics[width=1\linewidth]{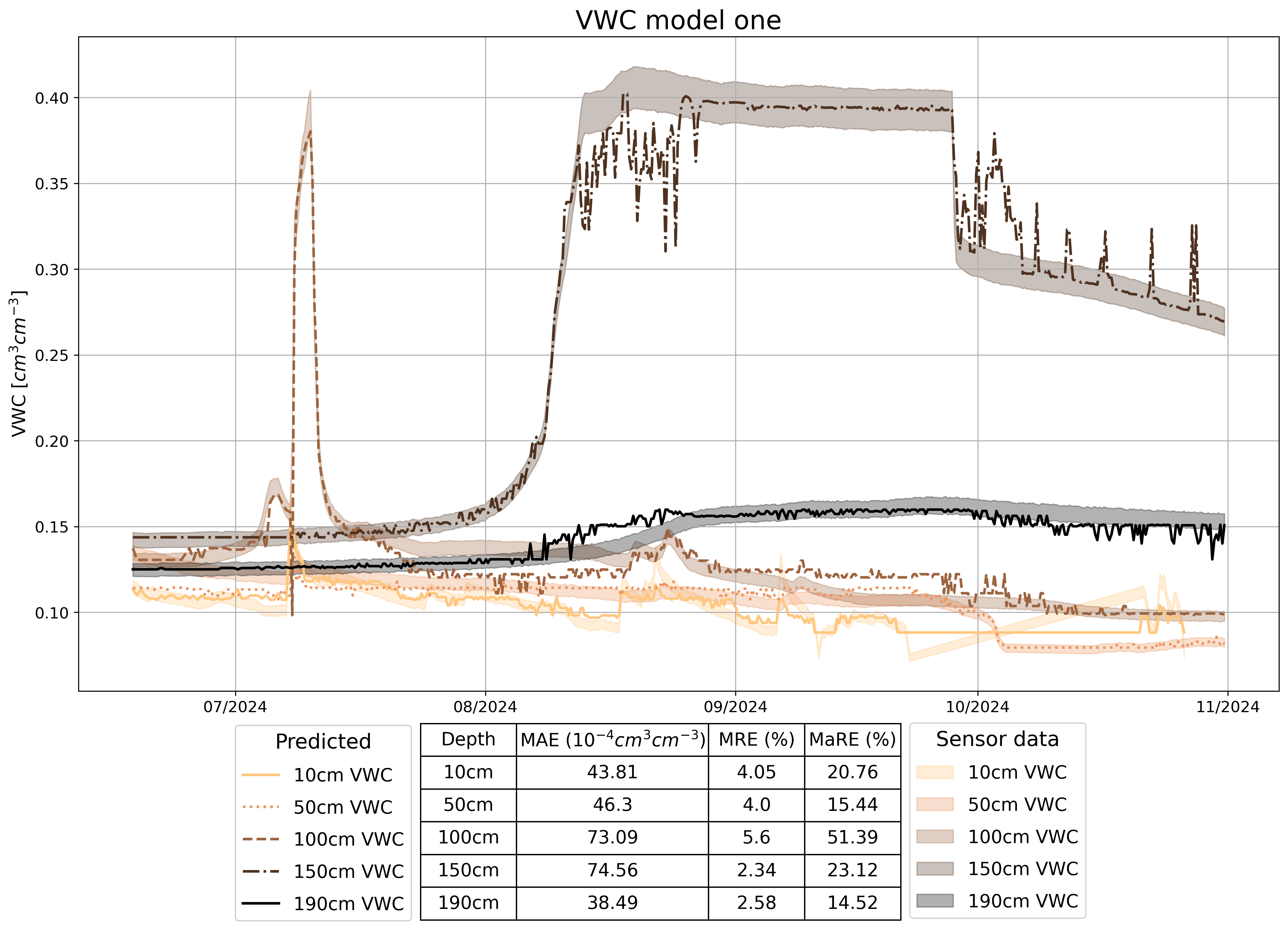}
    \caption[Model one, correlating ER and VWC]{Naive \acrlong{rf} approach correlating \acrshort{er} and \acrshort{vwc}; \acrshort{mae} is \acrlong{mae} and \acrshort{mre} is \acrlong{mre}.}
    \label{fig:ml_excl_temp_prec_lag}

    \vspace{0.5cm}

    \includegraphics[width=1\linewidth]{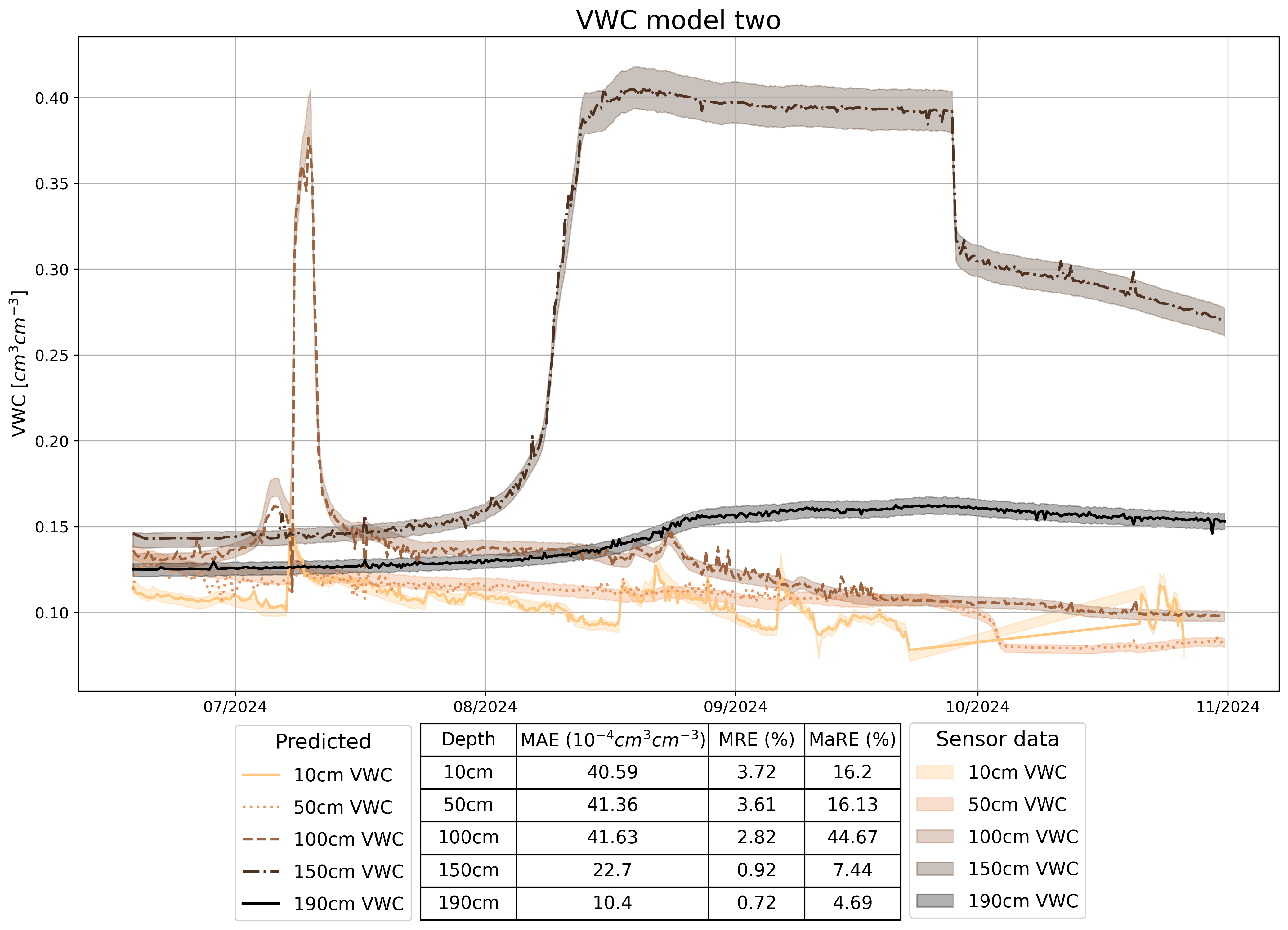}
    \caption[Model two, including temperature and precipitation]{Random forest approach correlating \acrshort{er} and \acrshort{vwc} including air temperature and precipitation data.}
    \label{fig:ml_incl_temp_prec_excl_lag}
\end{figure}

To quantify the correlation between \acrshort{er} and \acrshort{vwc}, a machine learning model based on the random forest concept was used. For all models, 100 decision trees were trained with 80\% of the input data and evaluated for the complement 20\%. As a split quality measure the mean squared error was minimized at each node. All data were recorded in the period from 18 June 2024 to 31 October 2024 by the sensors S 2.1. The error of \acrshort{vwc} was chosen as the weights for all such models. For model one (Fig.~ \ref{fig:ml_excl_temp_prec_lag}) only \acrshort{er} and its error were selected as features. Separate models were created for all depths. 

\begin{figure}[H]
    \centering
    \includegraphics[width=1\linewidth]{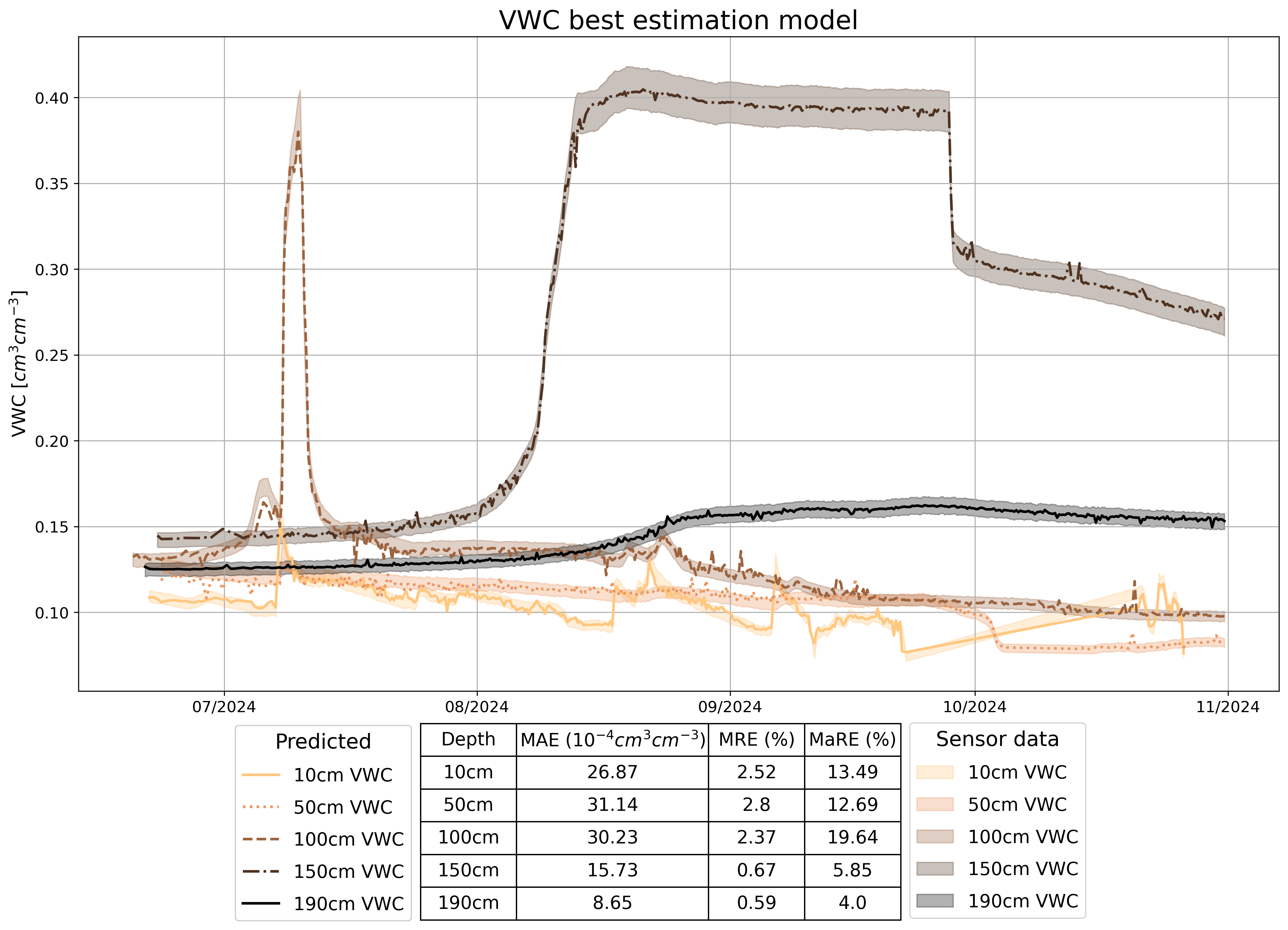}
    \caption[Model of best \acrshort{vwc} estimation]{Model of best \acrshort{vwc} estimation.}
    \label{fig:ml_vwc_best_estimation}

    \vspace{0.5cm}

    \includegraphics[width=1\linewidth]{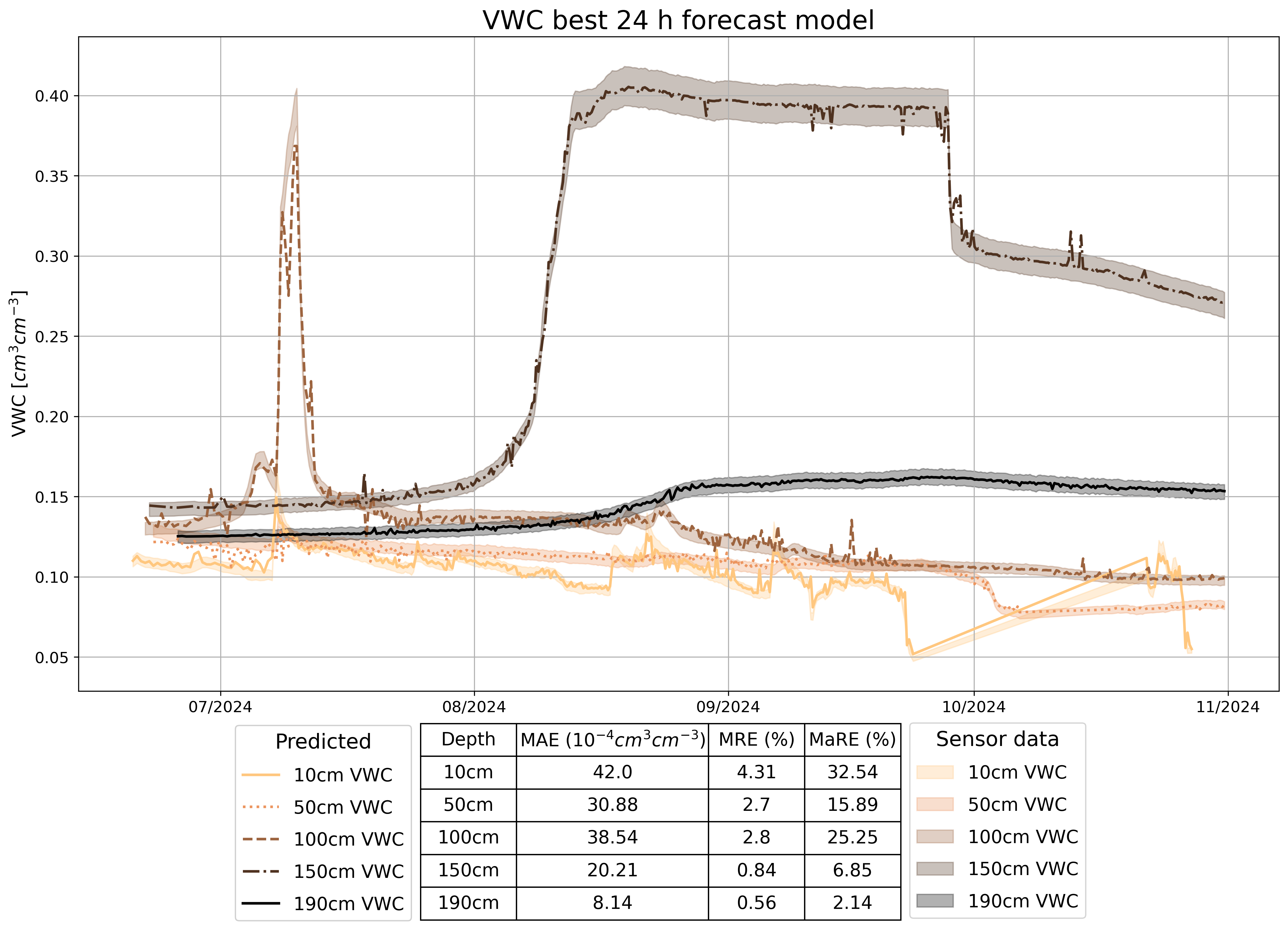}
    \caption[Model of best 24~h \acrshort{vwc} forecast]{Model of best 24~h \acrshort{vwc} forecast.}
    \label{fig:ml_best_24h_forecast}
\end{figure}

In the following Figs. (\ref{fig:ml_excl_temp_prec_lag}, \ref{fig:ml_incl_temp_prec_excl_lag}, \ref{fig:ml_vwc_best_estimation}, \ref{fig:ml_best_24h_forecast}, \ref{fig:ml_best_36h_48h_forecast}) the shaded area depicts the \acrshort{vwc} measurement $\pm 3 \%$. The lines indicate the \acrshort{vwc} value predicted by processing the \acrshort{er} sensor data point. Generally, the \acrshort{vwc} trends were met by the predictions, lacking in details for some time points. The \acrfull{mae} ($\frac{1}{n} \sum_{i=1}^{n} |y_i - \hat{y}_i|$) is below 0.01~$cm^3cm^{-3}$ for all depths while the \acrfull{mare} ($\max \left( \frac{|y_i - \hat{y}_i|}{y_i} \right)$) varies between 14\% to -52\%. The \acrfull{mre} ($\frac{1}{n} \sum_{i=1}^{n} \frac{|y_i - \hat{y}_i|}{y_i}$) is below 10\% for all depths ranging between 2.3\% to 6\%. 
\newline
A more sophisticated approach used in model two, includes air temperature and precipitation (Fig.~ \ref{fig:ml_incl_temp_prec_excl_lag}). Weather data were available at 10~min intervals from the Longyeardalen Central Station (SN99857) of the Norwegian Centre for Climate Services. As \acrshort{er} and \acrshort{vwc} were only measured every six hours (3:30, 9:30, 15:30, 21:30), the temperature measurements for every 6 hours between two measurements were averaged and related to the later value, the precipitation was accumulated. 
\newline
Adding weather data decreased the \acrshort{mae} and \acrshort{mare} significantly for every depth . The reduction in \acrshort{mae} to below 0.005 $cm^3 cm^{-3}$ for all depths is particularly evident. This is equivalent to a decrease of the \acrshort{mae} of 7.3\% and 10.7\% for 10~cm and 50~cm and 43.0\% for 100~cm and even close to 70\% to 75\% for 150~cm and 190~cm. The \acrshort{mare} for 100~cm remains relatively high at 51.4\%.  Nevertheless, the predictions are mainly in the $\pm \ 3 \ \%$ range of the \acrshort{vwc} measurements, with a few exceptions. This is also shown by \acrshort{mre} values between 0.7 and 3.7\%. For 10~cm from late September until mid October there were no data because \acrshort{ec} reached zero and these were excluded. 
\newline
As the influence of air temperature and especially precipitation is assumed to be delayed differently for each depth due to an infiltration effect, a depth-dependent delay has been introduced. The delay was constraint between 0-72~h before the measurement to reduce coincidental minima which could result in overfitting the training data and reducing the prediction accuracy. In addition to temperature and precipitation, \acrshort{er} was also varied and delayed to enable potential forecasting. A summary of all delay times and associated errors is presented in the Tab.  \ref{tab:best_parameters_ml}. The errors can also be displayed as a function of the three delay times in a three-dimensional plot (App. \ref{AppendixC}). The cube can be reduced in size by limiting the potential delay time from 0-72~h to 24-72~h, 36-72~h, and so forth, which may be applicable to forecasts of the \acrshort{vwc}. The \acrshort{mae} minimizing delay times for all depths are summarized in Tab. \ref{tab:best_parameters_ml} and Fig.~ \ref{fig:ml_vwc_best_estimation}. The optimal results are obtained, when the temperature is incorporated into the estimation with immediate responsiveness, extending to a depth of 100~cm. For greater depths, the preferred temperature delay increases to 30~h and 36~h between the depth of 150~cm to 190~cm, respectively. The precipitation delay for shallow depths is initially significantly delayed (60~h), but then decreases with increasing depth to 12~h. 
\newline
The optimal estimation of \acrshort{vwc} uses the current \acrshort{er} measurements and combines this with depth-dependent delayed air temperatures and precipitation in Tab. \ref{tab:best_est_delays}. The \acrshort{er} delay was left zero despite there being similarly small errors for delays reaching up to 66~h (combined with other delay times for precipitation and temperature than Tab. \ref{tab:best_est_delays}). Nonetheless, these minima were neglected. Introducing these delays reduced \acrshort{mae} further by 16.8\% to 33.8\%. The \acrshort{mre} could also be reduced for all depths and 2.8\% were never exceeded. The \acrshort{mare} also decreased below 20\% for all depths. 
\newline
The best delay times for available parameters from 24-72~h are calculated and shown in Tab. \ref{tab:best_forecast_parameters}. The temperature dependence for the 24~h forecast demonstrates a gradual elevation with depth between the 24~h and 66~h. Conversely, the delay for precipitation exhibits considerable variability across all depths. A comparable trend is observed in the optimal delay of the \acrshort{er} measurements. In contrast to precipitation, a longer delay is identified as optimal for greater depths, with the variations being less pronounced (see table \ref{tab:best_forecast_parameters}). 
\newline
The \acrshort{mae} increased and so did the \acrshort{mare}. For 50~cm both \acrshort{mae} and \acrshort{mare} stayed similar while for 190~cm both errors decreased even further. The \acrshort{mare} increased significantly for 10 and 100~cm. Both, the best estimation and 24~h forecast remain mostly in the $\pm \ 3 \ \%$ range of the \acrshort{vwc} measurements, as quantitatively supported by the \acrshort{mre}, which stays well below 3\% except for 10~cm. There, it slightly increases from 2.52\% to 4.31\%. Nonetheless, some inaccuracies are visible for the 24~h predictions. This becomes more prominent for  36~h forecast models and 48~h forecasts (Fig.~ \ref{fig:ml_best_36h_48h_forecast}). The \acrshort{mae} and \acrshort{mare} increase partially significant.

\subsection{VWC estimation and forecast from ERT}

\begin{figure}[H]
    \centering
    \includegraphics[width=0.45\textwidth]{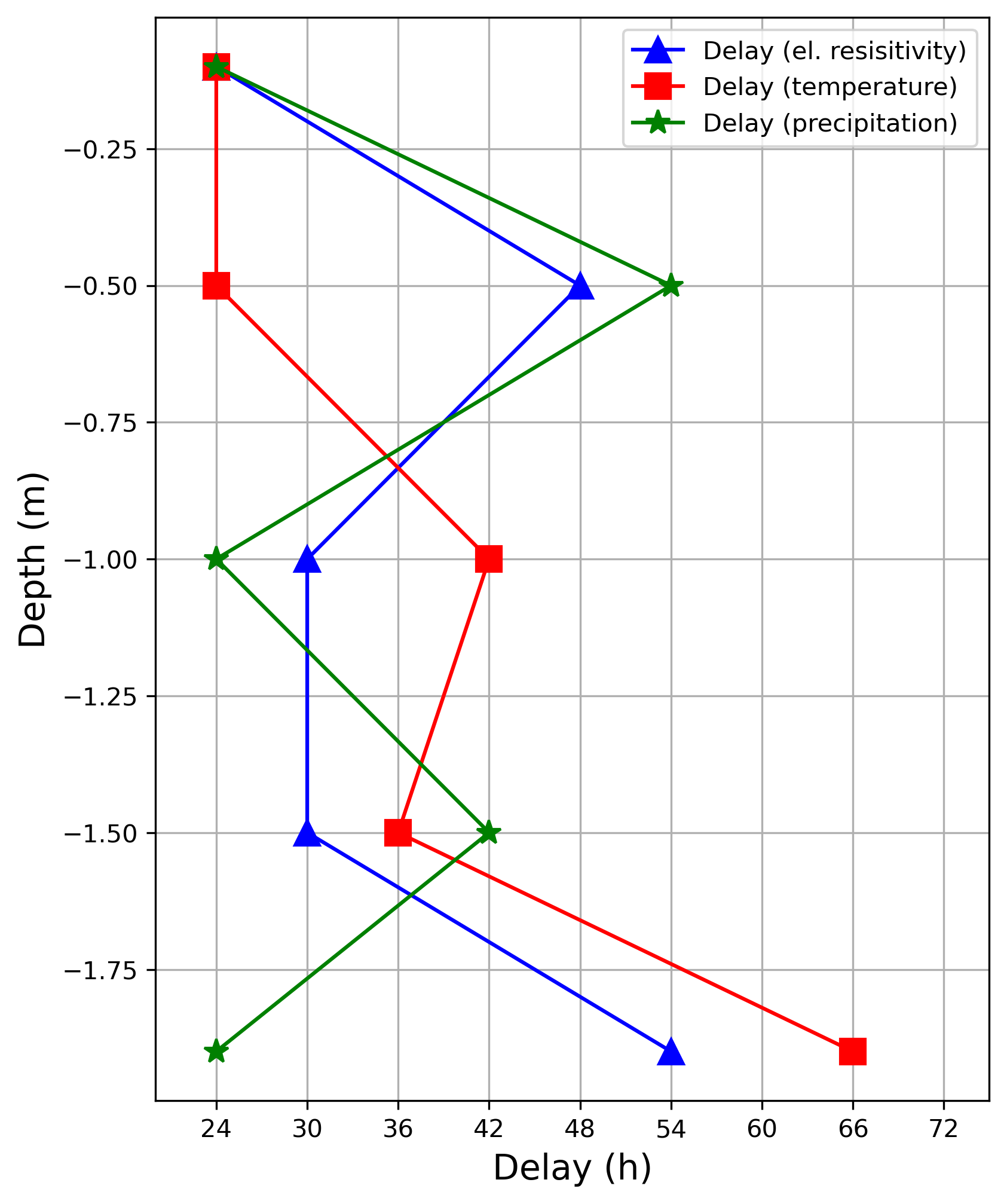} 
    \hfill
    \includegraphics[width=0.45\textwidth]{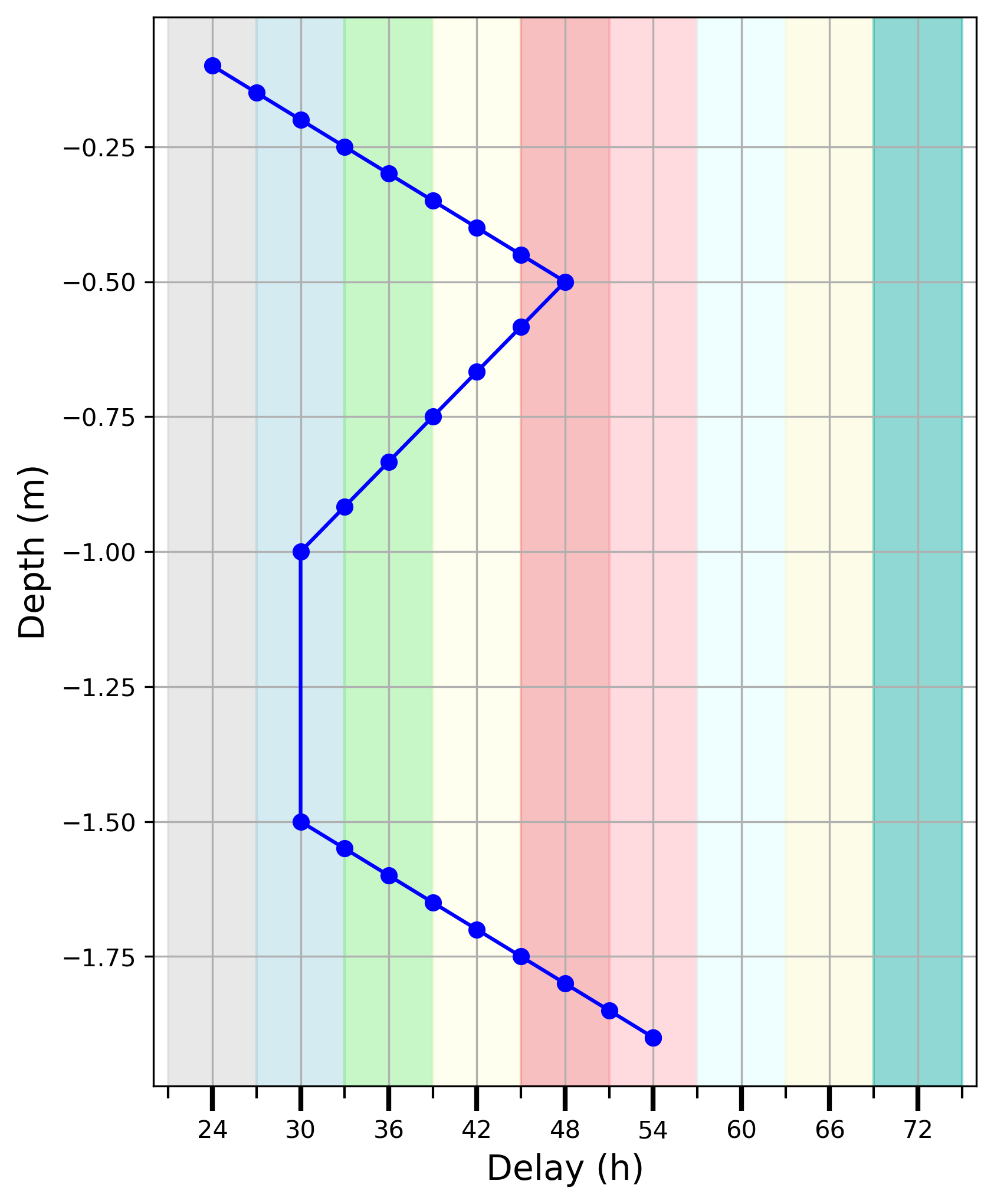}
    
    \caption[Depth-dependent delays]{Left: Interpolated 24~h forecast delays for all features depending on depth. Right: Interpolated 24~h forecast delays for \acrshort{er} measurements and shaded 6~h delay intervals.}
    \label{fig:delays_interpol}
\end{figure}

Up until this point, a distinct model has been developed for each depth, with unique delay times yielding the lowest possible error. To extend the predictions to the entire \acrshort{ert} area and to obtain a tomographic reconstruction of the \acrshort{vwc} estimation based on the \acrshort{er} measurement, it is necessary to interpolate the delay times for each depth. In the case of temperature and precipitation, the delays can be divided into 10~min intervals, as this is the smallest measured interval. For the best estimation, \acrshort{er} was taken into consideration without any delay, while for the forecast, \acrshort{er} was delayed by at least 24~h. Since \acrshort{er} is measured by the sensors every 6~h, the depth intervals are therefore generated summarizing every 6~h delays as indicated by the shaded regions in right Fig.~ \ref{fig:delays_interpol}.

\begin{figure}[H]
    \centering
    \includegraphics[width=1\linewidth]{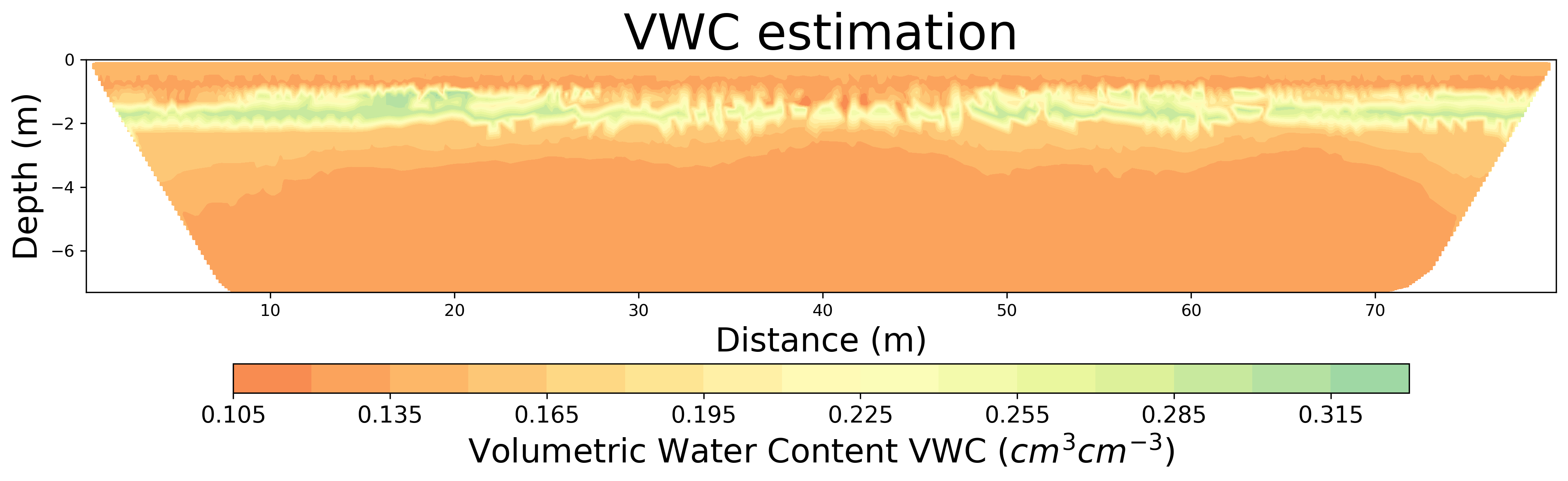}
    \caption[\acrshort{vwc} tomography]{\acrshort{vwc} estimation across the whole \acrshort{ert} area.}
    \label{fig:ert_vwc_est}    
\end{figure}

The interpolated delays and the different models for all depths can now be applied to both the \acrshort{ert} and the sensor \acrshort{er} measurements. Additionally, predictions were made with delays in accordance with the 24~h forecast. In the evaluation of the \acrshort{ert} values, only the most accurate estimation was used, given that no subsequent \acrshort{ert} were recorded at prior 6~h intervals.
\newline
It can be observed that both the predictions based on the \acrshort{er} sensor data and their best estimation agree well with the \acrshort{vwc} measurements. The deviations between the \acrshort{vwc} sensor measurements and the \acrshort{er} sensor \acrshort{vwc} estimations are scattered below 0.6~$\sigma$ and except for the 10~cm and 100~cm data also below 0.2~$\sigma$. For the \acrshort{vwc} predictions from the sensors, only 10~cm and 50~cm exhibited larger deviations with just under 1.5~$\sigma$ while the other depths deviated only minimally (<~1.0~$\sigma$). Contrary, the discrepancies between the optimal estimation and the prediction from the \acrshort{ert} are evident, with the former exhibiting a narrower range of error (below 8.2~$\sigma$), while the latter displays a somewhat broader range (between 0.2~$\sigma$ and 39~$\sigma$) (see also Tab. \ref{tab:borehole_vwc_est_pred_dev}). Nevertheless, in comparison to the sensor estimation and prediction, the error margins associated with the \acrshort{ert} estimations and predictions are significantly more pronounced. Furthermore, it is evident that the predicted \acrshort{vwc} of the \acrshort{ert} is in some instances significantly divergent from the measured values. The occurrence of discontinuities can be attributed to the use of a specific depth model in each of the shaded boundaries. Nevertheless, the \acrshort{vwc} for the entirety \acrshort{ert} area can be estimated in Fig.~ \ref{fig:ert_vwc_est}. It can be reasonably presumed that the discrepancies between the predicted and actual values are likely to be as significant in other areas as they are for the virtual borehole. Nonetheless, a moisture zone can be identified at depths between one and two meters, exhibiting relatively constant characteristics across the entire \acrshort{ert} line.

\begin{figure}[H]
    \centering
    \includegraphics[width=1\linewidth]{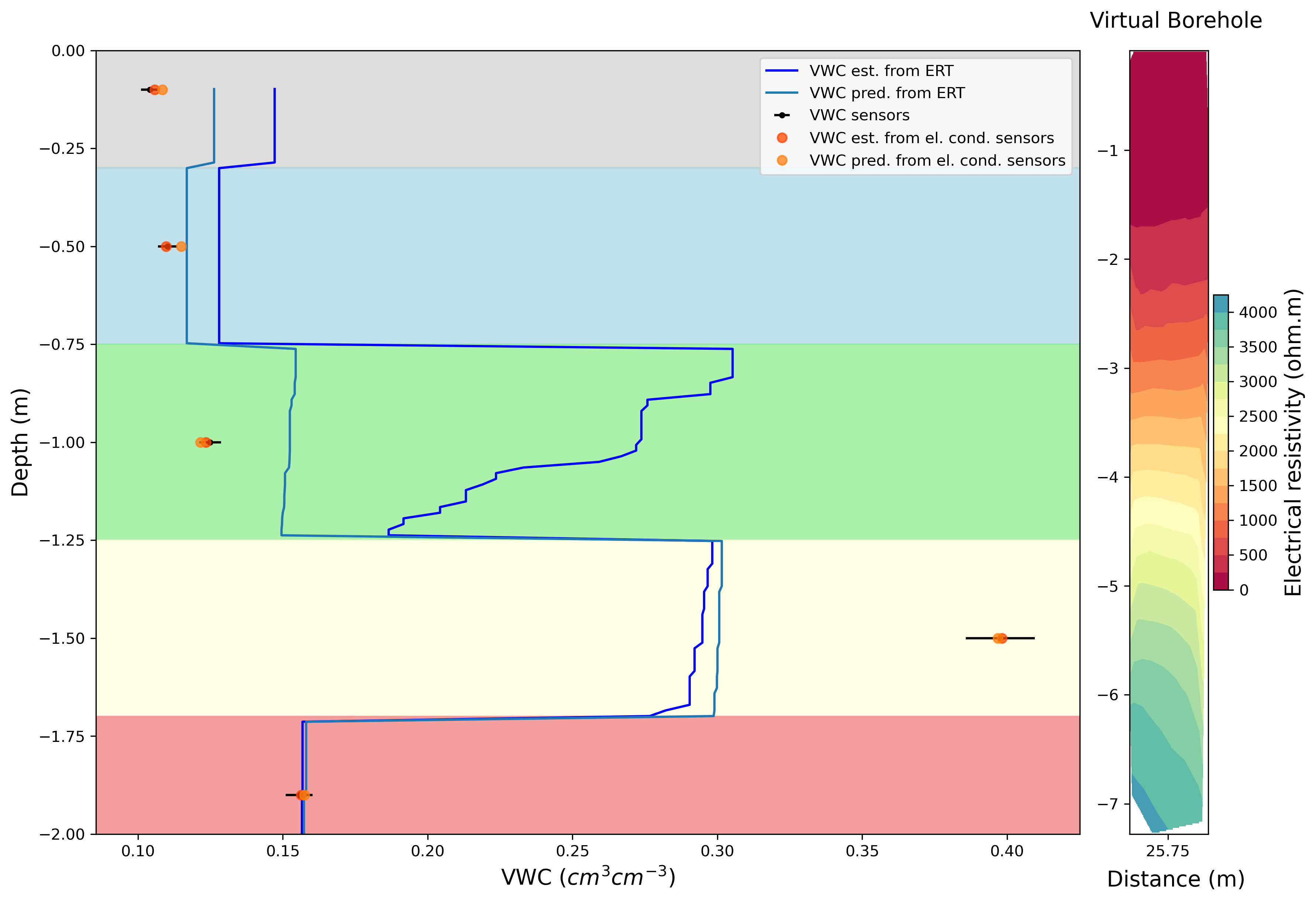}
    \caption[Comparison of VWC derived from ER and ERT]{Left: Comparison of the \acrshort{vwc} sensors, \acrshort{er} sensor estimated and predicted \acrshort{vwc} and \acrshort{ert} estimated and predicted \acrshort{vwc}.}
    \label{fig:vwc_borehole_est}
\end{figure}

The best \acrshort{vwc} estimation of site 2 on September 04th. 2024 is shown in Fig.~ \ref{fig:ert_vwc_est}.

\section{Survey one: short-term temporal variations}

The objective of this survey was to assess the efficacy of the models in response to fluctuations or short-term changes in temperature and precipitation. Two measurement series were conducted, the first without significant changes in precipitation and air temperature and a second one including heavy rainfall. 
\newline
The total precipitation of the first series of measurements was 0.7~mm at an average temperature of 2 to just under 13~$^{\circ}C$ as shown in Fig.~ \ref{fig:temp_prec_st_ert}. The change in \acrshort{ert} is between -0.5\%~log $\Omega \cdot m$ and 0.7\%~log $\Omega \cdot m$ and is limited to small scale anomalies (see also Fig. \ref{fig:log_change_resis_st}). The \acrshort{er} means do not change significantly (Fig.~ \ref{fig:st_means}). As expected, the \acrshort{vwc} also changes little, with a maximum change of -0.15\% log $cm^3 cm^{-3}$ to 0.35\% log $cm^3 cm^{-3}$ (Fig. \ref{fig:log_change_vwc_st}). 
\newline
For the second period, a control measurement was conducted 44~h before a large precipitation event. About 1~mm of rain fell between the control measurement and the heavy rainfall, the remaining 8.2~mm fell immediately before and during the first measurement. The average temperature was limited to 4.5-11~$^{\circ}C$. In contrast, the first period, there was a larger change of -0.7-0.6\%~log~$\Omega \cdot m$ of the \acrshort{er}, but only locally limited, as can be seen from the constancy of the mean values (Fig.~s \ref{fig:sst_means} and \ref{fig:log_change_resis_sst}). The means remained almost unchanged for all depths. At 10~cm the \acrshort{er} increased by about 15~$\Omega \cdot m$ in the last measurement. The resulting \acrshort{vwc} estimates were also constant. Depending on the depth, the used delays of the precipitation data are as shown above (Tab. \ref{tab:best_est_delays}) between 24~h and 66~h, which cancels the influence of the rain shower on the estimation of the \acrshort{vwc} so close after the rain.

\section{Survey two: spatial overview and validation}

\begin{figure}[H]
    \centering
    \includegraphics[width=0.49\textwidth]{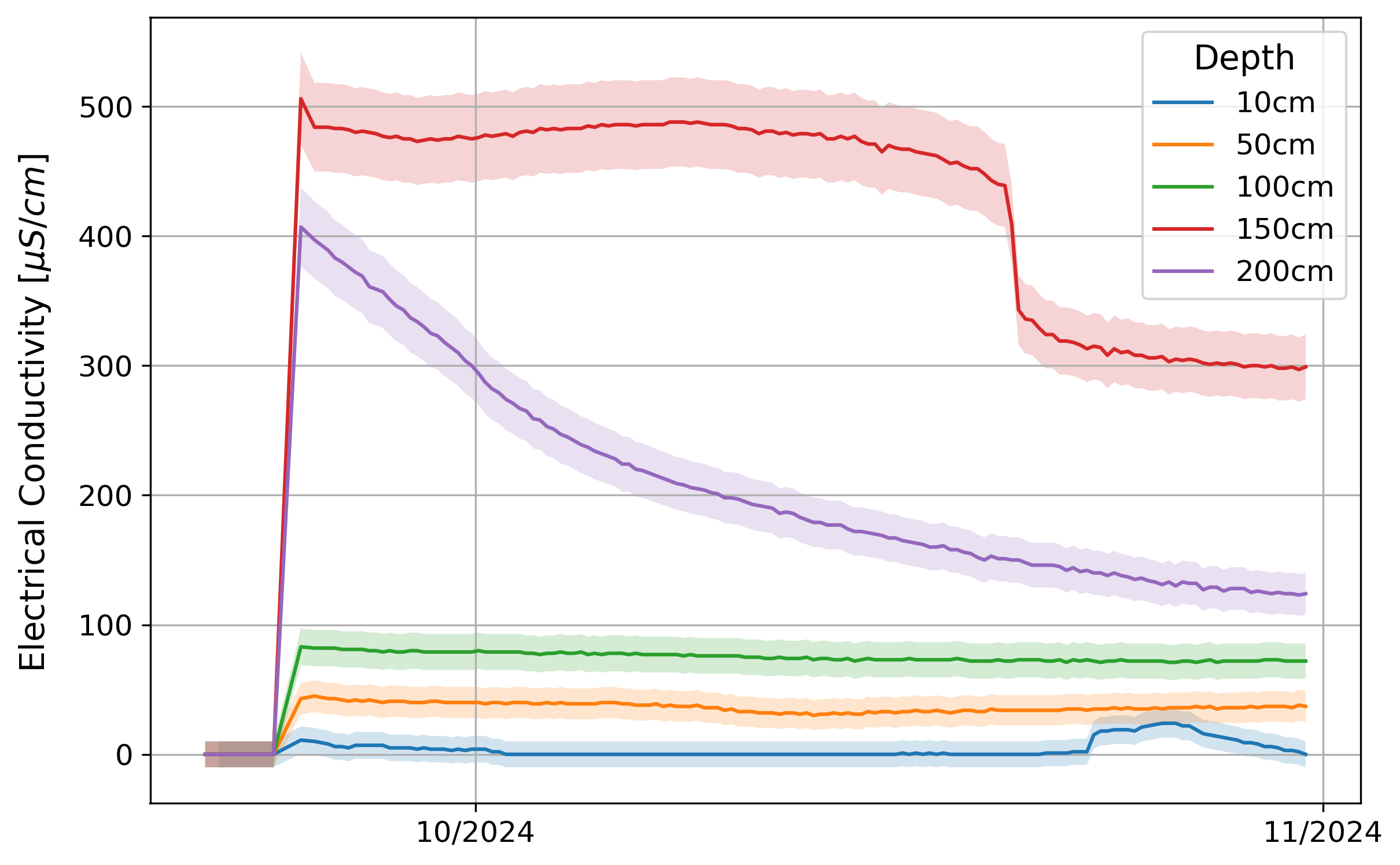} 
    \hfill
    \includegraphics[width=0.49\textwidth]{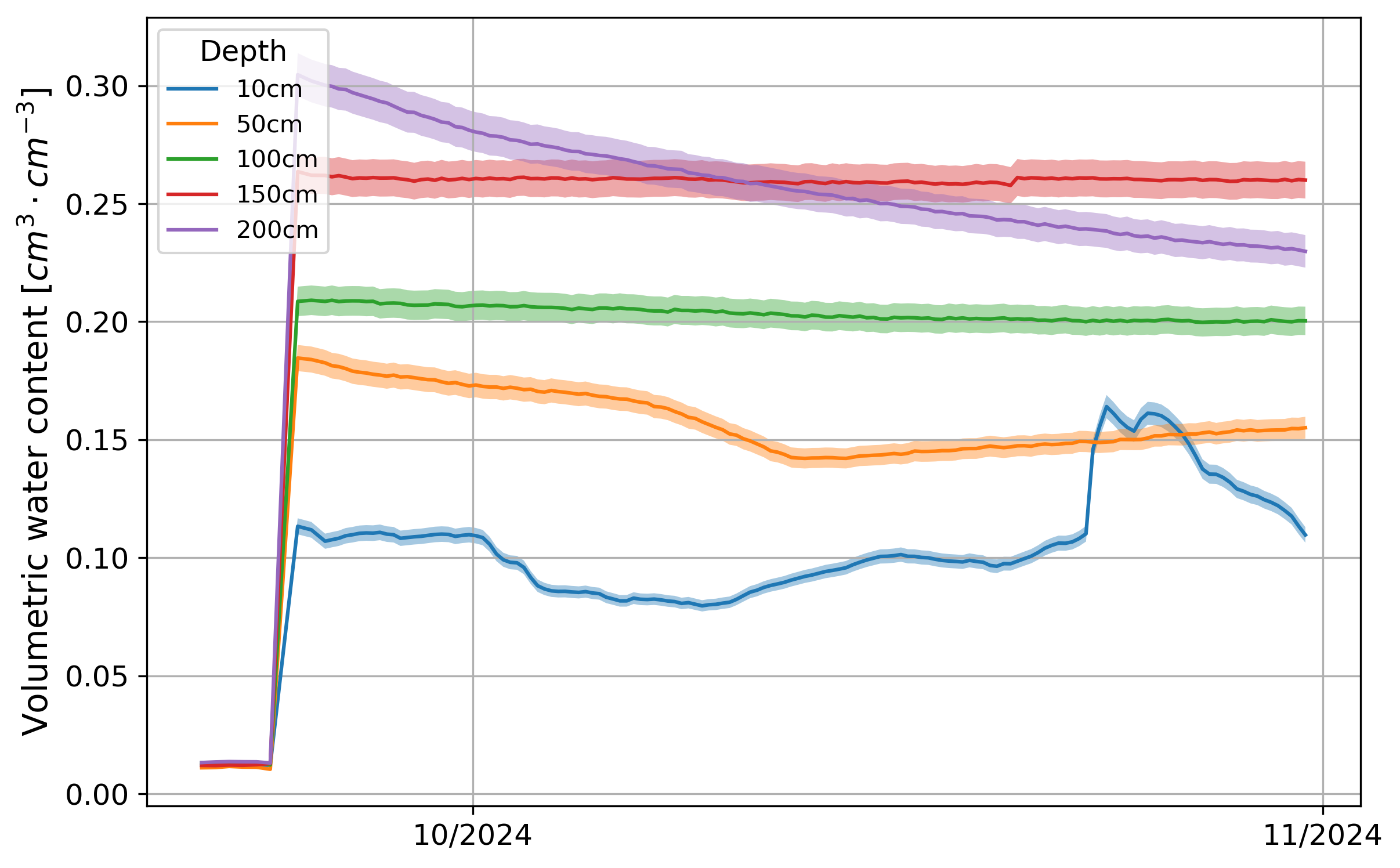}
    
    \caption[ER and VWC data for ID S 3.12]{\acrshort{er} (left) and \acrshort{vwc} (right) measurements for ID S 3.12.}
    \label{fig:gy_sensor}
\end{figure}

\begin{figure}[H]
    \centering
    \includegraphics[width=1\linewidth]{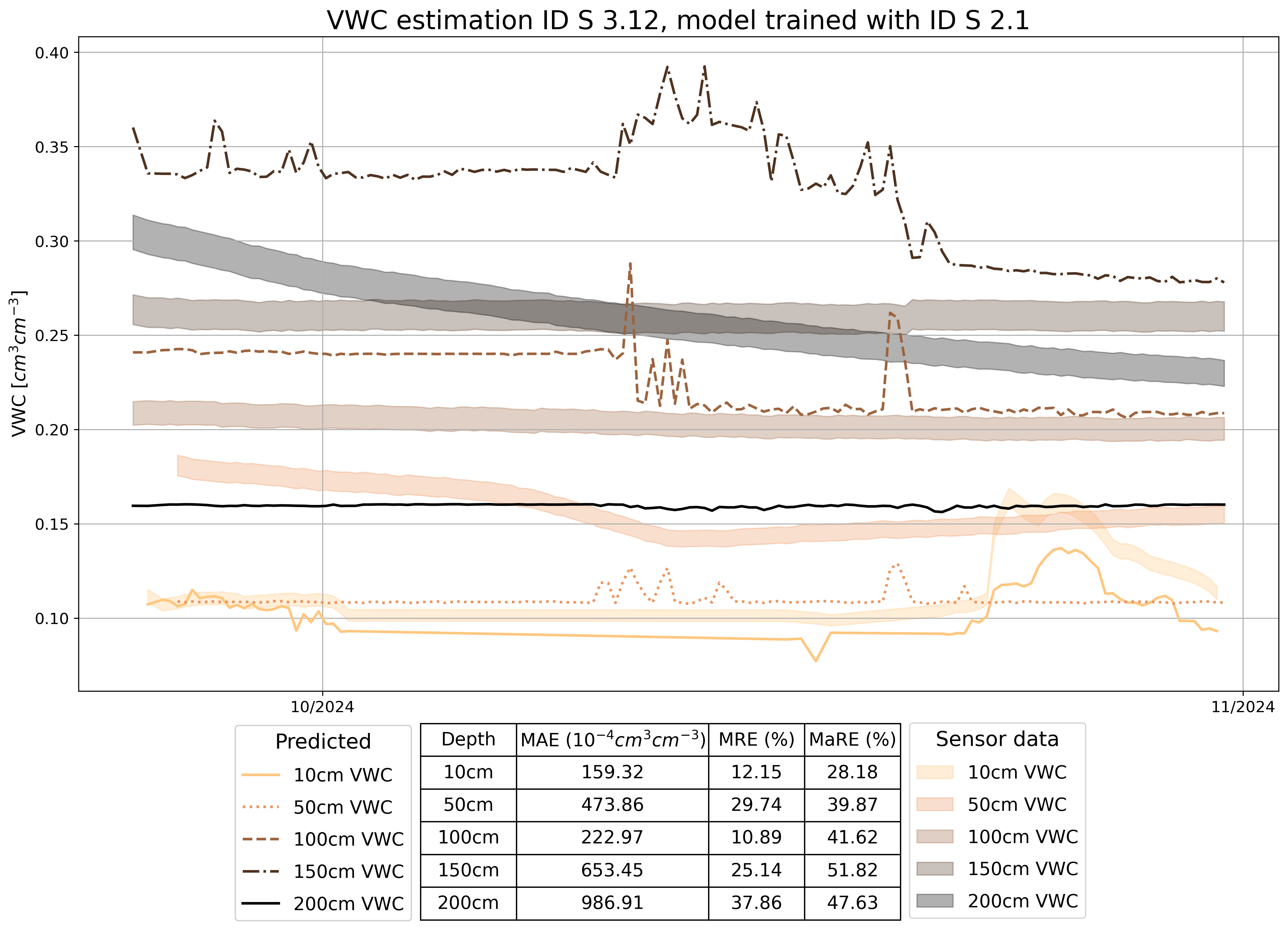}
    \caption[VWC location validation with ID S 3.12]{The trained \acrshort{rf} models with ID S 2.1 data were applied to the sensor values for sensor ID S 3.12.}
    \label{fig:gy_vwc_est}

    \vspace{0.5 cm}
    
    \includegraphics[width=1\linewidth]{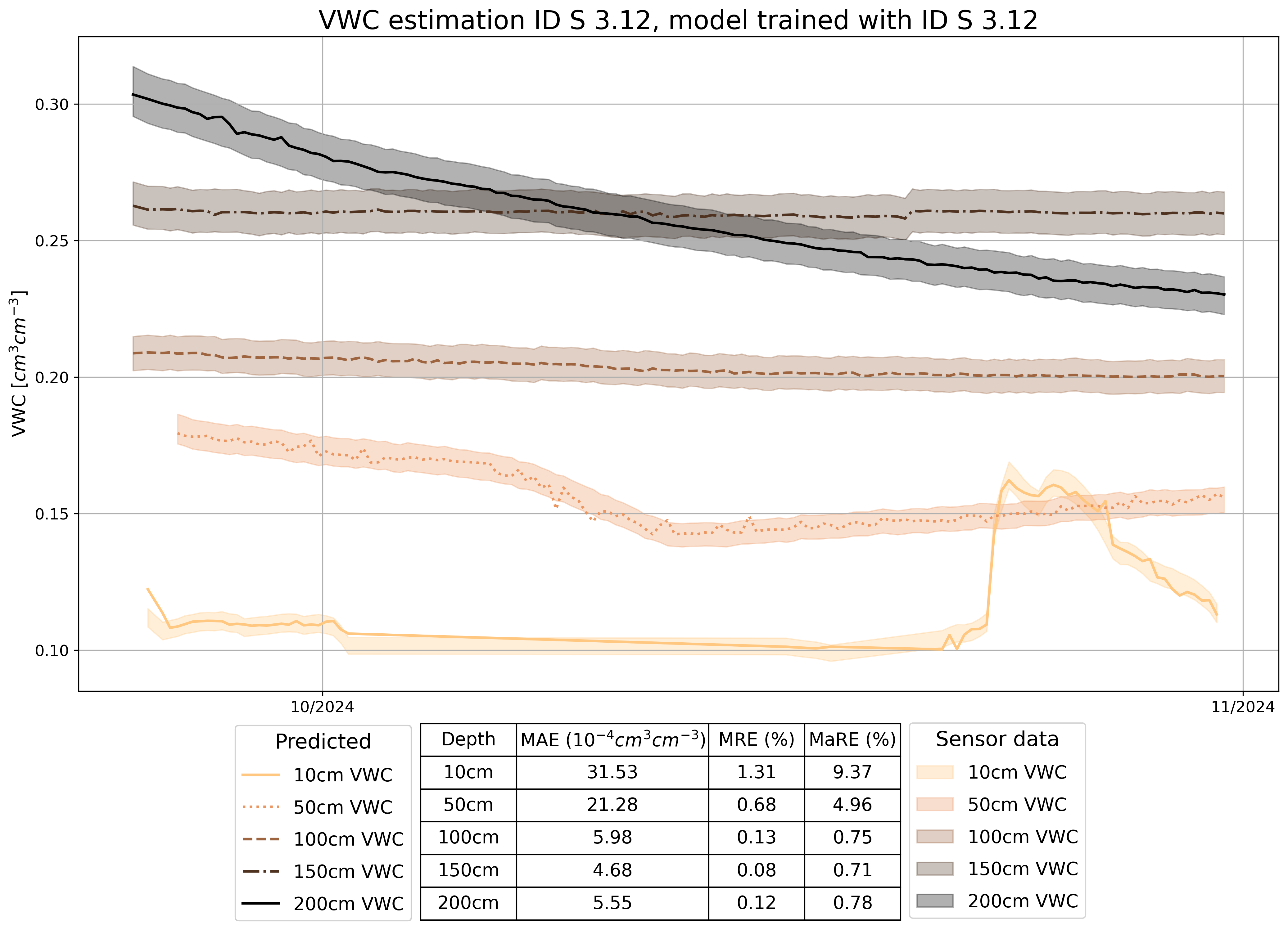}
    \caption[VWC training and est. with ID S 3.12]{Separate \acrshort{rf} models were trained based on ID S 3.12 data and tested with ID S 3.12.}
    \label{fig:gy_vwc_best_est}
\end{figure}

The second survey was conducted with the objective of transferring the results obtained for location 2 to the other research areas. The ERT and sensor measurements were performed at all sites in accordance with the methodology employed at site 2. 
\newline
In order to create a unified representation of the recorded \acrshort{ert}, the sensor temperature data was processed to identify the depth at which the temperature reached 0 $^{\circ}C$. This value was then marked as a white line in Fig.~ \ref{fig:locations_ert} (at the end of the section). The deepest thawing was observed at study site 3 at $D_{S31}(T=0^{\circ}C) = (-2.037 \pm 0.005) \ m$ whereas ID 4 and 5 lied at $D_{S4}(T=0^{\circ}C) = (-1.900 \pm 0.005) \ m$ and $D_{S5}(T=0\ ^{\circ}C) = (-1.958 \pm 0.015) \ m$. The shallowest depths, nearly 40~cm above $D_{S31}$ were the areas 1 and 6 with $D_{S11}(T=0 \ ^{\circ}C) = (-1.61 \pm 0.21)\ m$ and $D_{S6}(T=0 \ ^{\circ}C) = (-1.634 \pm 0.003)\ m$. The sensors S 1.21 and 1.22 were not installed at that time. The \acrshort{er} at these depths ranged from just below 200~$\Omega \cdot m$ at ID 3 to nearly 500~$\Omega \cdot m$ at ID 5. Precipitation between the measurements accumulated to 2.40~mm and the air temperature was stable between the measurements. Differences due to meteorological influences were minimized (see Fig.~ \ref{fig:prec_temp_locations}). 
\newline
Significant variations in \acrshort{er} values between different locations were observed. Generally, study sites 1 and 3 show low \acrshort{er} values, followed by ID 2 and 4 whereas 5 and 6 exhibit very high \acrshort{er}.  
\newline
Where available (S 3.12 and S 3.2), the estimated and predicted \acrshort{vwc} based on the models trained with measurements from site 2 was tested against the sensors of another site. A comparison of Fig. \ref{fig:gy_sensor} with Fig. \ref{fig:resis_long_term} and Fig. \ref{fig:vwc_long_term} shows that the \acrshort{ec} and \acrshort{vwc} ranges of the two sites differ significantly. Corresponding deviations for the \acrshort{vwc} estimations become visible in Fig. \ref{fig:gy_vwc_est}. 
The trained model for 190~cm was used to estimate the 200~cm sensor data. All evaluation metrics indicated a significant deviation between the estimated and measured values. This was also confirmed qualitatively. The estimates did not fit the measurements well at any point. The 10~cm and 100~cm estimates were the best, with an \acrshort{mre} of just under 12\% and 8\%. The estimates at other depths exhibited substantially higher absolute and relative errors. The estimates for ID S 3.2 and S 1.2 showed similar deviations. Reverse training with sensor data from ID S 3.12 also led to unsatisfactory results for estimates of ID S 2.1 values.

\noindent\begin{minipage}{\textwidth}
    However, if separate models are trained on the data from ID S 3.12, this data can also be visualized with high accuracy, resulting in minimal \acrshort{mae} and \acrshort{mre} (see Fig. \ref{fig:gy_vwc_best_est}). There are also no outliers (\acrshort{mare} below 1\% except 10~cm (1.3\%)). The limited available time interval and meteorological variations may have been smaller as the ground was generally colder. The data were trained with the best delays from the best estimate for ID S 2.1.

    \centering
    \includegraphics[width=0.99\linewidth]{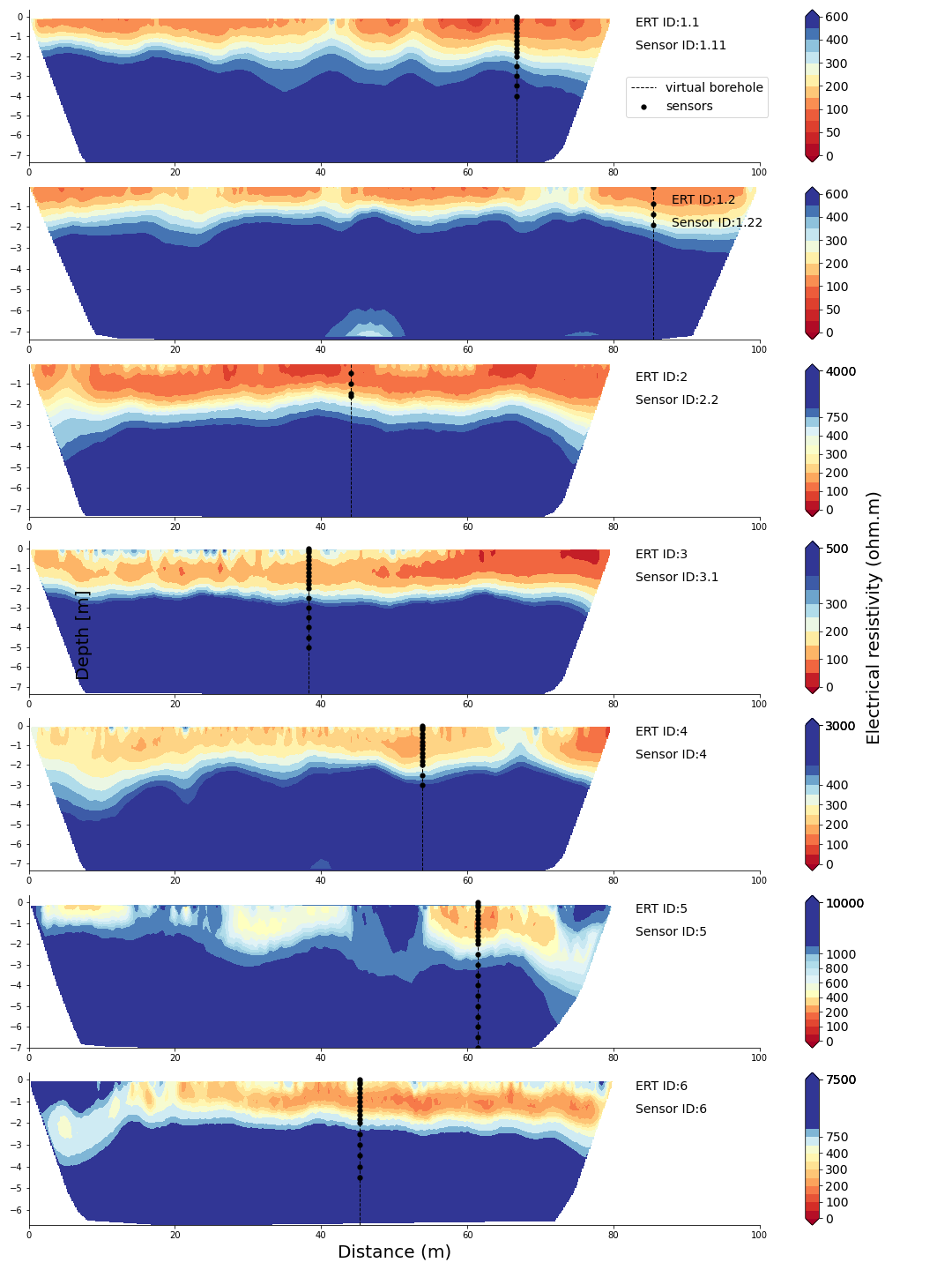}
    \captionof{figure}[\acrshort{ert} from the different study sites across Longyearbyen]{\acrshort{ert} from the different study sites across Longyearbyen, calibrated by their depth at 0~$^{\circ}C$.}
    \label{fig:locations_ert}
\end{minipage}

\section{Survey three: temporal long-term temporal variations}

\begin{figure}[h!]
    \centering
    \includegraphics[width=1\linewidth]{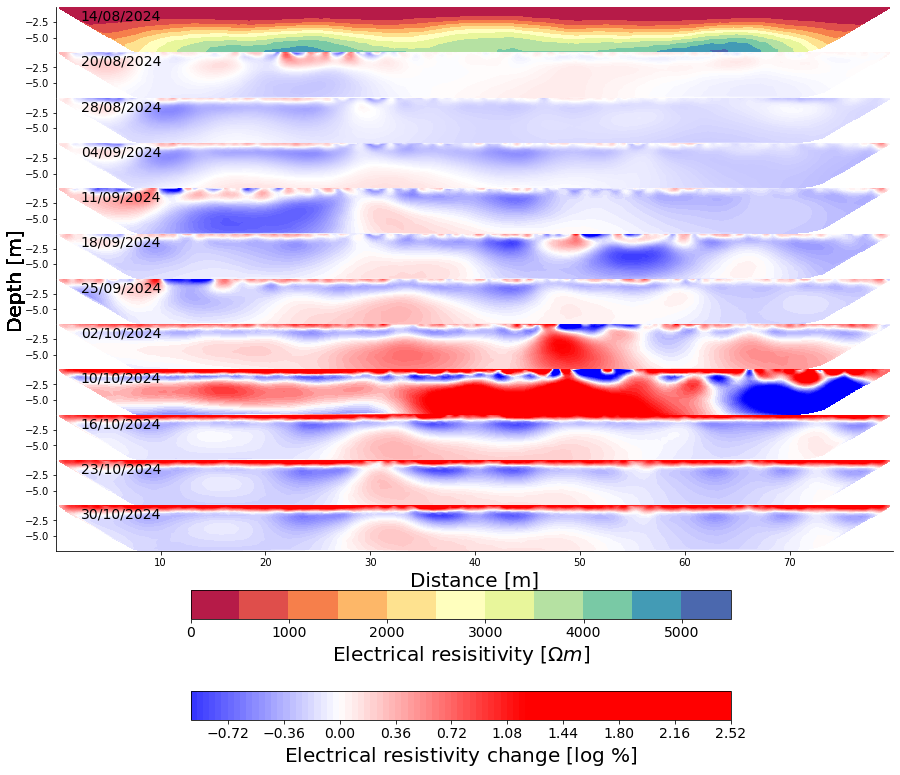}
    \caption[Seasonal long-term variations (ID2)]{Seasonal long-term variations for study site 2.}
    \label{fig:long_term}
\end{figure}

The aim of this third survey is to investigate seasonal differences in the \acrshort{er} and, in particular, the transition from summer and the thawed active layer to the freezing of the entire ground. The \% ~log~$\Omega \cdot m$ differences are shown in Fig.~ \ref{fig:long_term} for the weekly measurements with the first \acrshort{ert} as reference.

Freezing becomes visible from the beginning of October. In the uppermost layer and in the permafrost, the \acrshort{er} increases strongly and reaches magnifications of 2-2.4 \% ~log~$\Omega \cdot m$. The \acrshort{er} in the permafrost also increases mostly, except for distances from 65~m to 80~m. A small layer for which the \acrshort{er} decreases, is evident at 1.5-2~m depths. There is still water in a liquid state (see also Fig.~ \ref{fig:vwc_long_term}). In the following week (16.10.2024), the air temperature increased again, whereupon the \acrshort{er} decreased again. The topmost layer was still frozen and stayed frozen until the measurement period end. 
The overall \acrshort{er} stayed in the following weeks relatively constant. 
\newline
Complete freezing was not recorded, as the measurements had to be canceled at the beginning of November after the risk of avalanches increased due to snowfall.

%% file: 5Discussion.tex
\chapter{Discussion}
\label{ch:discussion}

In this work, sub-surface \acrshort{er} and \acrshort{vwc} structures in Longyearbyen were investigated using \acrlong{ert}. The \acrshort{ert} data was fitted to temperature data, the thaw depth and permafrost estimation were visualized. Next, the \acrshort{ert} was compared with the \acrshort{er} sensors.
\newline
A random forest model was trained using \acrshort{er} and meteorological data to predict \acrshort{vwc}. To validate the general accuracy and robustness of the resulting model, different scenarios were selected and evaluated, representing temporal, meteorological and local variations. Finally, the seasonal change of permafrost from summer to winter was observed.

\section{\acrshort{ert} monitoring}
\subsection{\acrshort{ert} acquisition}

The active layer was not yet frozen at the beginning of August when the first measurements started and neither at the beginning of September when the other sites were measured. Therefore, it was relatively easy to insert the electrodes into the soil. However, in some places, particularly at sites 5 and 6, there were a lot of stones, which made it difficult to place the electrodes. This was also reflected in the higher contact resistance. Nevertheless, all \acrshort{ert} could be carried out as planned. Straight \acrshort{ert} lines could be measured with 101 and 81 electrodes respectively for study site 1.2 and the others. 
It was not always feasible to maintain a strict spacing of one meter from neighboring electrodes, as is evident from the GPS data. Height data acquisition with the Leica GNSS 16 device was not successful, the measured height differences between adjacent electrodes were sometimes obviously higher than in reality. A satisfactory explanation for this has not yet been found. Instead, heights were determined using GPS data and a \acrshort{dem} model with a 2.5~m resolution. There were also some large differences between adjacent electrodes due to the resolution of the \acrshort{dem}, which had to be interpolated. This involved averaging 3 adjacent electrode heights. This inevitably leads to a linearisation of the \acrshort{dem} line, which is only an averaged representation of reality. 
\newline
The filtering of the data was successful without major limitations, the eliminated data was always less than 5-10\% of the total data. The criteria defined for sorting out the erroneous data followed the conventions of current research \cite{1_herring_2023_bestpractice, 35_Wicki_2022_landslides}. The data inversion was carried out with a least-squared data misfit norm instead of an L1 because it was not implemented yet in the current ResIPy version (3.3.5). Although an L1 norm is advantageous for characterizing the boundary between the active layer and permafrost, however, since the focus for the \acrshort{vwc} estimates lied on the active layer, these possible uncertainties are insignificant.

\subsection{Characterizing \acrshort{ert} and permafrost soil estimation}

The presented \acrshort{ert} data appear homogeneous, both qualitatively and quantitatively. The maximum relative  error for the meter by meter depth subdivision of the \acrshort{ert} was 40\%, with deviations between 15\% and 25\% calculated for over 50\% of the total depth. This is relatively low compared to \acrshort{ert} in other areas \cite{36_meng2024rf_ert_boss, 35_Wicki_2022_landslides, 21_Scapozza_2014_ert_alpine}. 
\newline
The comparison of the sensor measurement with the \acrshort{ert} showed no significant deviations, except for the measurement at 150~cm. The insignificant deviations result from the large combined errors, which in turn are largely determined by the errors of the sensors. According to the manufacturer, the \acrshort{ec} accuracy is $\pm (5 \% + 0.01 dS/m)$ from 0-10~dS/m and $\pm \ 8 \ \%$ for values between 10-20 dS/m. For very low values (about 5-50 \ $\mu S / cm$), as measured at each site, the error is in the range of 10\% to 150\% of the measurement. In this context, it seems reasonable to assume that the actual deviations for all depths are as high as the ones for 150~cm. The \acrfull{rmse} of the \acrshort{ert} for all measurements is around 1.0. 
\newline
The depths $D_{S21}(T=0 \ ^{\circ}C) = (-1.82 \pm 0.65)\ m$ and $D_{S22}(T=0 \ ^{\circ}C) = (-1.48 \pm 0.07)\ m$ have been determined as temperature zero depths from the temperature sensor data at study site 2. These are accurate for 28 August 2024 as they are continuously influenced by meteorological conditions. The error of $D_{1}$ is relatively high at 35.9\%, which is due to the fact that only three sensors are placed in the relevant area, one of which, at 2.5~$^{\circ}C$, was already far away from $D(T=0)$. It is possible that a linear fit was applied to the exponential part of the permafrost temperature curve. An adjustment would require an exponential fit to the given data or a further discrimination of the upper sensor, which in turn would lead to a higher uncertainty as only two sensor measurements would be available for estimation.
\newline
As the average temperature decreased in the following period, it can also be assumed that the measured thaw depth also represents the thickness of the permafrost. The assumption includes, that the thaw depth was not deeper last year. The depths correspond to the surrounding measured annual maxima of the active layer \cite{SESS2020}.
Herring et al. \cite{0_herring2021temp_correction} argued for an improved interpretation of \acrshort{ert} with temperature correction, however, it was highlighted that there is still much research to be done in this field, especially concerning partially frozen ground. That is why the data were not temperature corrected, one should be aware of interpretation.

\subsection{Sensor data}

The seasonal temperature variations decrease with depth, which is expected, as shallower depths are more influenced by seasonal and daily air temperature variations than greater depths (compare Fig.~ \ref{fig:temp_long_term}). There is a sharp rise in temperature at the end of May 2024 for all depths. The hypothesis is, that when the topmost layers thaws, the water penetrates along the sensor cable into deeper layers and washes around the sensors. Since the water has just thawed, it has a temperature of around 0~$^{\circ}C$. This is supported by the observation that the sudden rise occurs when 10~cm is at 0~$^{\circ}C$. Moreover, the \acrshort{vwc} at all depths increases drastically simultaneously (Fig.~ \ref{fig:vwc_long_term}). 
\newline
The \acrshort{vwc} measurements are also relatively accurate with a $\pm$~3\% deviation. The peaks at 10~cm are attributable to precipitation, while the peaks at greater depths are due to subsequent successive infiltration. A decrease as rapid as the increase in the \acrshort{vwc} curve suggests that the water continues to infiltrate downwards. The constant high level for 150~cm from mid-August is probably due to the fact that the frozen part of the ground begins between 150~cm and 190~cm, so that deeper penetration of the water is no longer possible and the water remains trapped. The decrease in \acrshort{vwc} in early September can be attributed to the freezing of the water, given the decreasing temperature trends during this period. 
\newline
The \acrshort{ec} sensor readings (see Fig.~ \ref{fig:resis_long_term}) are difficult to interpret due to large errors. The Teros 12 sensors are clearly not suitable for such poorly conductive soils, so this data should be used with caution. Therefore, the quantitative data, especially in the following sections, are only recognized as reasonable under these circumstances.
\newline
On 18 September 2024, an \acrfull{aert} system was installed next to the \acrshort{ert} ID 2 line, so that the \acrshort{ec} sensors can be replaced by repeated \acrshort{ert} in the future. Successful commissioning of the \acrshort{aert} was achieved on 07 November 2024, after the field laptop arrived. A detailed description can be found in Ch. \ref{ch:conclusion}.

\section{Estimating and predicting \acrshort{vwc} from \acrshort{er} measurements}

In the research field of hydrology, one current focus is the search for a relationship between \acrshort{er} and \acrshort{vwc} \cite{36_meng2024rf_ert_boss, 0_terry2023field}. \acrshort{ert} offers a cost-effective, minimally invasive and real-time alternative to the locally very limited, invasive sensor measurements \cite{36_meng2024rf_ert_boss}. Common approaches to correlate \acrshort{er} and \acrshort{vwc} include curve fits, which requires the use of additional rock physics models and sample data \cite{35_Wicki_2022_landslides}. Moreover, this approach may be applicable only at the local level due to the specification of local parameters \cite{0_schwartz2008quantifying}. Two common curve fits include the Archie equation (equation \ref{eq:archie}) and the Waxman-Smits equation. Both equations are based on empirical parameters and are only reliable for certain soil types \cite{28_archie_1942_beginning_ert}. Furthermore, \acrshort{er} is not only influenced by \acrshort{vwc} but also and not limited to mineral composition, temperature, precipitation and microbial activity \cite{0_schwartz2008quantifying}. 
\newline
The combination of remote sensing and machine learning has previously facilitated significant advances in the estimation of near-surface soil moisture content from satellite imagery \cite{0_adab2020machine, 0_ahmad2010estimating}. However, the spatial and depth resolution is insufficient to provide added value for local applications. 
\newline 
A compromise between point sensor measurements and area-wide analyses on a kilometer scale is represented by \acrshort{ert}. The present research employs machine learning models to correlate \acrshort{ert} and \acrshort{vwc} \cite{0_terry2023field, 0_cussei2020estimation}. In a study published in November 2024, Meng et al. \cite{36_meng2024rf_ert_boss} compared various machine learning models for estimating \acrshort{vwc} based on \acrshort{ert} data. The results demonstrated that the \acrshort{rf} model outperformed all other tested models, exhibiting the lowest \acrshort{rmse}, the highest $R^2$, and the most robust performance in the validity check. The random selection of possible features for splitting on a node has been set to two for all models. This provides greater robustness and a further reduction in variance due to the additional randomness \cite{0_murphy2012machine_learning}. However, this is achieved at the potential compromise of further accuracy, as the best split parameter is not always included and thus the best split possible is not realized. 
\newline
In this thesis a separate \acrshort{rf} model was trained for each depth.  This ensures a better visualization of the \acrshort{vwc} depending on the height, while depth-dependent delays can be implemented separately. In addition, better output stability can be expected as the respective input spaces are more centered. However, a favorable advantage still needs to be validated. The training and testing data from 18. June 2024 to 31. October 2024 yielded 370 to 500 data points after excluding measurement times with recorded electrical conductivities of 0~$S/m$, as otherwise the electrical resistance would approach infinity. Comparable models were trained and tested with 270 data points, with further data used for validation purposes \cite{36_meng2024rf_ert_boss}. The onset of the training data from 18 June 2024 was selected to ensure the inclusion of the entirety of the available data set, while simultaneously excluding any instances of inaccuracy pertaining to the thawing of the upper layer and the subsequent infiltration of water around the sensors, as previously described. 
\newline
This thesis employs an extensive temporal validation with survey one and a comprehensive spatial validation with survey two. Three evaluation metrics were selected to quantify the prediction outcomes. The \acrfull{mae} was chosen as a measure of the overall predictive power, whereas the \acrfull{mre} was chosen to represent the discrepancy between the predicted and actual values in relation to their magnitudes. In addition, the \acrfull{mare} was considered as a further measure of outliers.
\newline
A detailed evaluation of the individual models revealed that on occasion the \acrshort{mare} exhibits a notably elevated level. The model for 100~cm exhibited significant difficulties in consistently providing reliable predictions. This is particularly evident after the pronounced peak observed in mid-July. Furthermore, the \acrshort{mae}, \acrshort{mre} and \acrshort{mare} values are also higher compared to those of other models. Additionally, substantial individual deviations (up to 52\%) are apparent. This is likely attributable to the simplicity of the models, which only receive two inputs (\acrshort{er} and \acrshort{vwc}) and are trained based on depth. Without this depth information, the models' performance is likely to be even more adversely affected, as demonstrated by Meng et al \cite{36_meng2024rf_ert_boss}. The 150~cm model also encountered difficulties in achieving a consistent estimate. However, the errors remain relatively low, due to the fact that only relatively short (in absolute terms) deviating peaks are present, and the total \acrshort{vwc} is also increased in comparison to the other depths. This results in lower relative errors. From late September until mid October the \acrshort{vwc} for 10~cm was affected by freezing. The models were programmed that data points with \acrshort{ec} measurements of zero were excluded. One solution would include to set a relatively low value for the \acrshort{vwc} in situations where the EC is zero. Nevertheless, in the cases of ground freezing, stability is enhanced, which makes the estimation of \acrshort{vwc} values an arguably superfluous endeavor.

\subsection{Influence of meteorological parameters}

Since air temperature and precipitation have a major impact on \acrshort{er} and \acrshort{vwc}, they were included in the second model \cite{35_Wicki_2022_landslides, 0_cao2021differential_infiltration}.
The meteorological data was taken from Longyeardalen Central Station (SN99857) approximately 300~m from the study site 2. It is plausible that local disparities may have been introduced as a result. Nevertheless, a decrease across all metric values at each depth was observed. This allows for the hypothesis, that, provided that meteorological influencing factors remain relatively constant, there is no requirement to install an additional temperature and precipitation measuring station in immediate proximity to the \acrshort{ert}. Nonetheless, the installation of such a measuring station would yield even more accurate results. Including the additional features ledt \acrshort{mre} values below 4\% for all depths. Outliers have not yet been effectively filtered. These are primarily observed in instances of pronounced peaks in the \acrshort{vwc}, which exhibit a high degree of correlation with the underlying trend. However, due to the pronounced gradients, the models are unable to provide precise quantitative values.

\subsection{Influence of depth-dependent delays}

The infiltration of precipitation can be calculated using a variety of equations, including Richard's and Horton's \cite{0_serrano2004modeling, 0_horton1940equation}. The solution of these equations frequently necessitates the utilization of numerical techniques and the incorporation of empirically derived supplementary data. However, since precipitation has a significant influence on \acrshort{vwc} \cite{35_Wicki_2022_landslides, 0_serrano2004modeling}, a different approach was considered: The precipitation, air temperature and \acrshort{er} were delayed depth-dependent. For example, a cumulative precipitation of 4~mm over the previous 6 hours in the model at 50~cm is only assigned to the data point with the \acrshort{vwc} measurement 12 hours later. 
In this method, precipitation, air temperature, and \acrshort{er} curves are processed accordingly. Prior to this, no modeling assumptions regarding an ideal depth-dependent delay were implemented. A delay range from 0-72~h was defined, as there is the possibility of random correlations arising for delays outside this range due to the size of the training data set. The combinations of delays for all three parameters were compared with one another (see Tab. \ref{tab:best_parameters_ml}), and the \acrshort{mae} were calculated. In the following, two combinations of delays were further analyzed. 
The optimal estimation of the \acrshort{vwc} was described with the parameters presented in Tab. \ref{tab:best_est_delays}. The \acrshort{er} delay was adjusted from 66~h to 0~h for 50~cm and 190 cm, effectively disregarding the total \acrshort{er} delay.  It seems probable that the two 66~h delays are the result of local minima. The optimal delay combination for both depths was calculated and resulted in only insignificantly higher \acrshort{mae}. 
\newline
The introduction of delays resulted in a further reduction in the \acrshort{mae}. It is debatable which of the metrics should be used for minimization. However, the \acrshort{mae} was deemed to be a clear measure of the accuracy of the model and therefore seemed appropriate. But it is also possible that minimization objectives could be set for \acrshort{rmse} or \acrfull{mare}, given that landslides are primarily caused by heavy rainfall, which results in a significant increase \acrshort{vwc} in a short time period. 
To further analyze the correlation between the estimated and measured values for the best estimate, the \acrfull{rmse} and the coefficient of determination ($R^2$) for the best model were calculated and are shown in Fig.~ 8. The optimal monotonically continuous linear curve was found to fit for the majority of the data points, which are largely within the sensor uncertainty range of $\pm$~3\% (shaded). 

\begin{figure}[H]
    \centering
    \includegraphics[width=0.8\linewidth]{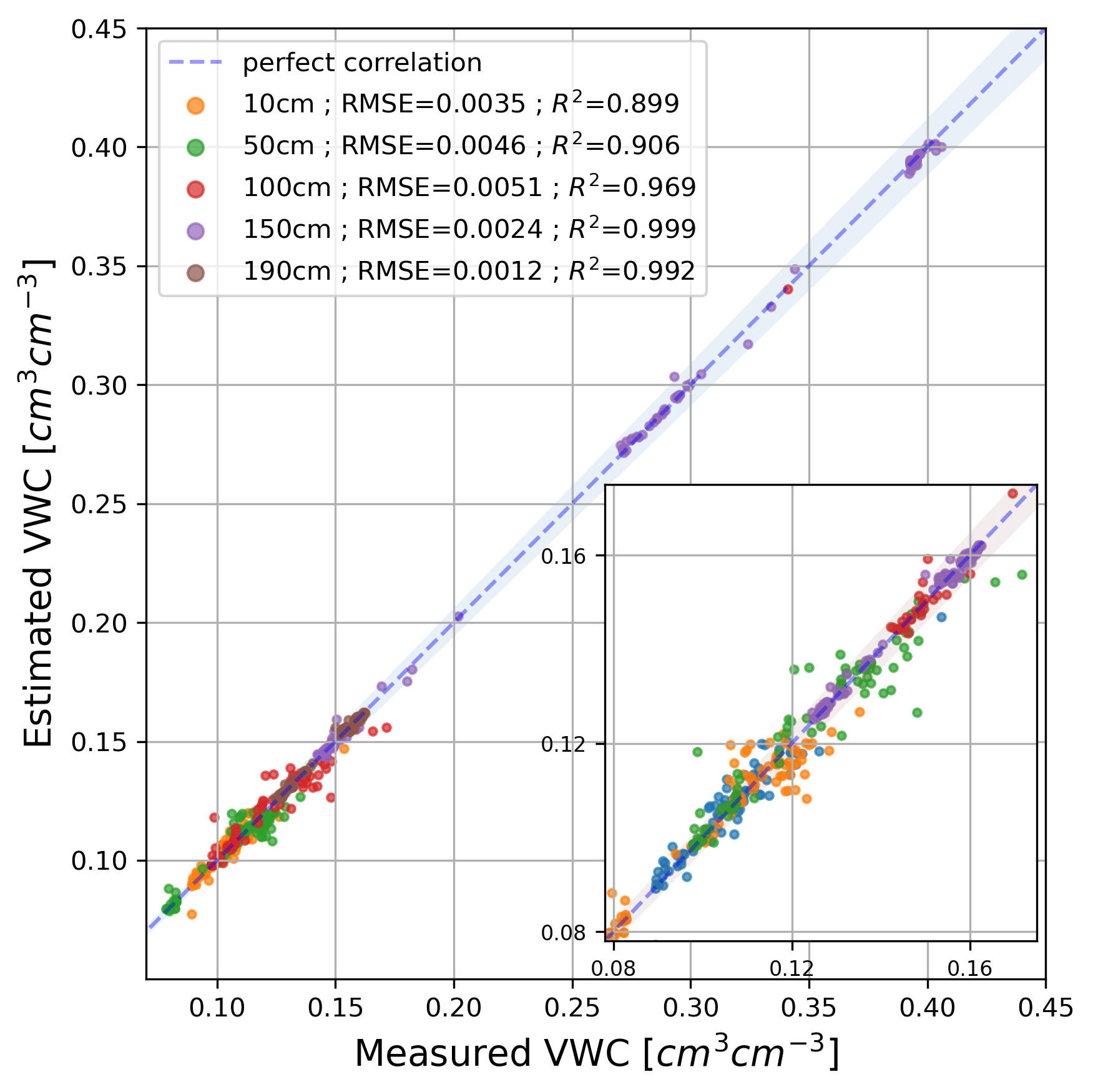}
    \caption[Correlation between \acrshort{vwc} and \acrshort{vwc} estimation]{Estimated and measured \acrshort{vwc} of the test data for the best estimation \acrshort{rf} model.}
    \label{fig:correlation_vwc}
\end{figure}

Compared to literature sources \cite{36_meng2024rf_ert_boss, 35_Wicki_2022_landslides, FU2021104948_vwc_est}, both the robustness and accuracy of all models are superior. The research conducted by Meng et al. \cite{36_meng2024rf_ert_boss} yielded $R^2$ values as high as 0.92 and \acrshort{rmse} of 0.0041 for their RF model. In comparison, the estimation proposed by Wicki and Hauck \cite{35_Wicki_2022_landslides}, based on Archie's Law, never exceeded 0.87 for $R^2$. Their \acrshort{rmse} was as low as 0.004. Fu et al \cite{FU2021104948_vwc_est} derived separate values for different soil types by applying another petrophysical formula. The combination of all soils resulted in an \acrshort{rmse} of 0.041 and $R^2$ of 0.95 with ranging values for different soil types (\acrshort{rmse}: 0.020-0.051; $R^2$: 0.91-0.99). 
\newline
A 24~h as well as 36~h and 48~h prediction was selected based on the restriction of potential delay values to the intervals between 24 (36/48) and 72 hours. For the 24~h forecast small deviation became knowledgeable, especially for the \acrshort{mare}. The choice to further investigate the 24~h forecast was a trade-off between temporal foresight and accuracy and robustness of the models. If desired one could vary minimization metrics and further forecast times.

\subsection{Comparison of forecast to estimation}

In implementing the depth-independent \acrshort{rf} model, which was trained on sensor data, to the \acrshort{ert} data, the delays obtained for the five heights were interpolated in an effort to encompass the entire height spectrum of the \acrshort{ert}. For \acrshort{er}, temperature and precipitation, an increase in the delay with increasing depth was to be expected. The best estimated as well as the predicted best temperature delays were more or less depth-independent. 
\newline
In contrast, the precipitation delays were higher, increased first but than scattered. The 24~h times did not agree the expectations. The forecast \acrshort{er} delays demonstrate an alternating pattern, initially exhibiting constant values and then increasing, before subsequently decreasing again after 100~cm. For all delays there was no recognizable pattern. 
\newline 
It is possible that no global minima but only local minima were reached and thus the delays do not increase with depth. Further discussion reveals that physical argumentation is not expedient.
One approach to address this issue is to formulate assumptions in advance about the restriction of possible delays depending on their height. Upon applying the trained models to the aforementioned \acrshort{ert} and simultaneous sensor measurements, it becomes evident that, on the one hand, the estimates and predictions of the \acrshort{er} sensor measurements align closely with those of the \acrshort{vwc} measurement. In contrast, the \acrshort{vwc} predictions and estimates of the \acrshort{ert} do not agree well with the sensor measurements. It is noteworthy that the \acrshort{ert} \acrshort{vwc} exhibits a sudden change at the boundaries of the shaded regions. The aforementioned models were trained within these areas. Each region is represented by a distinct model. One potential solution to align these depth boundaries is to integrate all models and consider depth as a parameter, as proposed by Meng et al \cite{36_meng2024rf_ert_boss}. Nonetheless, it is mainly driven by the amount of depth \acrshort{vwc} point available. Therefore, incorporating additional \acrshort{vwc} measurement points would the depth variation more precisely. Furthermore, it is plausible that the \acrshort{er} training data of the models were not within the range of those employed for validation (\acrshort{ert}, domain shift). Consequently, a possible solution would be to train and test the models with \acrshort{ert} data as the \acrshort{er} measurement data sometimes differ greatly from those of the \acrshort{ert}, which was already shown in Fig.~ \ref{fig:virtual_borehole}. A significant improvement can be expected when the \acrshort{aert} is put into operation and the \acrshort{er} sensors can be neglected. A shift of the training and test data from \acrshort{er} sensor data towards \acrshort{ert} data is exprected to further improve \acrshort{vwc} estimation from \acrshort{ert} similar to the current sensor data alignment. The fact that the prediction of the \acrshort{ert} measurement gives a better estimate of the \acrshort{vwc} than the best estimate may be due to the way the \acrshort{rf} models are constructed. For each node, the model decides on the optimal split parameter to minimize the error. In the forecast, \acrshort{er} may be chosen less frequently as split parameter. As previously explained, the best \acrshort{vwc} estimate does not include a delay in the \acrshort{er}. However, the forecast includes a delay, hence \acrshort{vwc} values for similar \acrshort{er} will be scattered, reducing the quality of the split.
\newline
Nevertheless, cautious estimation of the \acrshort{vwc} can be conducted. A layer with a high \acrshort{vwc} is apparent (Fig.~ \ref{fig:ml_vwc_best_estimation}), although it is likely to be overestimated in thickness (see Fig.~ \ref{fig:vwc_borehole_est}). This layer extends from 75~cm to a depth of approximately 175~cm. The underlying permafrost is dry, as are the top few centimeters of the soil. It seems probable that precipitation infiltrate the upper layers of soil and then stagnates above the permafrost, since the ice establishes a natural boundary to the depth of infiltration.

\section{Testing the \acrshort{rf} model for short temporal changes}

Since neither large temperature differences nor precipitation could have influenced \acrshort{vwc} in the first measurement interval, any differences in \acrshort{vwc} were due solely to differences in \acrshort{er}. No large differences were found either in the mean values or in the total \acrshort{ert}. Furthermore, the estimation of the \acrshort{vwc} was apparently good throughout the period shown in Fig.~ \ref{fig:ml_vwc_best_estimation} and largely in the $\pm$~3\% range for all depths. Not all 6~h \acrshort{ert} were carried out, which makes it difficult to check the predictive capabilities, as the delays of the \acrshort{er} would then always need to be rounded to the next possible \acrshort{ert} times. This could possibly deviate by up to 9~h from the actual delay value. For this reason, this approach was rejected as it would have introduced large uncertainties. Nevertheless, it can be assumed that the predictions have the same uncertainties as previous \acrshort{vwc} estimates of \acrshort{ert} considering estimation and prediction capabilities for the sensors. 
\newline
The heavy precipitation in the second interval did not lead to a significant change in the mean \acrshort{er}. This is probably because there are large but locally heterogeneous differences in \acrshort{er}. Since the regions of large increases and decreases in \acrshort{er} due to the depression are very close together, it is reasonable to assume that the water missing in one region is present in the other. It is possible that the water was able to penetrate the soil more easily. The \acrshort{vwc} estimates and their changes are generally considered insignificant. Regarding the delay times previously obtained, it is plausible that the \acrshort{ert} was recorded too soon after the downstroke and that there is an error in the methodology. It is advisable to repeat the measurement of the influence of a large rainfall event with adjusted measurement times, taken further after the precipitation.

\section{Testing the \acrshort{rf} model at various different locations}

The seemingly inappropriate estimates from all the other sites on the models from study site 2 are based on the fact that the \acrshort{er} and \acrshort{vwc} values of the validation data fall outside the training ranges of the models (domain shift). As a result, they encounter \acrshort{er} values that they were not trained with, i.e. higher or lower than all the values they know, and then logically choose the highest or lowest \acrshort{vwc} value that seems to fit best. Since all \acrshort{er} values are consequently higher than those for the study site, the same \acrshort{vwc} value is always selected and a constant prediction is generated. This can be solved by providing additional data with a wider range of \acrshort{er} and \acrshort{vwc} as training data. By increasing the available data it could be possible to make more accurate predictions. Therefore, it is useful to implement a study, in which all the test sites to be investigated (or at least the ones with the most extreme \acrshort{er} and \acrshort{vwc} values in addition to some data in between) in one model.
\newline
Another source of uncertainty is that \acrshort{er}, temperature and precipitation are not the only factors influencing \acrshort{vwc}. For a more detailed consideration, factors such as soil conditions and soil type must also be taken into account. However, this becomes very complex quickly, as all parameters need to be measured relatively frequently, at least for training the models. Training the models with a wide range of \acrshort{er} and \acrshort{vwc} data to get a good estimate seems sufficient. It is worth considering whether \acrshort{vwc} and \acrshort{aert} should be installed for many areas to be monitored for a period of time - in order to collect enough training data - with the \acrshort{vwc} sensors removed after the training period if the deviations have reduced sufficiently.

\section{Active layer development throughout seasonal changes}
A clear change is visible only from the beginning of October. This is reasonable because only the transition from water to ice causes a large change in \acrshort{er} and at this point the temperatures dropped drastically, yielding freezing of the topmost layer. This is enhanced in the following weeks. The measurements would have had to be continued to see an even more drastic increase in total \acrshort{er}. It is plausible that this would have occurred one week or two weeks after the end of the survey, as air temperatures were consistently around -5~$^{\circ}C$. 

%% file: 6Conclusion.tex
\chapter{Summary and Outlook}
\label{ch:conclusion}

\begin{figure}[H]
    \centering
    \includegraphics[width=0.46\linewidth]{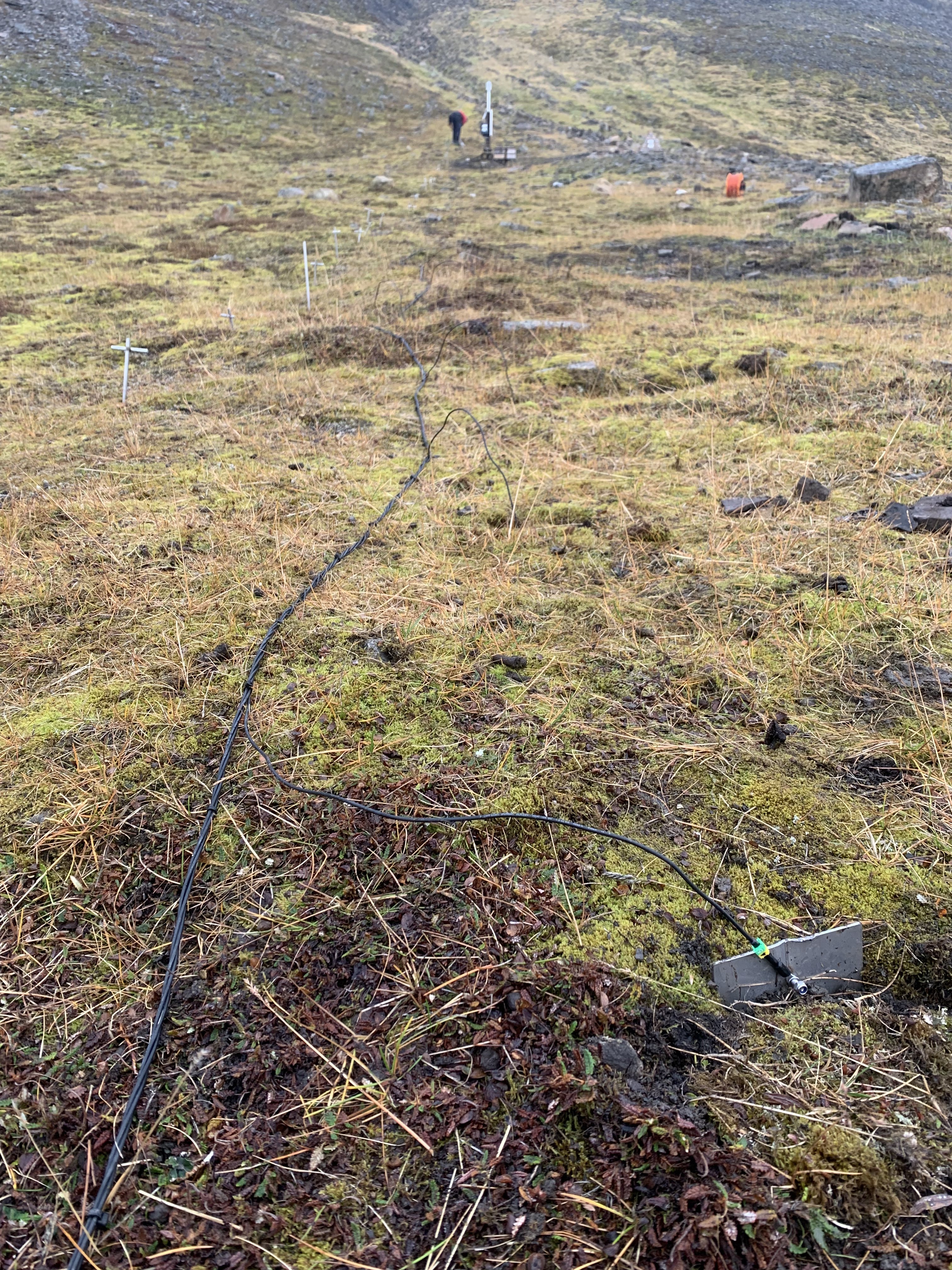}
    \hfill
    \includegraphics[width=0.46\linewidth]{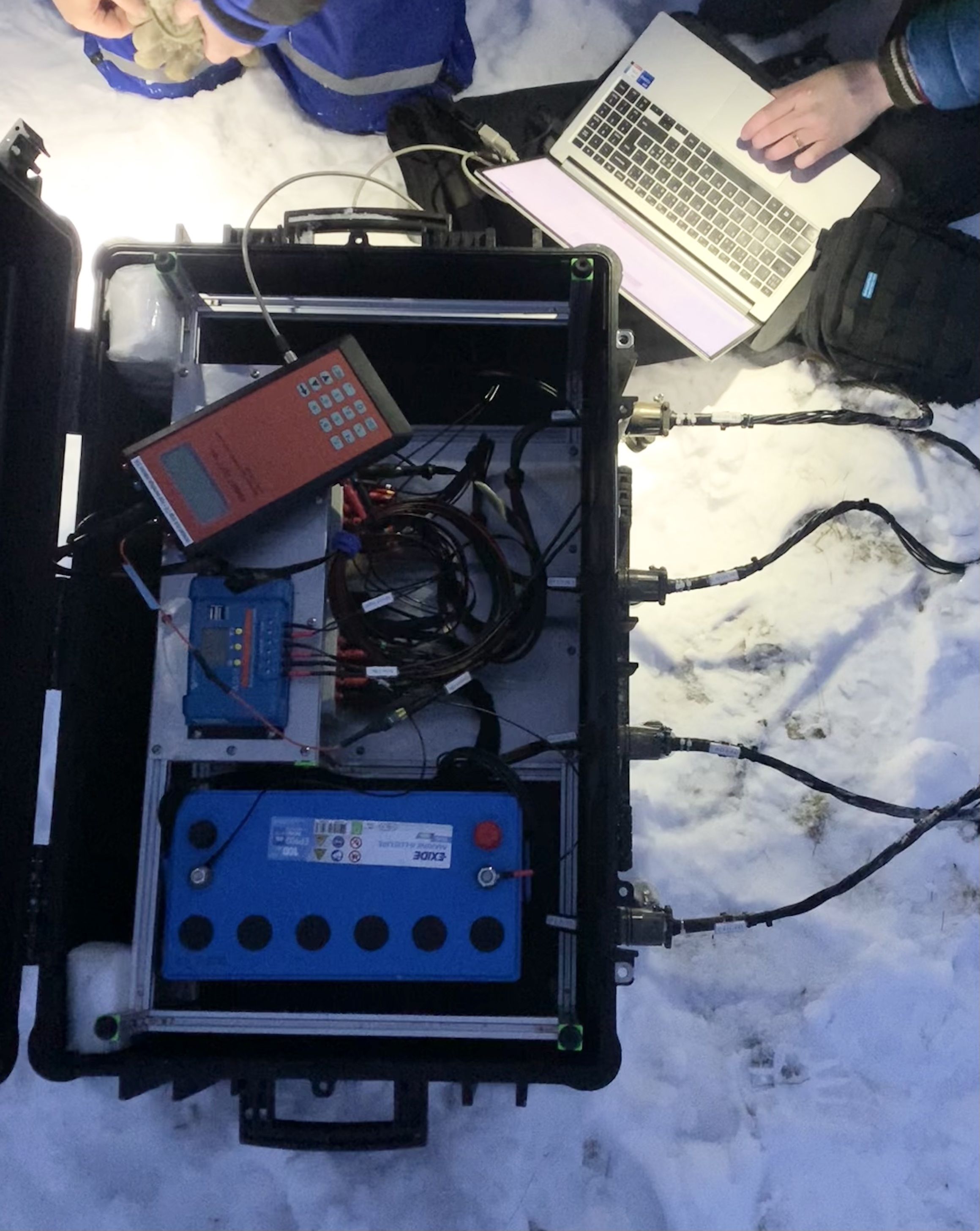}
    \caption[\acrshort{aert}]{Installation of the \acrshort{aert} at study site 2 in the beginning of October (left) and instrument deployment early November (right).}
    \label{fig:aert}
\end{figure}

Due to the extreme changes in the climate, especially in arctic permafrost soil regions, landslides are becoming increasingly likely on steep slopes \cite{0_patton2019landslide_permafrost}. Permafrost soil observations distinguished active layer depths and their seasonal development. This work helps to improve the assessment of \acrshort{vwc} as a key parameter for estimating landslide risk. Random forest models were trained to estimate and predict \acrshort{vwc} on the basis of \acrshort{er} measurements as well as air temperature and precipitation. Depth-dependent delaying  \acrshort{er}, temperature and precipitation led to further error reductions below \acrshort{mre} of 2.8\%. It was shown that \acrshort{vwc} can be reliably estimated and predicted from \acrshort{er} for the location the models were trained on. Local variations resulted in great deviations because the \acrshort{er} validation data located outside of the training data range. Moreover, estimation of \acrshort{vwc} from \acrshort{ert} revealed significant deficits. 
\newline
The implementation of an \acrfull{aert} system and training of the models on \acrshort{ert} data at various locations is expected to eliminate these differences. This allows the \acrshort{er} sensors to be neglected and the training, test and validation data to be acquired using the same measurement method. By training on various sites, the different correlations between \acrshort{er} and \acrshort{vwc} due to other parameters such as porosity and soil type are learned. 
At the beginning of October, 49 rectangular electrodes, with one meter spacing, were buried in the ground at study site 2 and connected with cables. The instrument was first deployed on 7 November 2024, after the final components had been received and integrated (Fig. \ref{fig:aert}). The initial preliminary data appears to be promising, indicating that more accurate predictions can be anticipated by next summer. 
\newline
For further model improvements, it is recommended to install \acrshort{vwc} sensors at much more depths at study site 2 next to the \acrshort{aert}. Moreover, it could be beneficial to ensure a coverage of a broader \acrshort{er} range to be sure that the validation data is not outside the training parameters. This could be achieved be including different sites with varying \acrshort{er} and \acrshort{vwc} ranges in one model. Furthermore, hidden parameters influencing the \acrshort{er}-\acrshort{vwc} correlation would be incorporated in a broader range. 
\newline
Finally, an introduction of the gained relation between \acrshort{er} and \acrshort{vwc} can be interpreted as a first step towards a sophisticated landslide risk hazard assessment. Following investigations could include parameters soil structures and pore water.

%% file: Poststuff/References.tex
\phantomsection

\bibliographystyle{unsrt} 

\pagebreak

\newpage
\bibliography{References.bib}

%% file: Poststuff/Appendix.tex
\appendix

\addchap{Appendix}

\section{Additional \acrshort{ert}} 
\label{AppendixA}  

\begin{figure}[H]
    \centering
    \includegraphics[width=0.8\linewidth]{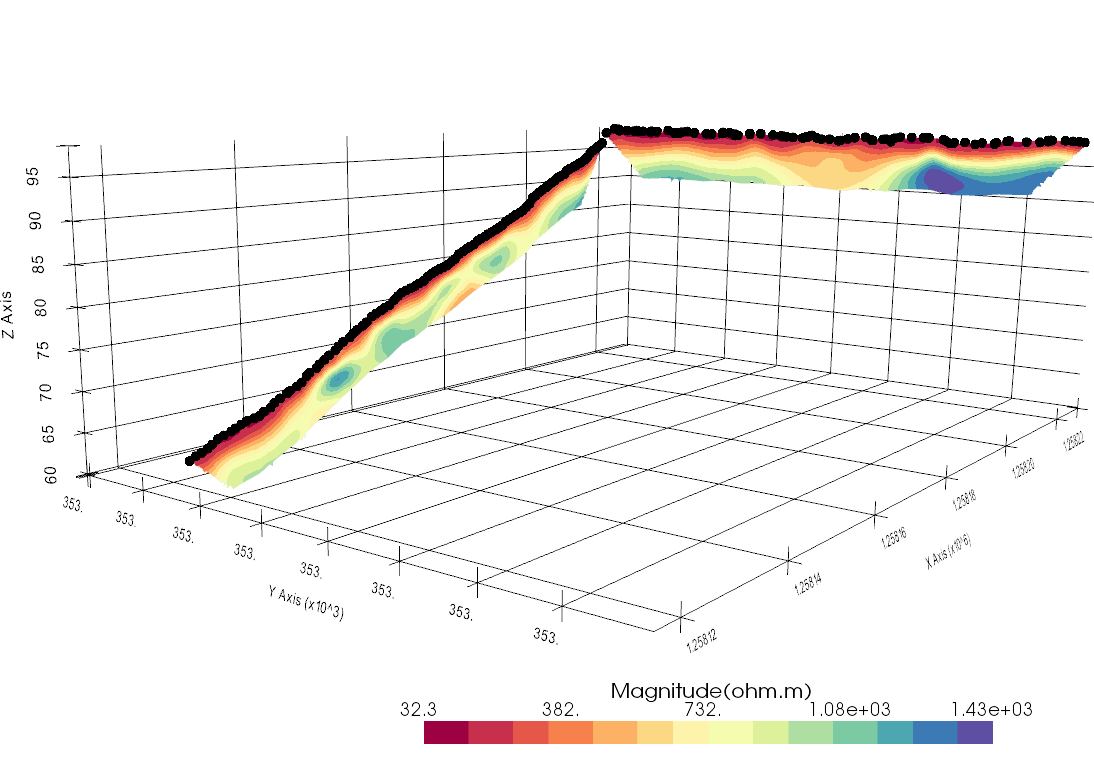}
    \includegraphics[width=0.8\linewidth]{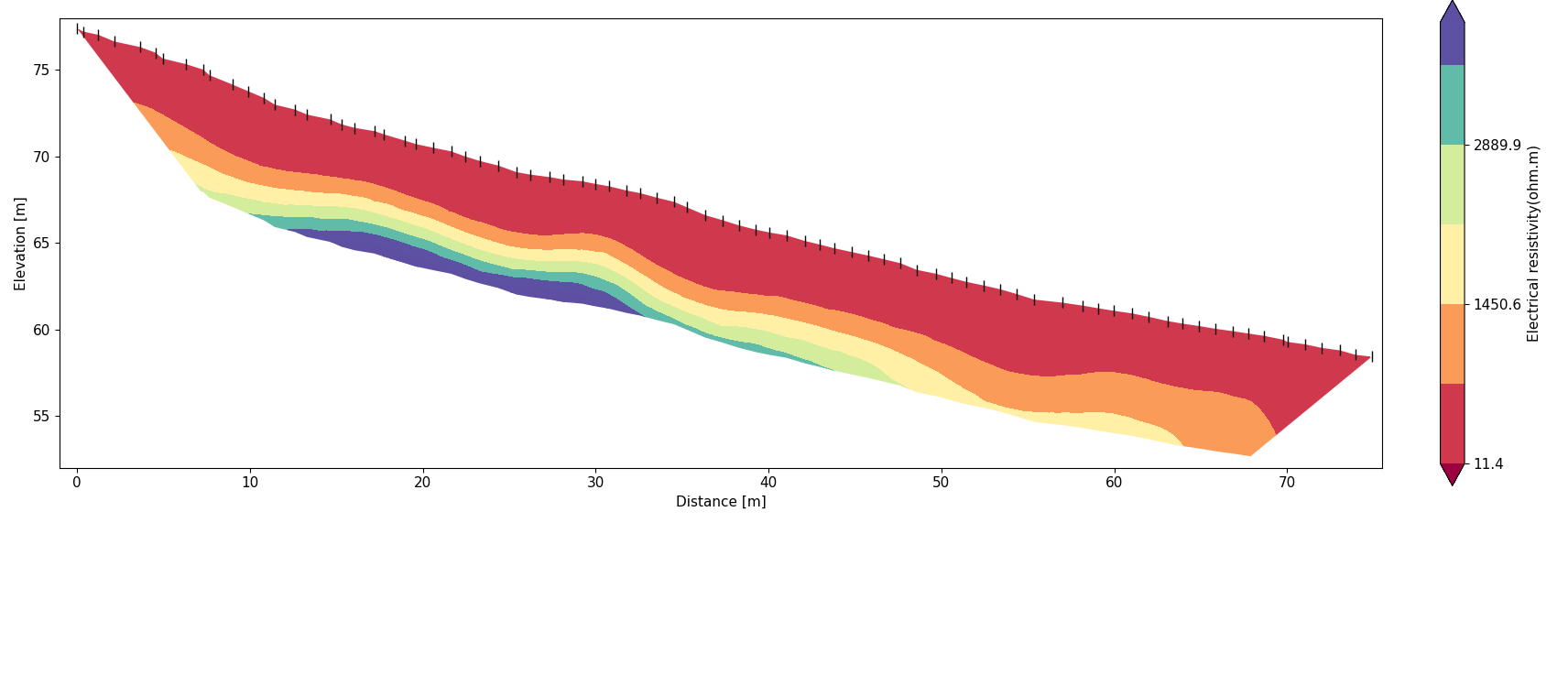}
    \includegraphics[width=0.8\linewidth]{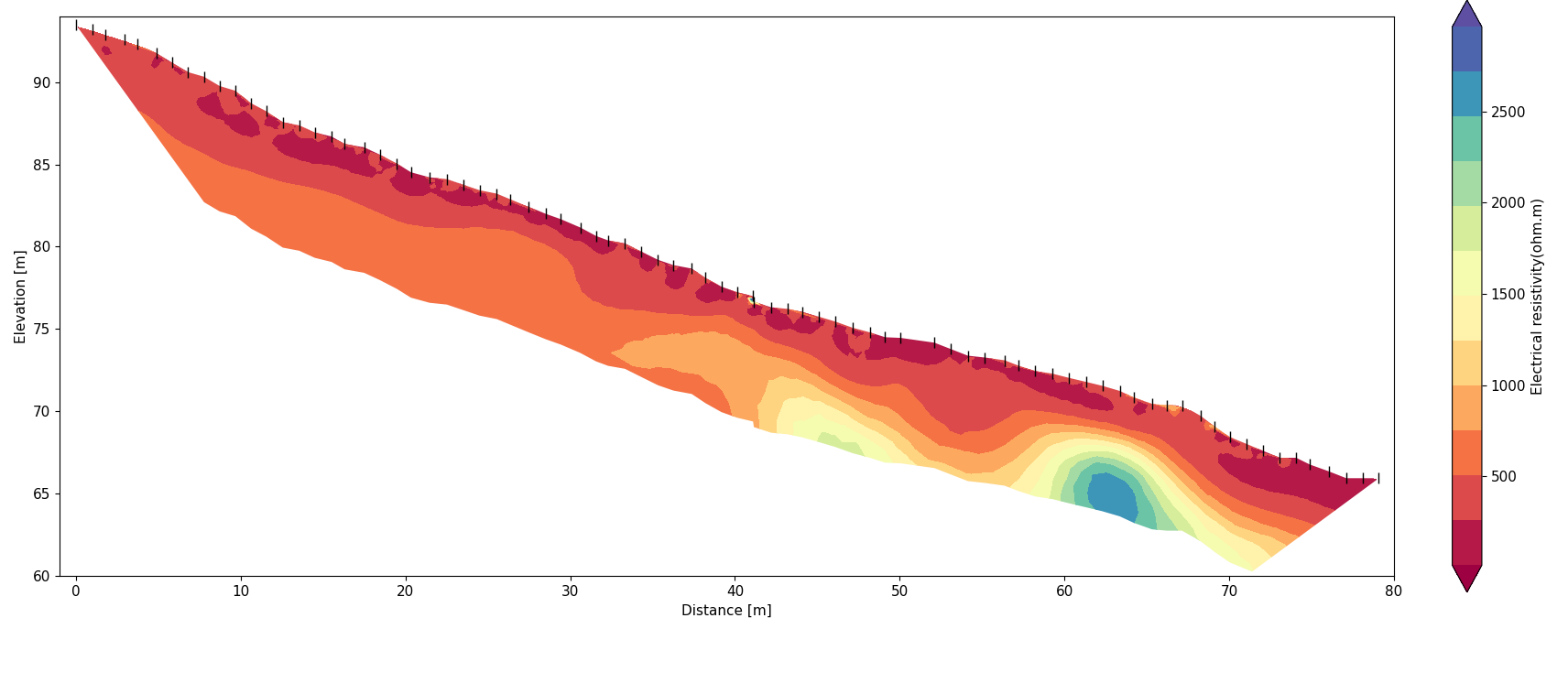}
    \caption[ERT for study sites 1, 3, 4]{Height corrected \acrshort{ert} of the different locations, acquired 02.-05.09.2024. The x and y axis for the 3d figure (sites 1.1 and 1.2) are cartesian coordinates. The middle \acrshort{ert} is site 3 and the lower one site 4. }
    \label{fig:ert_diff_locations_1_3}
\end{figure}

\begin{figure}[H]
    \centering
    \includegraphics[width=1\linewidth]{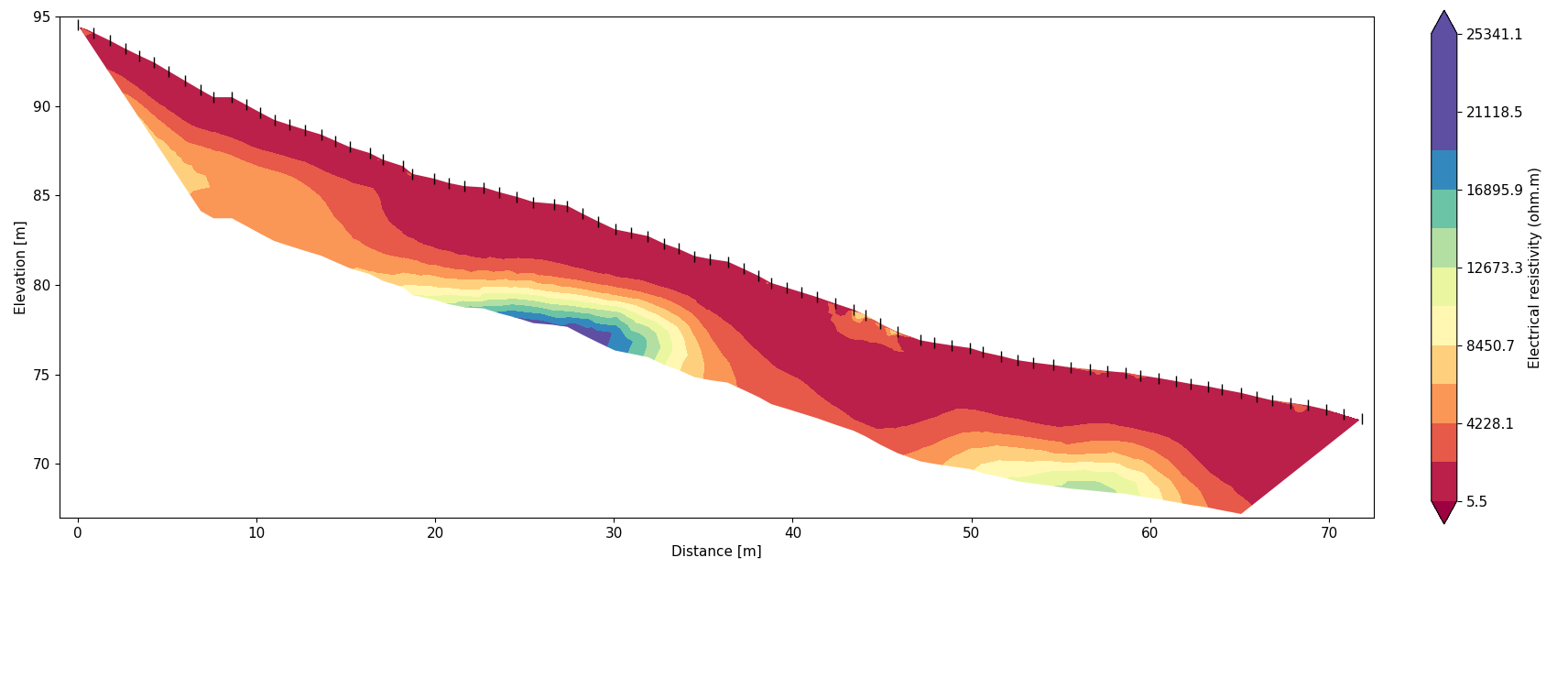}
    \includegraphics[width=\linewidth]{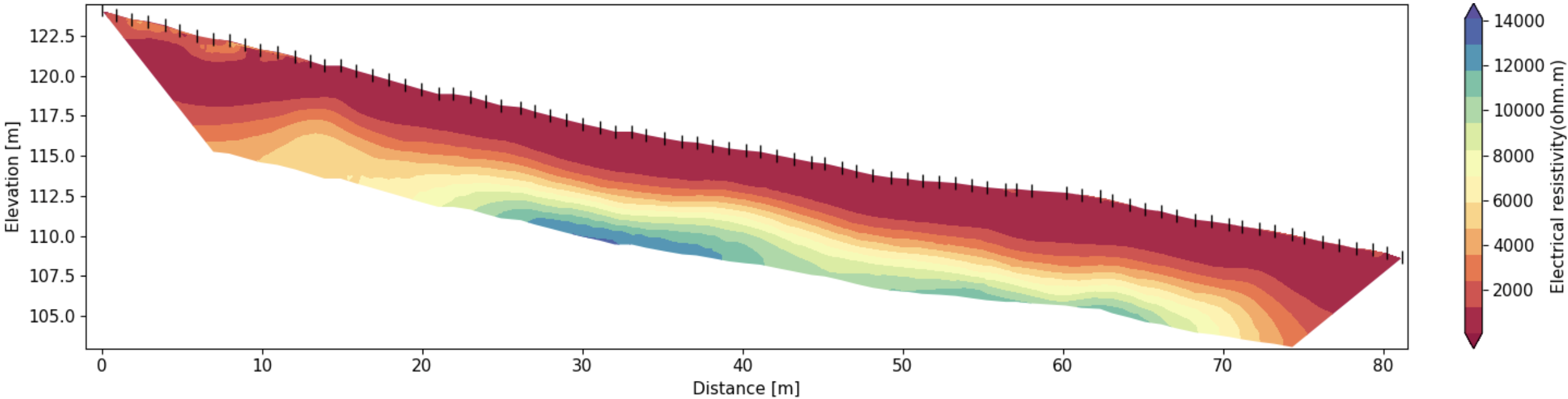}
    \caption[ERT for study sites 5, 6]{Height corrected \acrshort{ert} of the different locations (top: 5, bottom: 6), acquired 02.-05.09.2024.}
    \label{fig:ert_diff_locations_4_5_6}
\end{figure}

\section{General \acrshort{ert}} 
\label{AppendixB}  

\begin{table}[h!]
\centering
\caption[Mean ER by meter]{The mean electrical resistivity, standard deviation, and the relative error for every meter in depth.}
\begin{tabular}{lccc}
\toprule
Depth range (m) & \makecell{Mean electrical resistivity \\ ($\Omega \cdot m$)} & \makecell{Standard deviation \\ of mean ($\Omega \cdot m$)} & \makecell{Relative error \\ (\%)} \\
\midrule
0.07 - 1.00  & 134.80  & 32.93  & 24.43 \\
1.00 - 2.00       & 223.20  & 90.32  & 40.47 \\
2.00 - 3.00       & 620.81  & 251.44 & 40.50 \\
3.00 - 4.00       & 1315.15 & 408.92 & 31.09 \\
4.00 - 5.00       & 2121.48 & 507.97 & 23.94 \\
5.00 - 6.00       & 2822.26 & 517.37 & 18.33 \\
6.00 - 7.36    & 3341.14 & 486.65 & 14.57 \\
\bottomrule
\end{tabular}
\label{tab:resistivity_mean_meter}
\end{table}

\begin{longtable}{lccc}
\caption[ERT with means for dynamic depth ranges]{Electrical resistivity means dynamically calculated with a relative error of 20\%. In addition, a constraint of a minimum depth range of 5~cm was applied. The blue rows indicate significantly larger depth ranges with similar large errors.} \\
\toprule
Depth range (m) & \makecell{Mean electrical resistivity \\ ($\Omega \cdot m$)} & \makecell{Standard deviation \\ of mean ($\Omega \cdot m$)} & \makecell{Relative error \\ (\%)} \\
\midrule
\endfirsthead

\toprule
Depth range (m) & \makecell{Mean electrical resistivity \\($\Omega \cdot m$)} & \makecell{Standard deviation \\ of mean ($\Omega \cdot m$)} & \makecell{Relative error \\ (\%)} \\
\midrule
\endhead

\midrule
\multicolumn{4}{r}{\textit{Continued on next page}} \\
\midrule
\endfoot

\endlastfoot

\rowcolor{blue} -7.36 - -4.66 & 3017.44 & 605.64 & 20.07 \\
4.66 - 4.61 & 2236.11 & 469.81 & 21.01 \\
4.61 - 4.56 & 2195.30 & 468.35 & 21.33 \\
4.56 - 4.51 & 2154.68 & 465.23 & 21.59 \\
4.51 - 4.46 & 2115.28 & 459.05 & 21.70 \\
4.46 - 4.41 & 2071.59 & 455.72 & 22.00 \\
4.41 - 4.36 & 2029.57 & 450.78 & 22.21 \\
4.36 - 4.31 & 1994.68 & 444.64 & 22.29 \\
4.31 - 4.26 & 1951.66 & 439.49 & 22.52 \\
4.26 - 4.21 & 1908.48 & 434.88 & 22.79 \\
4.21 - 4.16 & 1867.81 & 428.00 & 22.91 \\
4.16 - 4.11 & 1825.26 & 422.40 & 23.14 \\
4.11 - 4.06 & 1782.35 & 417.65 & 23.43 \\
4.06 - 4.01 & 1741.64 & 410.39 & 23.56 \\
4.01 - 3.96 & 1705.52 & 404.79 & 23.73 \\
3.96 - 3.91 & 1661.92 & 399.31 & 24.03 \\
3.91 - 3.86 & 1620.77 & 391.48 & 24.15 \\
3.86 - 3.81 & 1578.66 & 384.22 & 24.34 \\
3.81 - 3.76 & 1535.70 & 377.66 & 24.59 \\
3.76 - 3.71 & 1494.94 & 369.67 & 24.73 \\
3.71 - 3.66 & 1453.79 & 361.61 & 24.87 \\
3.66 - 3.61 & 1416.59 & 356.28 & 25.15 \\
3.61 - 3.56 & 1376.06 & 348.44 & 25.32 \\
3.56 - 3.51 & 1335.54 & 340.15 & 25.47 \\
3.51 - 3.46 & 1293.61 & 333.58 & 25.79 \\
3.46 - 3.41 & 1253.66 & 325.73 & 25.98 \\
3.41 - 3.36 & 1214.36 & 318.03 & 26.19 \\
3.36 - 3.31 & 1173.71 & 311.17 & 26.51 \\
3.31 - 3.26 & 1140.80 & 304.63 & 26.70 \\
3.26 - 3.21 & 1103.00 & 296.72 & 26.90 \\
3.21 - 3.16 & 1063.78 & 289.61 & 27.22 \\
3.16 - 3.11 & 1026.36 & 282.01 & 27.48 \\
3.11 - 3.06 & 989.83 & 273.85 & 27.67 \\
3.06 - 3.01 & 957.81 & 267.40 & 27.92 \\
3.01 - 2.96 & 921.53 & 259.76 & 28.19 \\
2.96 - 2.91 & 886.54 & 251.30 & 28.35 \\
2.91 - 2.86 & 851.18 & 243.59 & 28.62\\
2.86 - 2.81 & 816.69 & 235.69 & 28.86 \\
2.81 - 2.76 & 783.39 & 227.64 & 29.06 \\
2.76 - 2.71 & 749.78 & 219.68 & 29.30 \\
2.71 - 2.66 & 721.24 & 212.86 & 29.51 \\
2.66 - 2.61 & 689.13 & 204.70 & 29.70 \\
2.61 - 2.56 & 657.39 & 196.28 & 29.86 \\
2.56 - 2.51 & 625.97 & 188.21 & 30.07 \\
2.51 - 2.46 & 596.28 & 179.47 & 30.10 \\
2.46 - 2.41 & 567.16 & 171.26 & 30.20 \\
2.41 - 2.36 & 538.73 & 163.14 & 30.28 \\
2.36 - 2.31 & 515.47 & 156.43 & 30.35 \\
2.31 - 2.26 & 489.28 & 148.77 & 30.41 \\
2.26 - 2.21 & 463.71 & 141.18 & 30.45 \\
2.21 - 2.16 & 439.36 & 134.06 & 30.51 \\
2.16 - 2.11 & 415.91 & 126.92 & 30.52 \\
2.11 - 2.06 & 393.10 & 120.07 & 30.54 \\
2.06 - 2.01 & 371.67 & 113.06 & 30.42 \\
2.01 - 1.96 & 353.89 & 107.19 & 30.29 \\
1.96 - 1.91 & 333.62 & 100.56 & 30.14 \\
1.91 - 1.86 & 314.57 & 94.16 & 29.93 \\
1.86 - 1.81 & 296.68 & 88.56 & 29.85 \\
1.81 - 1.76 & 279.58 & 83.00 & 29.69 \\
1.76 - 1.71 & 263.89 & 77.89 & 29.52 \\
1.71 - 1.66 & 249.71 & 72.77 & 29.14 \\
1.66 - 1.61 & 238.79 & 68.61 & 28.73 \\
1.61 - 1.56 & 227.23 & 63.77 & 28.06 \\
1.56 - 1.51 & 216.44 & 59.15 & 27.33 \\
1.51 - 1.46 & 206.27 & 54.81 & 26.57 \\
1.46 - 1.41 & 196.56 & 50.52 & 25.70 \\
1.41 - 1.36 & 187.55 & 46.59 & 24.84 \\
1.36 - 1.31 & 179.07 & 42.70 & 23.84 \\
1.31 - 1.26 & 172.29 & 39.49 & 22.92 \\
1.26 - 1.21 & 165.16 & 36.02 & 21.81 \\
1.21 - 1.16 & 158.79 & 32.88 & 20.70 \\
\rowcolor{blue} 1.16 - 0.31 & 135.27 & 27.67 & 20.46 \\
0.31 - 0.26 & 139.44 & 41.55 & 29.80 \\
0.26 - 0.21 & 141.35 & 44.22 & 31.28 \\
0.21 - 0.16 & 142.87 & 46.64 & 32.65 \\
0.16 - 0.11 & 143.92 & 48.46 & 33.67 \\
0.11 - 0.07 & 145.73 & 48.70 & 33.41 \\ \hline

\label{tab:dynamic_heights_mean_ert}
\end{longtable}

\begin{table}[h!]
\centering
\caption[ERT means at sensor positions]{Electrical resistivity, standard deviation (Std), and relative error for different sensor depths of S 2.1.}
\begin{tabular}{lccc}
\toprule
Sensor (x, depth) (m) & \makecell{Electrical resistivity \\ ($\Omega \cdot m$)} & \makecell{Std deviation \\ ($\Omega \cdot m$)} & \makecell{Relative error \\ (\%)} \\ \midrule
(25.75, -0.1) & 108.61 & 5.36  & 4.93 \\
(25.75, -0.5) & 109.46 & 3.58  & 3.27 \\
(25.75, -1.0) & 123.41 & 11.12 & 9.01 \\
(25.75, -1.5) & 230.24 & 24.71 & 10.73 \\
(25.75, -1.9) & 333.10 & 78.16 & 23.46 \\ \bottomrule
\end{tabular}

\label{tab:sensor_resistivity}
\end{table}

\begin{figure}[H]
    \centering
    \includegraphics[width=0.8\linewidth]{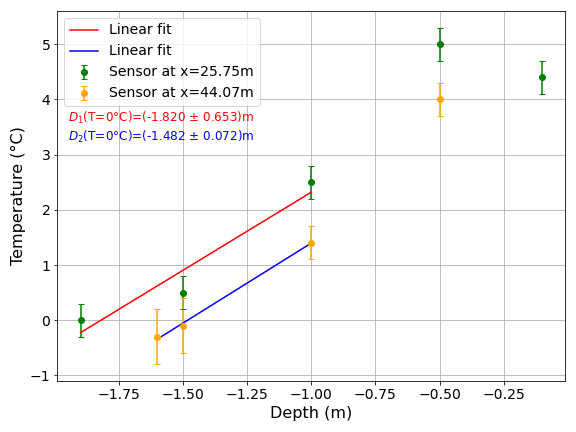}
    \caption[\acrshort{ert} temperature calibration]{The temperature calibration of the \acrshort{ert} in the general \acrshort{ert} analysis (\acrshort{ert} ID 2, 28.08.2024). Here, the temperatures of the sensors S 2.1 are interpolated to find the depth at which the temperature reaches 0~$^{\circ}C$. }
    \label{fig:temp_cali}
\end{figure}

\section{Correlating \acrshort{vwc} and \acrshort{er}}
\label{AppendixC}

\begin{longtable}{ccccccc}
\caption[Best delay parameters for \acrshort{rf} models]{Searching for the best delay parameters for each depths and combination of delays between 0 and 72~h.} \\
\toprule
Depth (cm) & \makecell{Observation \\ period (h)} & \makecell{$\text{Delay}_\text{R}$ \\(h)} & \makecell{$\text{Delay}_\text{T}$ \\(h)} & \makecell{$\text{Delay}_\text{P}$ \\(h)} & \makecell{MAE  ($10^{-4}$) \\ $cm^3 cm^{-3}$}  & \makecell{Relative \\ error (\%)} \\
\midrule
\endfirsthead

\toprule
Depth (cm) & \makecell{Observation \\ period (h)} & \makecell{$\text{Delay}_\text{R}$ \\(h)} & \makecell{$\text{Delay}_\text{T}$ \\(h)} & \makecell{$\text{Delay}_\text{P}$ \\(h)} & \makecell{MAE  ($10^{-4}$) \\ $cm^3 cm^{-3}$} & Relative error (\%) \\
\midrule
\endhead

\midrule
\multicolumn{7}{r}{\textit{Continued on next page}} \\
\midrule
\endfoot

\endlastfoot

10  & 0-72  & 0   & 6   & 42  & 26.87 & 0.00  \\
10  & 6-72  & 6   & 12  & 48  & 30.96 & 15.21 \\
10  & 12-72 & 12  & 72  & 18  & 36.23 & 34.81 \\
10  & 18-72 & 18  & 18  & 18  & 39.50 & 46.99 \\
10  & 24-72 & 24  & 24  & 24  & 41.99 & 56.26 \\
10  & 30-72 & 30  & 30  & 30  & 47.34 & 76.16 \\
10  & 36-72 & 48  & 36  & 42  & 51.03 & 89.87 \\
10  & 42-72 & 72  & 54  & 42  & 55.01 & 104.69 \\
10  & 48-72 & 54  & 54  & 48  & 60.24 & 124.14 \\
50  & 0-72  & 36  & 18  & 60  & 29.89 & 0.00  \\
50  & 6-72  & 36  & 18  & 60  & 29.89 & 0.00  \\
50  & 12-72 & 36  & 18  & 60  & 29.89 & 0.00  \\
50  & 18-72 & 36  & 18  & 60  & 29.89 & 0.00  \\
50  & 24-72 & 48  & 24  & 54  & 30.88 & 3.32  \\
50  & 30-72 & 72  & 30  & 42  & 31.62 & 5.82  \\
50  & 36-72 & 48  & 54  & 72  & 33.24 & 11.22 \\
50  & 42-72 & 48  & 54  & 72  & 33.24 & 11.22 \\
50  & 48-72 & 48  & 54  & 72  & 33.24 & 11.22 \\
100 & 0-72  & 0   & 0   & 18  & 30.23 & 0.00  \\
100 & 6-72  & 18  & 18  & 36  & 32.72 & 8.23  \\
100 & 12-72 & 18  & 18  & 36  & 32.72 & 8.23  \\
100 & 18-72 & 18  & 18  & 36  & 32.72 & 8.23  \\
100 & 24-72 & 30  & 42  & 24  & 38.54 & 27.48 \\
100 & 30-72 & 30  & 42  & 36  & 42.40 & 40.24 \\
100 & 36-72 & 66  & 72  & 54  & 44.52 & 47.27 \\
100 & 42-72 & 66  & 72  & 54  & 44.52 & 47.27 \\
100 & 48-72 & 66  & 72  & 54  & 44.52 & 47.27 \\
150 & 0-72  & 0   & 30  & 24  & 15.73 & 0.00  \\
150 & 6-72  & 6   & 36  & 24  & 16.86 & 7.15  \\
150 & 12-72 & 30  & 36  & 42  & 20.21 & 28.47 \\
150 & 18-72 & 30  & 36  & 42  & 20.21 & 28.47 \\
150 & 24-72 & 30  & 36  & 42  & 20.21 & 28.47 \\
150 & 30-72 & 30  & 36  & 42  & 20.21 & 28.47 \\
150 & 36-72 & 54  & 42  & 72  & 23.66 & 50.39 \\
150 & 42-72 & 54  & 42  & 72  & 23.66 & 50.39 \\
150 & 48-72 & 54  & 60  & 72  & 25.20 & 60.17 \\
190 & 0-72  & 36  & 18  & 54  & 8.12  & 0.00  \\
190 & 6-72  & 36  & 18  & 54  & 8.12  & 0.00  \\
190 & 12-72 & 36  & 18  & 54  & 8.12  & 0.00  \\
190 & 18-72 & 36  & 18  & 54  & 8.12  & 0.00  \\
190 & 24-72 & 54  & 66  & 24  & 8.14  & 0.30  \\
190 & 30-72 & 54  & 66  & 36  & 8.44  & 3.93  \\
190 & 36-72 & 54  & 66  & 36  & 8.44  & 3.93  \\
190 & 42-72 & 66  & 60  & 66  & 8.44  & 3.94  \\
190 & 48-72 & 66  & 60  & 66  & 8.44  & 3.94  \\
\hline

\label{tab:best_parameters_ml}
\end{longtable}

\begin{figure}[H]
    \centering
    \includegraphics[width=0.75\linewidth]{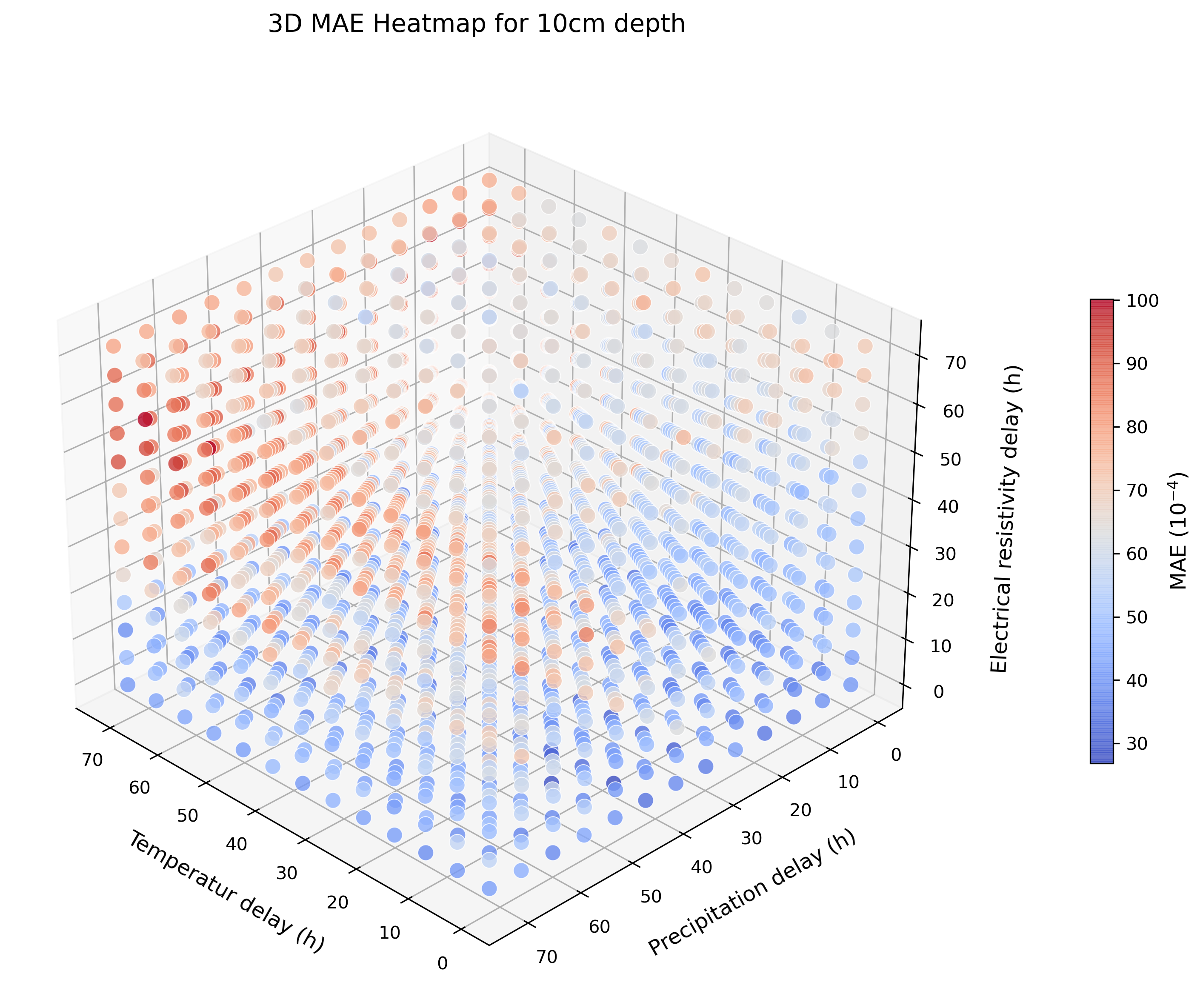}
    \includegraphics[width=0.75\linewidth]{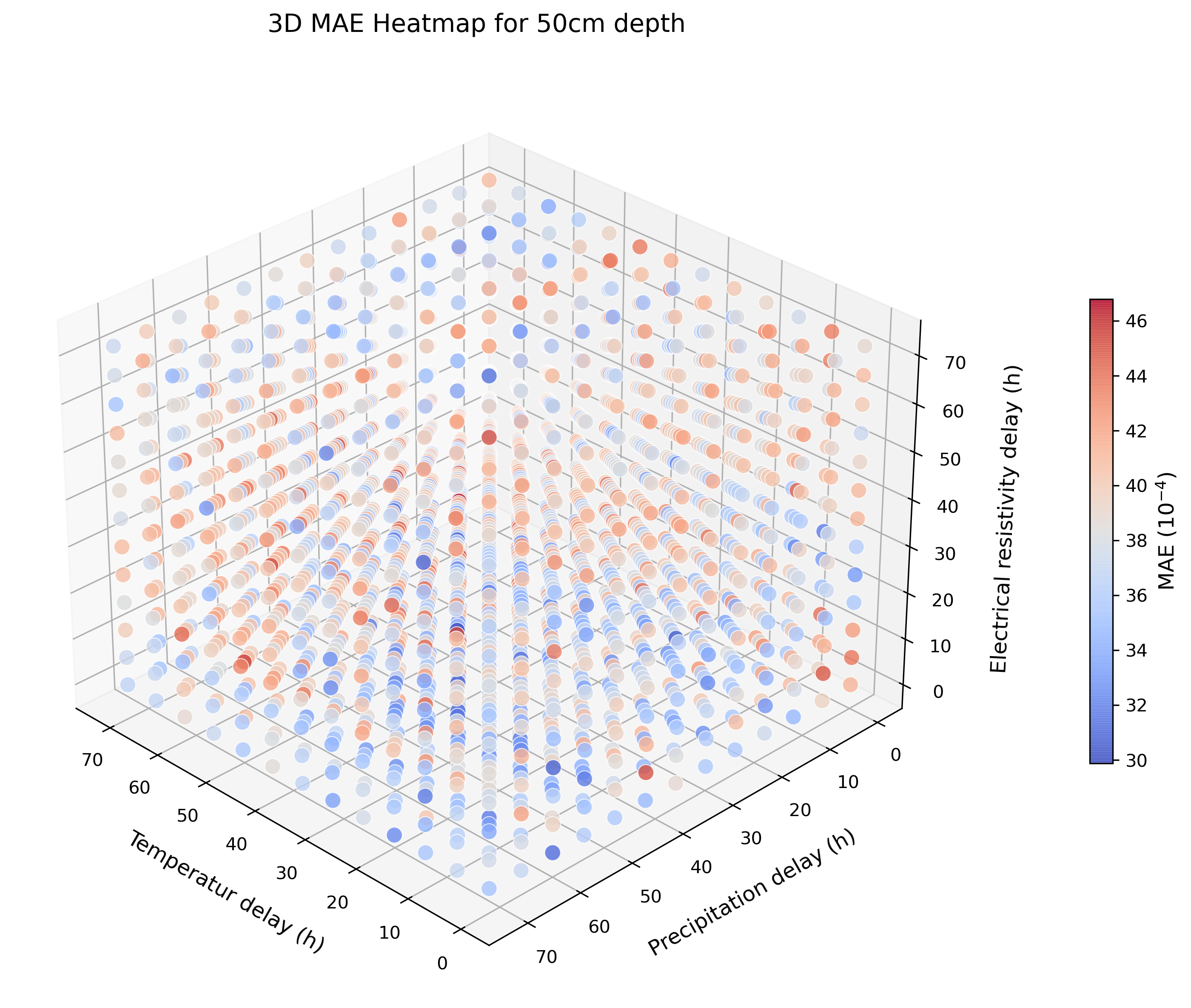}

    \caption[\acrshort{mae} depending on delays for 10~cm and 50~cm model]{\acrshort{mae} depending on delay times for \acrshort{er}, air temperature and precipitation at depth of 10~c and 50~cm.}
    \label{fig:3d_10_50_delays}
\end{figure}

\begin{figure}[H]
    \centering
    \includegraphics[width=0.75\linewidth]{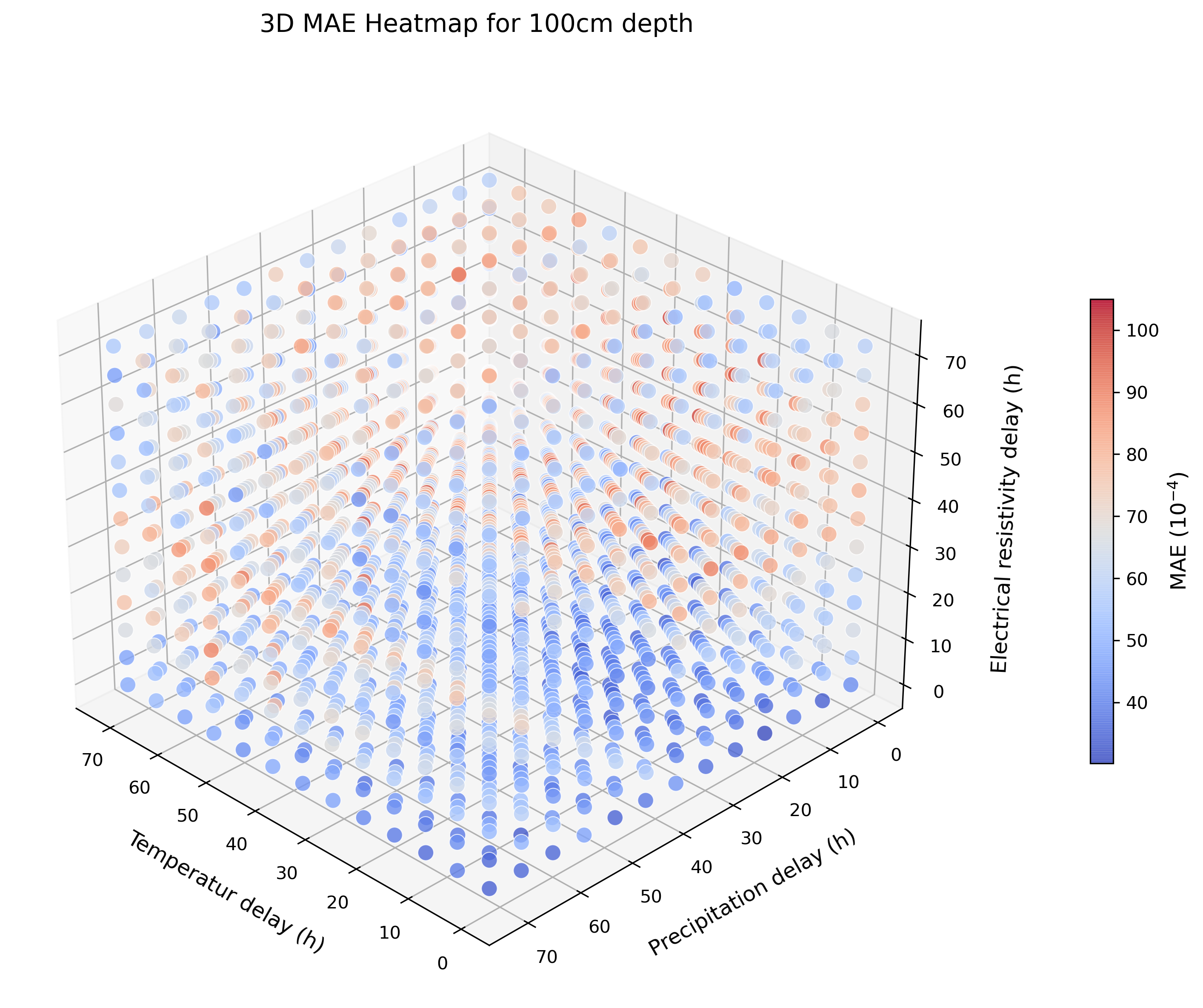}
    \includegraphics[width=0.75\linewidth]{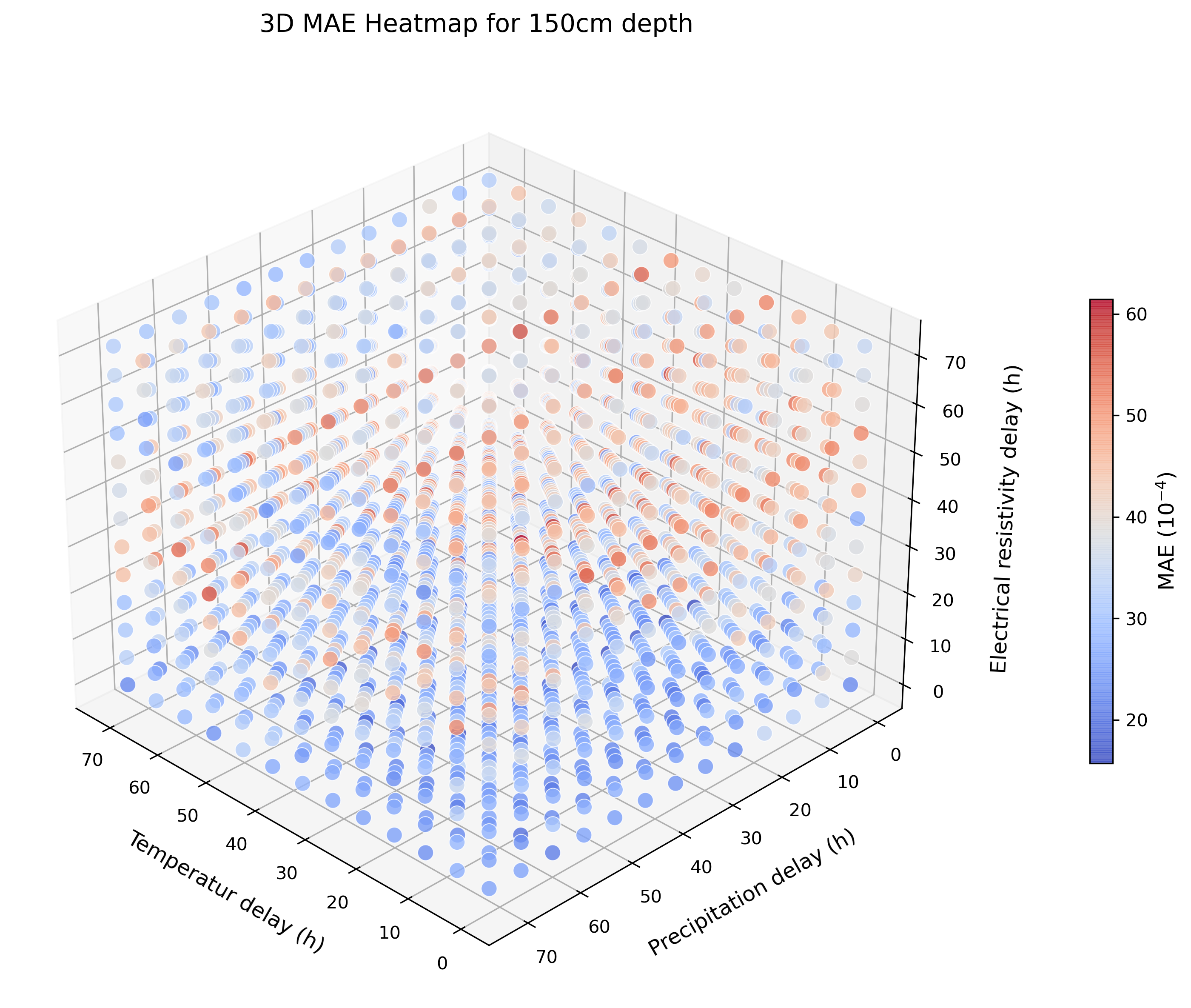}
    \caption[\acrshort{mae} depending on delays for 100 and 150~cm models]{\acrshort{mae} depending on delay times for \acrshort{er}, air temperature and precipitation at depth of 100~cm and 150~cm.}
    \label{fig:3d_100_150_delays}
\end{figure}

\begin{minipage}{\linewidth} 

    \begin{figure}[H]
        \centering
        \includegraphics[width=0.75\linewidth]{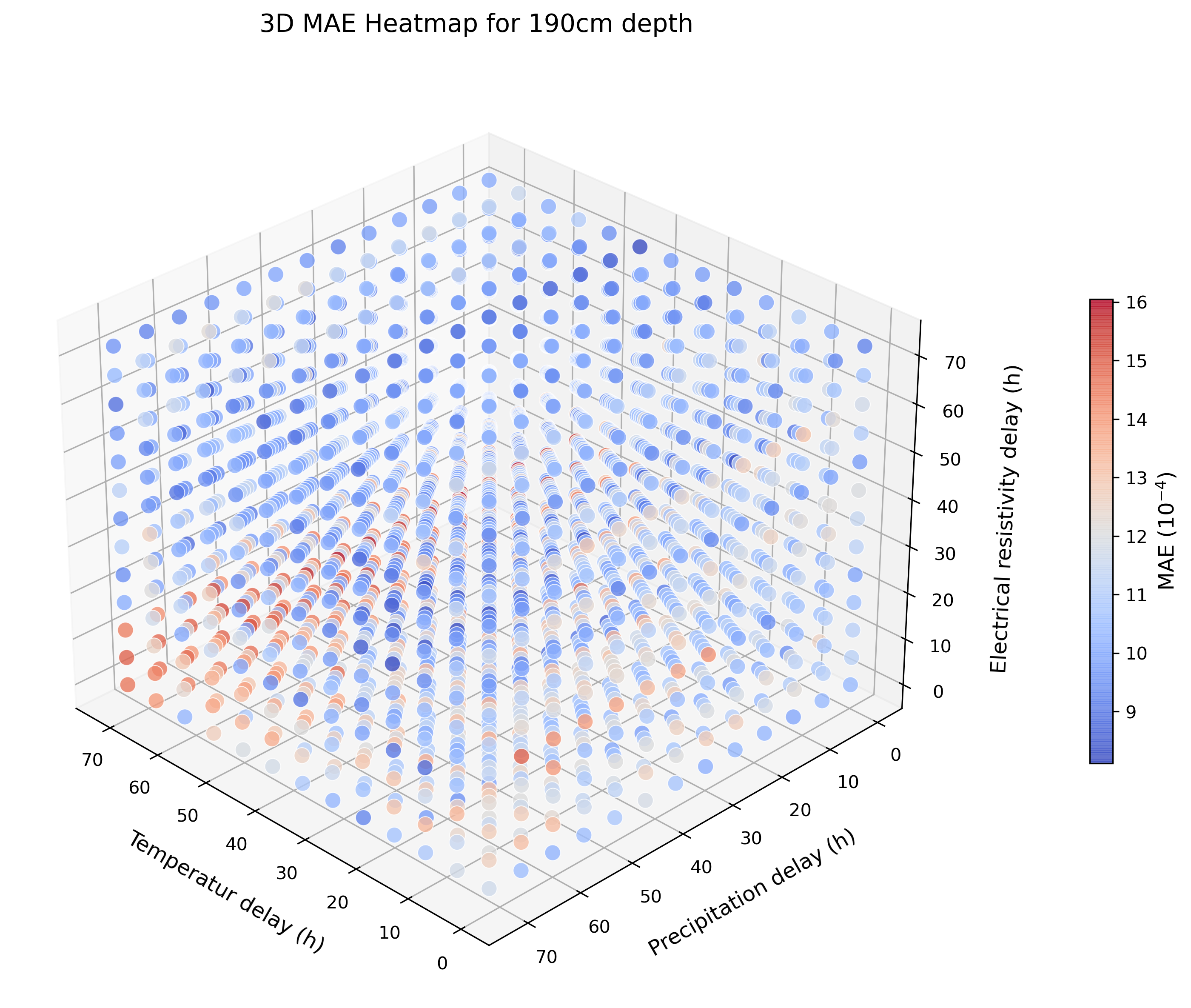}
        \caption[\acrshort{mae} depending on delays for 190~cm models]{\acrshort{mae} depending on delay times for \acrshort{er}, air temperature and precipitation at depth of 190~cm.}
        \label{fig:3d_190_delays}
    \end{figure}
    
    \begin{table}[H]
    \centering
    \caption[Delays for best VWC estimation]{Delay parameters for the best \acrshort{vwc} estimation, the electrical resistivity is left without any delay for all depths. For the delay time 'R' indicates electrical resistivity, 'T' stands for the air temperature and 'P' is precipitation. }
    \begin{tabular}{cccccc}
    \toprule
    Depth & 10~cm & 50~cm & 100~cm & 150~cm & 190~cm \\ \midrule
    $\text{Delay}_\text{R}$ (h) & 0 & 0 & 0 & 0 & 0 \\
    $\text{Delay}_\text{T}$ (h)  & 6 & 0 & 0 & 30 & 36 \\
    $\text{Delay}_\text{P}$ (h)  & 42 & 60 & 18 & 24 & 12 \\
    MAE ($10^{-4} cm^3 cm^{-3}$) & 26.9 & 31.1 & 30.2 & 15.7 & 8.6 \\ \bottomrule
    \end{tabular}

    \label{tab:best_est_delays}
    \end{table}
    
    \begin{table}[H]
    \centering
    \caption[Delays for best VWC 24~h prediction]{Delay parameters and mean absolute error  for the best \acrshort{vwc} 24~h forecast.}
    \begin{tabular}{cccccc}
    \toprule
    Depth & 10 cm & 50 cm & 100 cm & 150 cm & 190 cm \\ \midrule
    $\text{Delay}_\text{R}$ (h) & 24 & 48 & 30 & 30 & 54 \\ 
    $\text{Delay}_\text{T}$ (h)  & 24 & 24 & 42 & 36 & 66 \\
    $\text{Delay}_\text{P}$ (h)  & 24 & 54 & 24 & 42 & 24 \\
    MAE ($10^{-4} cm^3 cm^{-3}$) & 42.0 & 30.9 & 38.5 & 20.2 & 8.1 \\ \bottomrule
    
    \end{tabular}
    \label{tab:best_forecast_parameters}
    \end{table}
\end{minipage}

\begin{figure}[H]
    \centering
    \includegraphics[width=\linewidth]{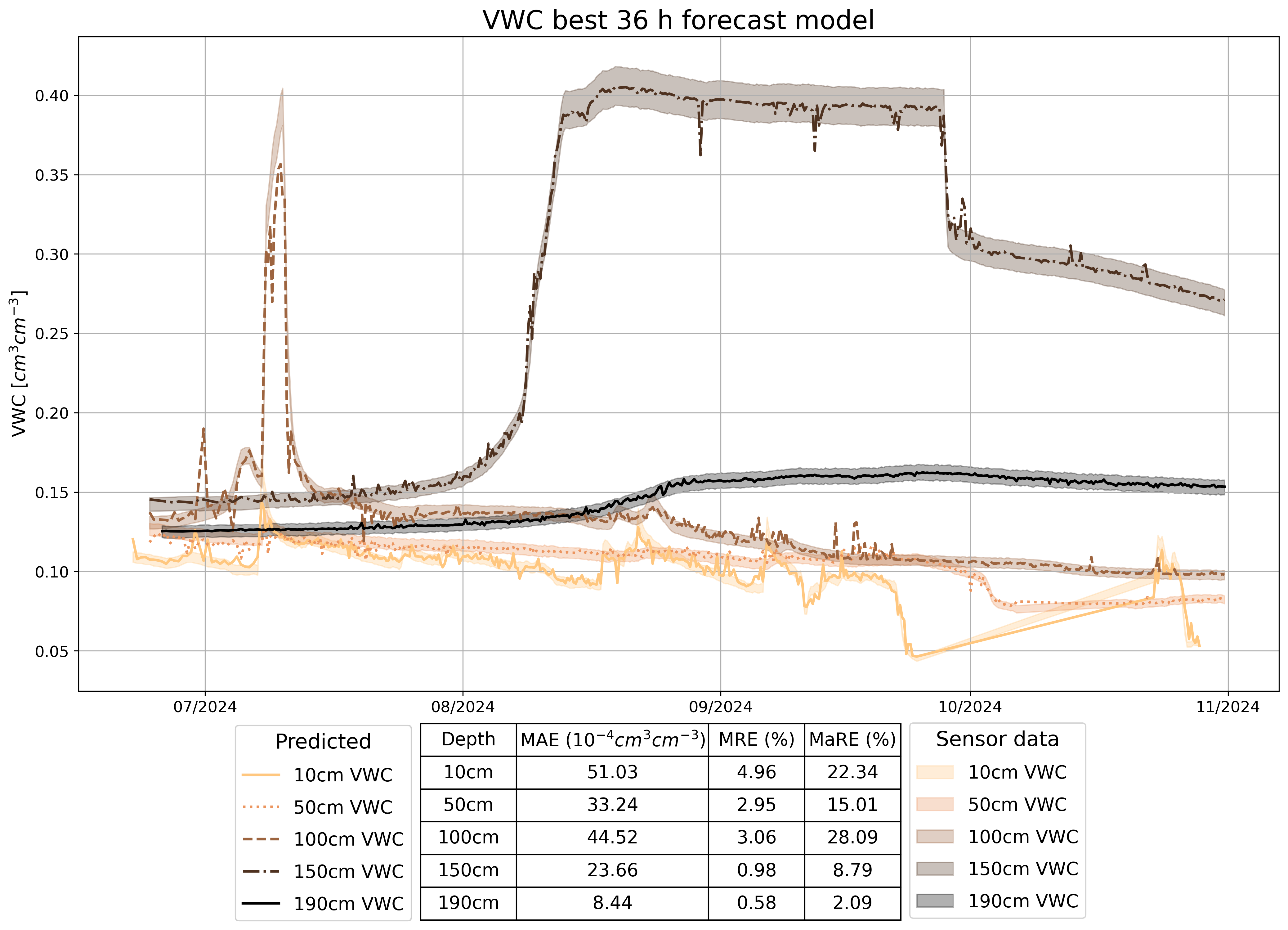}
    \includegraphics[width=\linewidth]{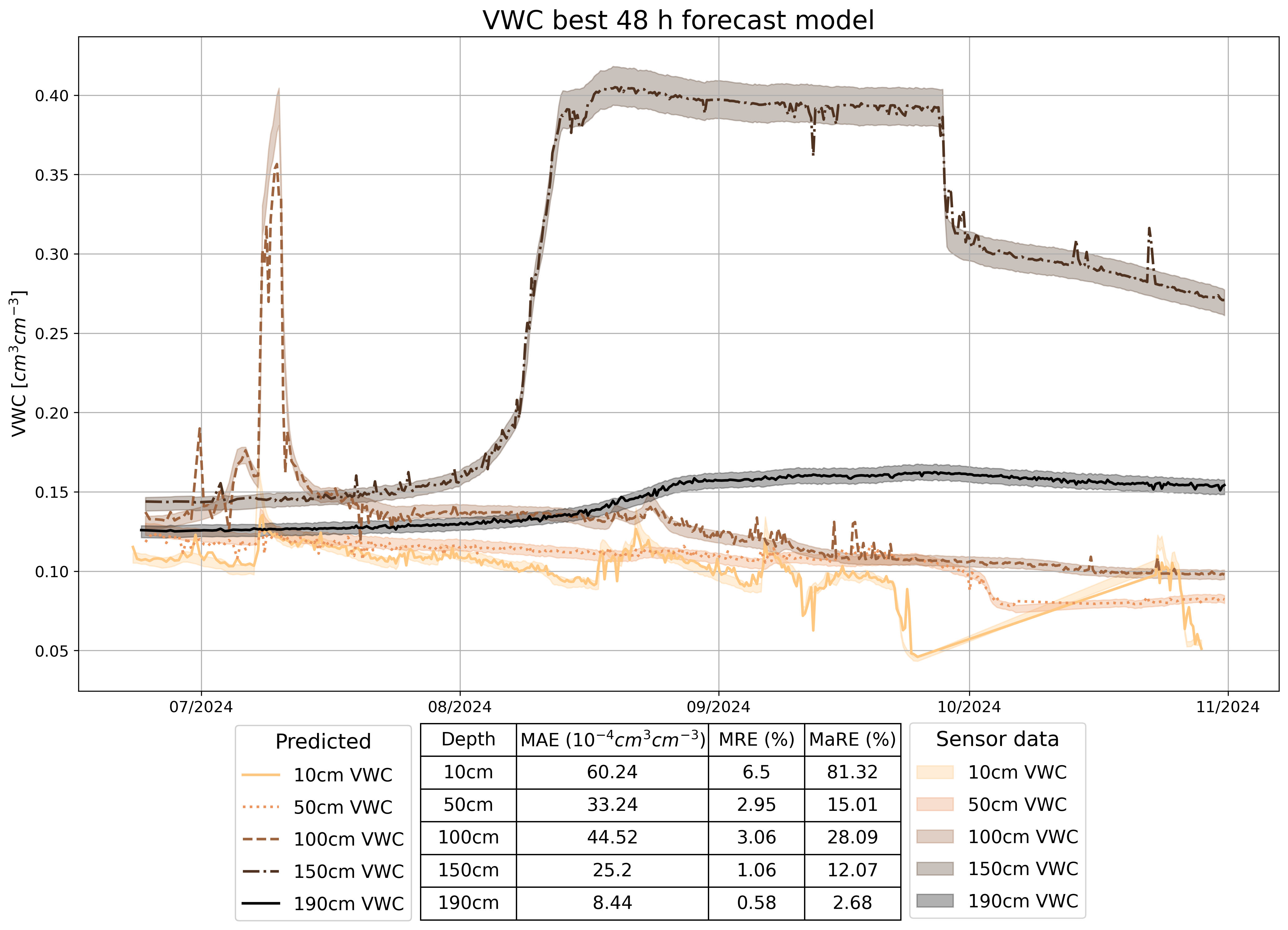}
    \caption[38/48~h forecasts]{Model and evaluation metrics of best 36~h and 48~h \acrshort{vwc} forecast.}
    \label{fig:ml_best_36h_48h_forecast}
\end{figure}

\begin{table}[H]
\centering
\caption[VWC comparison derived from sensors and ERT]{Upper part: Depth-dependent \acrshort{vwc} measurements, estimations and predictions from sensors and \acrshort{ert}; lower part: comparisons between the estimations and predictions to the true measured values with the \acrshort{vwc} sensors; est. means estimation and pred. describes the predictions.}

\begin{tabular}{c c c c c c c}
\toprule
\makecell{Depth \\ (cm)} & \makecell{\acrshort{vwc} \\ sensor  \\ ($cm^3 cm^{-3}$)} & \makecell{\acrshort{vwc} sensor \\ error\\ ($cm^3 cm^{-3}$)} & \makecell{\acrshort{er} sensor \\est.  \\($cm^3 cm^{-3}$)} & \makecell{\acrshort{er} sensor \\pred. \\($cm^3 cm^{-3}$)} & \makecell{\acrshort{ert} est. \\($cm^3 cm^{-3}$)} &  \makecell{\acrshort{ert} pred. \\($cm^3 cm^{-3}$)} \\
\midrule
10 & 0.1041 & 0.0031 & 0.1057 & 0.1085 & 0.1472 & 0.1263 \\
50 & 0.1102 & 0.0033 & 0.1096 & 0.1148 & 0.1280 & 0.1168 \\
100 & 0.1249 & 0.0037 & 0.1234 & 0.1215 & 0.2720 & 0.1524 \\
150 & 0.3977 & 0.0119 & 0.3982 & 0.3969 & 0.2948 & 0.3007 \\
190 & 0.1556 & 0.0047 & 0.1563 & 0.1574 & 0.1566 & 0.1573 \\
\bottomrule
\end{tabular}
\raggedleft

\begin{tabular}{c c c c c}
\multicolumn{5}{c}{Deviation to \acrshort{vwc} sensor} \\
\toprule
\makecell{Depth \\ (cm)} & \makecell{\acrshort{er} sensor \\ est. ($\sigma$)} & \makecell{\acrshort{er} sensor \\ pred. ($\sigma$)} & \makecell{\acrshort{ert} \\ est. ($\sigma$)} & \makecell{\acrshort{ert} \\ pred. ($\sigma$)} \\
\midrule
10 & 0.5 & 1.4 & 14 & 7 \\
50 & 0.18 & 1.4 & 5 & 2.0 \\
100 & 0.4 & 0.9 & 39 & 7 \\
150 & 0.04 & 0.07 & 9 & 8 \\
190 & 0.14 & 0.4 & 0.20 & 0.4 \\
\bottomrule
\end{tabular}
\label{tab:borehole_vwc_est_pred_dev}
\end{table}


\section{Short-term temporal variations} 
\label{AppendixD}

\begin{figure}[H]
    \centering
    \includegraphics[width=0.9\linewidth]{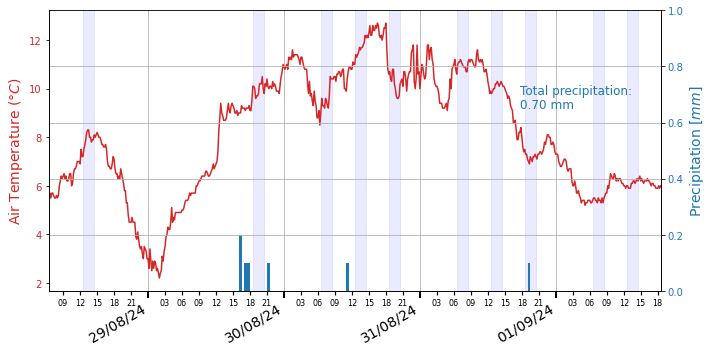}
    \caption[Meteorological parameters for survey one without rain]{Air temperature and precipitation between the measurement interval (28.08.-01.09.2024) for short-term temporal variations without heavy rainfall. The shaded regions symbolize the ERT measurement times. }
    \label{fig:temp_prec_st_ert}
\end{figure}

\begin{figure}[H]
    \centering
    \includegraphics[width=0.9\linewidth]{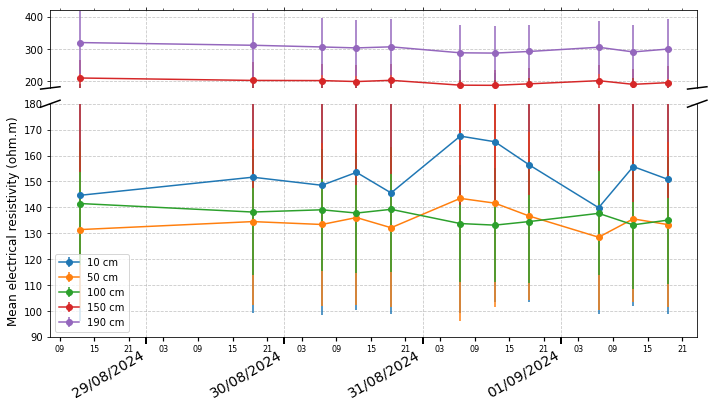}
    \caption[Mean ER development for survey one without rain]{The mean electrical resistivity development at the sensor points ($\pm 5 cm$) without heavy rainfall. }
    \label{fig:st_means}
\end{figure}

\begin{figure}[H]
    \centering
    \includegraphics[width=1\linewidth]{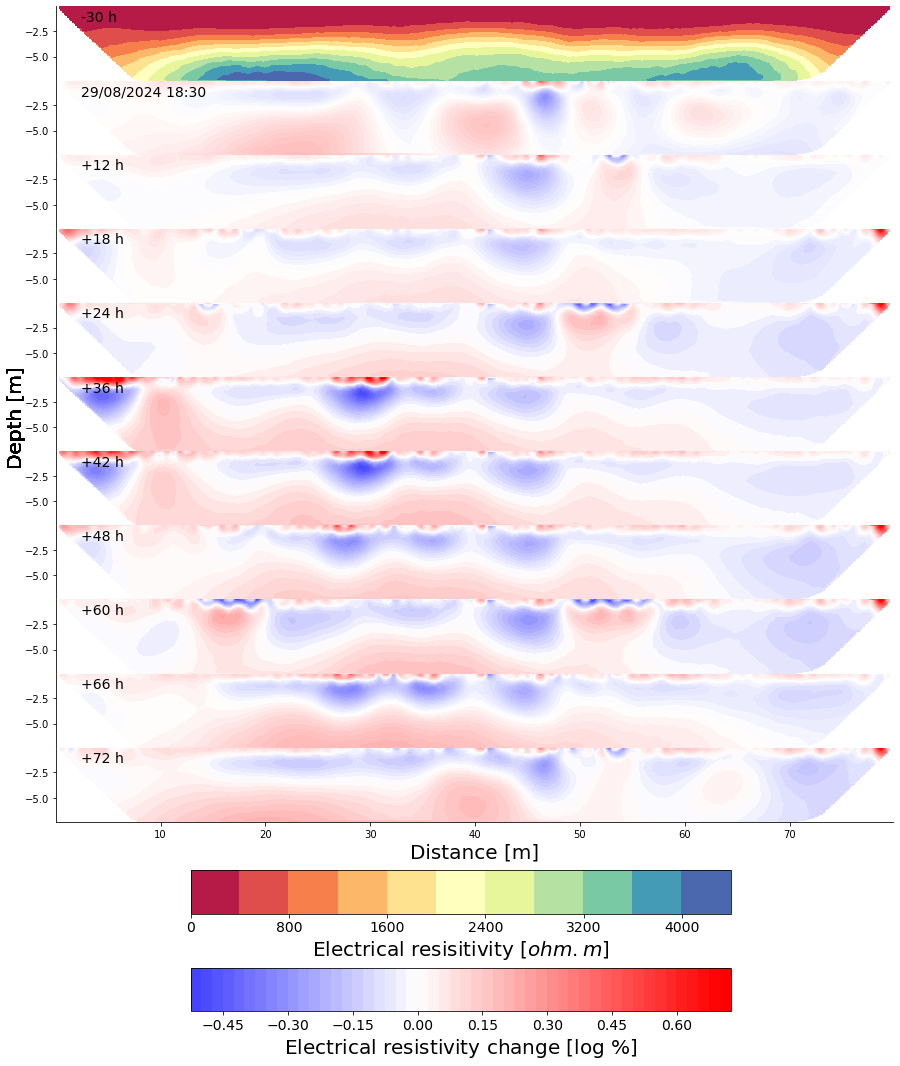}
    \caption[ERT development for survey one without rain]{Top: The acquired \acrshort{ert} as a reference measurement (Ref), Following: the logarithmic change ($log(ER)/Log(Ref)$) between subsequent measurements and the reference. }
    \label{fig:log_change_resis_st}
\end{figure}

\begin{figure}[H]
    \centering
    \includegraphics[width=1\linewidth]{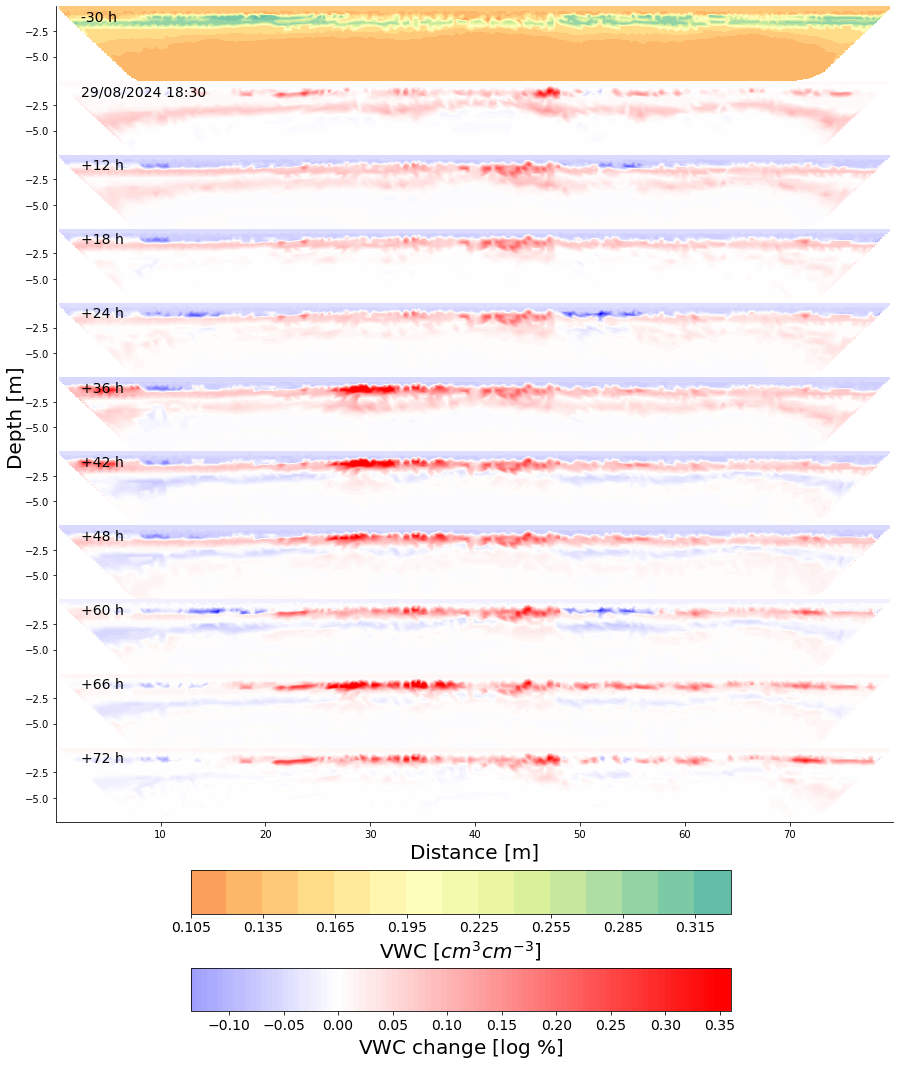}
    \caption[VWC development for survey one without rain]{Top: The \acrshort{vwc} estimation of the first \acrshort{ert} as a reference measurement (Ref), Following: the logarithmic change ($log(ER)/Log(Ref)$) between subsequent measurements and the reference. There is no heavy rainfall in between.}
    \label{fig:log_change_vwc_st}
\end{figure}

\begin{figure}[H]
    \centering
    \includegraphics[width=1\linewidth]{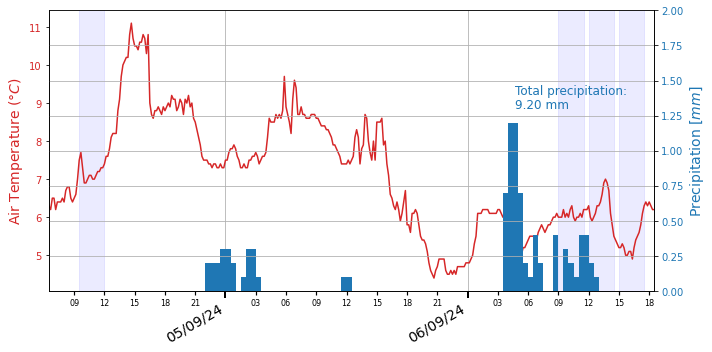}
    \caption[Meteorological parameters for survey one with rain]{Air temperature and precipitation between the measurement interval (04.09.-06.09.2024) for short-term temporal variations with heavy rainfall. }
    \label{fig:temp_prec_sst_ert}
\end{figure}

\begin{figure}[H]
    \centering
    \includegraphics[width=1\linewidth]{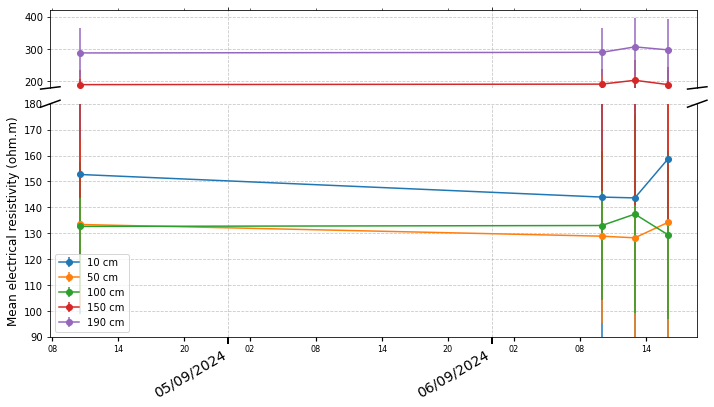}
    \caption[Mean ER development for survey one with rain]{The mean electrical resistivity development at the sensor points ($\pm 5cm$) with heavy rainfall. The shaded regions symbolize the ERT measurement times.}
    \label{fig:sst_means}
\end{figure}

\begin{figure}[H]
    \centering
    \includegraphics[width=0.9\linewidth]{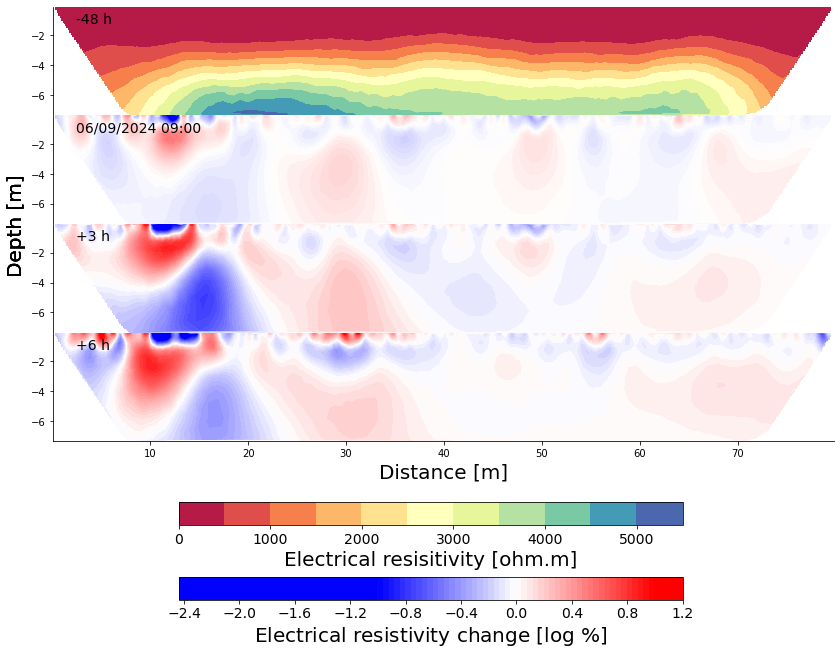}
    \includegraphics[width=0.9\linewidth]{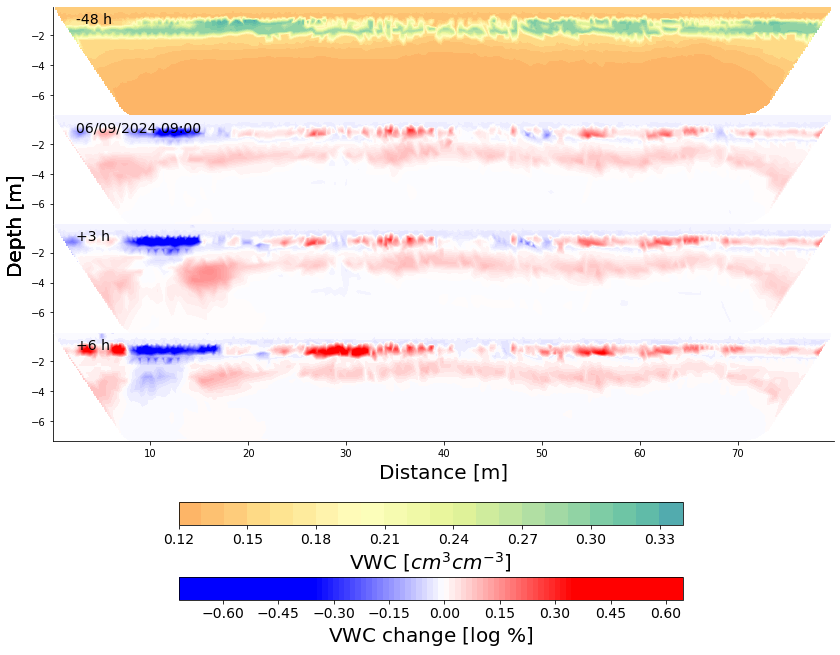}
    \caption[ERT and VWC development for survey one with rain]{Top: The acquired \acrshort{ert} as a reference measurement (Ref) and he logarithmic change ($log(ER)/Log(Ref)$) between subsequent measurements and the reference. Bottom: The \acrshort{vwc} estimation of the first \acrshort{ert} as a reference measurement (Ref) and the logarithmic change ($log(VWC)/Log(Ref)$) between subsequent measurements and the reference. There is heavy rainfall in between.}
    \label{fig:log_change_resis_sst}
\end{figure}

\section{Spatial variations}

\begin{figure}[H]
    \centering
    \includegraphics[width=\linewidth]{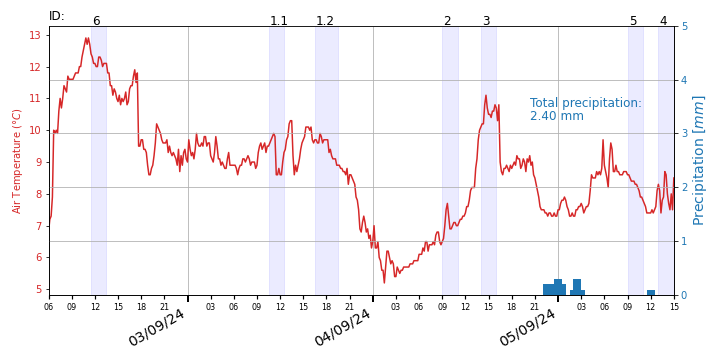}
    \caption[Meteorological parameters for survey two]{Air temperature and precipitation between the measurement interval (02.09.- 05.09.2024) for spatial variations. The IDs and shaded areas depict respective duration ERT recording duration.}
    \label{fig:prec_temp_locations}
\end{figure}

\section{Long-term temporal variations}

\begin{figure}[H]
    \centering
    \includegraphics[width=0.9\linewidth]{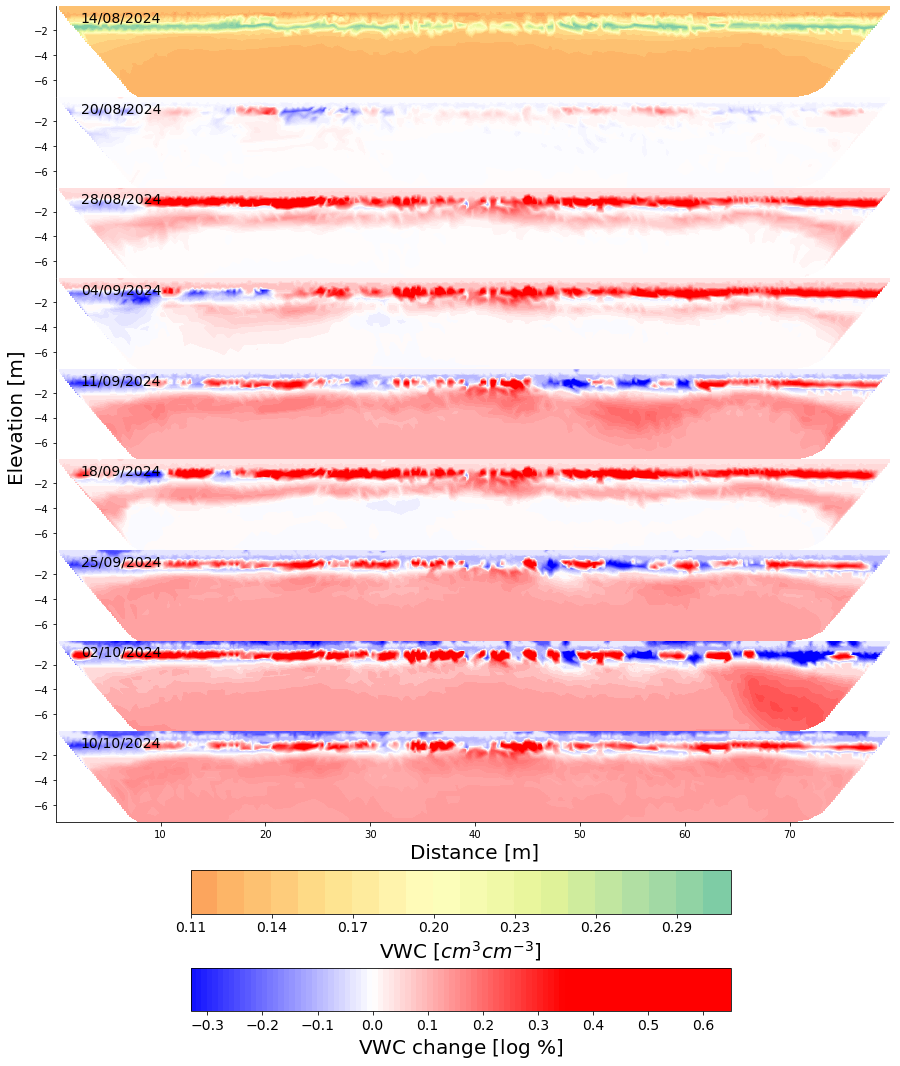}
    \caption[VWC development for survey three]{The \acrshort{vwc} estimation of the first \acrshort{ert} as a reference measurement (Ref) and the logarithmic change ($log(VWC)/Log(Ref)$) between subsequent measurements and the reference.}
    \label{fig:long_vwc_changes}
\end{figure}